\documentclass[aps,prb,amsmath,amssymb,footinbib,showpacs,twocolumn,superscriptaddress,longbibliography]{revtex4-2}
\usepackage{amsmath}
\usepackage{amssymb}
\usepackage{amsthm}
\usepackage{setspace}
\usepackage{graphicx}
\usepackage{braket}
\usepackage{mathrsfs}
\usepackage{float}
\usepackage[colorlinks = true,linkcolor = red,urlcolor  = blue,citecolor = blue,anchorcolor = blue]{hyperref}
\usepackage[utf8]{inputenc}
\usepackage[english]{babel}
\usepackage{bm}
\usepackage[caption=false]{subfig}

\begin{document}

\title{Stabilizing topological superconductivity in disordered spin-orbit coupled semiconductor-superconductor heterostructures}

\author{Binayyak B. Roy}
\affiliation{Department of Physics and Astronomy, Clemson University, Clemson, SC 29634, USA}

\author{R. Jaiswal}
\affiliation{Department of Physics, University of California, Santa Barbara, CA 93106, USA}

\author{Tudor D. Stanescu}
\affiliation{Department of Physics and Astronomy, West Virginia University, Morgantown, WV 26506, USA}

\author{Sumanta Tewari}
\affiliation{Department of Physics and Astronomy, Clemson University, Clemson, SC 29634, USA}

\begin{abstract}
We investigate theoretically a one-dimensional semiconductor-superconductor (SM-SC) heterostructure with Rashba spin-orbit coupling and parallel Zeeman field in the presence of disorder generated by random charged impurities and identify the optimal regimes for realizing topological superconductivity and Majorana zero modes. Using a Green's function approach, we show that upon increasing the disorder strength the stable topological superconducting phase characterized by robust end-to-end Majorana correlations ``migrates'' toward larger values of the Zeeman field and can be stabilized by increasing the effective SM-SC coupling. Based on these findings, we propose a strategy for accessing a regime characterized by well-separated Majorana zero modes that is based on (a) enhancing the strength of the effective SM-SC coupling (e.g., through interface engineering) and (b) expanding the range of accessible Zeeman fields (e.g., by enhancing the gyromagnetic ratio or optimizing the parent superconductor, to enable the application of larger magnetic fields). While this strategy may still require some reduction of the disorder strength, this requirement is significantly less strict than the corresponding requirement in a strategy that focuses exclusively on disorder reduction. 
\end{abstract}

\maketitle

\section{Introduction}

Majorana fermions were first introduced in high-energy physics \cite{Majorana1937} as an exotic class of quantum particles that are indistinguishable from their own anti-particles. In condensed matter physics, the analog of Majorana fermions, the so-called Majorana zero modes (MZMs), were proposed as building blocks for topological quantum computation (TQC) \cite{Kitaev2003, Nayak2008} due to their ability to encode quantum information non-locally and because one can exploit their fundamentally new type of particle statistics --- the so-called non-Abelian statistics \cite{wilczek1982quantum, Moore1991, Read2000, Nayak1996} --- to perform topologically protected gate operations. Despite being rather exotic quasiparticles, it was proposed that MZMs could be realized in a conceptually simple and experimentally realizable heterostructure consisting of a (quasi) one-dimensional (1D) spin-orbit coupled semiconductor (SM) proximity-coupled to an ordinary $s$-wave superconductor (SC) in the presence of a parallel Zeeman field \cite{sau2010generic, sau2010non, oreg2010helical, lutchyn2010majorana}. Above a certain critical field, the system is predicted to enter a topological superconducting (TSC) phase that supports mid-gap Majorana zero modes localized at the ends of the 1D structure. Because of its conceptual simplicity and its potential for transformative technological applications, the proposal has spurred tremendous experimental activity in the past few years \cite{mourik2012signatures, Deng2012, Das2012, rokhinson2012fractional, churchill2013superconductor, finck2013anomalous, deng2016majorana, zhang2017ballistic, chen2017experimental, nichele2017scaling, albrecht2017transport, o2018hybridization, shen2018parity, 
sherman2017normal, vaitiekenas2018selective, albrecht2016exponential, Yu_2021, zhang2021, kells2012near, PhysRevB.107.245423}. 

Despite extensive experimental efforts and significant progress, it has now become increasingly clear that the practical implementation of this proposal (which is theoretically straightforward and should work under the ``prescribed conditions'') faces serious challenges due to the presence of disorder.  
On the one hand, disorder induces trivial low-energy states that mimic the phenomenology of MZMs \cite{Mi2014, bagrets2012class, pikulin2012zero, prada2012transport, pan2020physical, moore2018two, Moore2018, vuik2018reproducing, stanescu2019robust, added_Loss_2018prb_abs, san2016majorana, ramon2019nonhermitian, Jorge2019supercurrent, ramon2020from, Jorge2021distinguishing}, which can render the unambiguous detection of topological superconductivity and  MZMs experimentally challenging. On the other hand, strong-enough disorder may even lead to the disappearance of the TSC phase from the experimentally accessible region of the phase diagram. 
In addition, in finite systems a combination of disorder and finite size effects can destabilize the MZMs even in a parameter regime that would be consistent with the presence of TSC in the thermodynamic limit. More specifically, in certain conditions (e.g., weak SM-SC coupling) disorder-induced low-energy states (that would be localized in a long-enough system) can have characteristic length scales comparable to the wire length \cite{sarma2023spectral}. In turn, these effectively delocalized disorder-induced states couple to the Majorana bound states at both ends of the system, ``removing'' them from zero energy. 
Unfortunately, the most likely current experimental situation appears to involve strongly disordered SM-SC nanostructures. Recent extensive numerical calculations suggest that, to demonstrate unambiguously the realization of topological superconductivity (and MZMs) in SM-SC hybrid wires, the disorder strength to superconducting gap ratio must be substantially reduced \cite{sarma2023spectral}. While we leave open the possibility of success in unambiguously observing the TSC phase by progressive improvement of the sample quality, in this paper we carry out an in-depth theoretical analysis of the Majorana physics in the presence of disorder to investigate what other key factors could (and perhaps should) be leveraged to defeat the dominance of the ubiquitous disorder-induced low energy states and help revealing the underlying MZMs.  

We carry out our calculations starting with a model Hamiltonian that describes the (coupled) semiconductor and superconductor subsystems. The low-energy properties of the heterostructure are investigated using an effective Green's function for the semiconductor, which is derived by ``integrating out'' the degrees of freedom associated with the SC. Using this Green's function, we calculate the zero-energy (total) density of states (DOS) and the local density of states (LDOS) in the SM wire. In addition, we introduce a new (experimentally accessible) quantity  that measures the end-to-end correlations of the low-energy LDOS peaks near the two ends of the wire, which serves as a useful marker of the topological phase. We note that this quantity can be determined experimentally based on differential conductance measurements at the ends of the hybrid system.
Based on the dependence of these quantities on various control (e.g., chemical potential, $\mu$, and Zeeman field, $\Gamma$) and system parameters (e.g., effective SM-SC coupling and disorder strength), we conclude that for a heterostructure with a strongly disordered SM wire, enhancing the coupling to the SC is generically beneficial for stabilizing a robust TSC phase. In essence, a stronger SM-SC coupling results in a weaker ``effective disorder,'' due to proximity-induced energy renormalization \cite{PhysRevB.96.014510}. In addition, enhancing the SM-SC coupling reduces the characteristic length scales of the low-energy states, which alleviates the finite size effects. We also find that, with increasing the strength of the disorder potential, the topologically robust regions move to areas of the phase diagram characterized by larger values of $\Gamma$ and $\mu$. These results suggest that, in addition to systematic efforts to reduce the disorder strength, a successful strategy for realizing robust topological superconductivity (and MZMs) should involve (a) optimizing the effective SM-SC coupling (e.g., by interface engineering) and (b) expanding the range of accessible Zeeman fields (e.g., by enhancing the gyromagnetic ratio or optimizing the parent superconductor, to enable the application of larger magnetic fields). This strategy can provide experimental access to robust topological superconducting regions after a relatively modest reduction of the disorder strength.  Accessing such regimes would enable the unambiguous demonstration of MZMs and would provide conditions for performing meaningful braiding experiments.

The remainder of the paper is organized as follows: In Sec. II we introduce the model Hamiltonian and describe the Green's function approach. In Sec. III we discuss in detail our numerical results showing the dependence of the density of states (DOS), local density of states (LDOS), and end-to-end correlation function on the control parameters -- chemical potential $\mu$ and Zeeman field $\Gamma$ -- for different disorder scenarios: clean heterostructure (which is used as a benchmark), disorder only in the semiconductor, disorder only in the superconductor, and disorder in both the semiconductor and the superconductor. We consider different disorder amplitudes and effective SM-SC coupling strengths. We illustrate our key findings by calculating the position (and energy) dependence of the LDOS, determining end-to-end correlation maps of the relevant parameter space, as well as generating representative cuts of the DOS as a function of energy and Zeeman field. In Sec. IV we provide a summary of the main results and present our conclusions.      

\section{Tight-binding model and Green's function formalism} 

\begin{figure}[t]
    \subfloat{%
    \includegraphics[clip,width=\columnwidth]{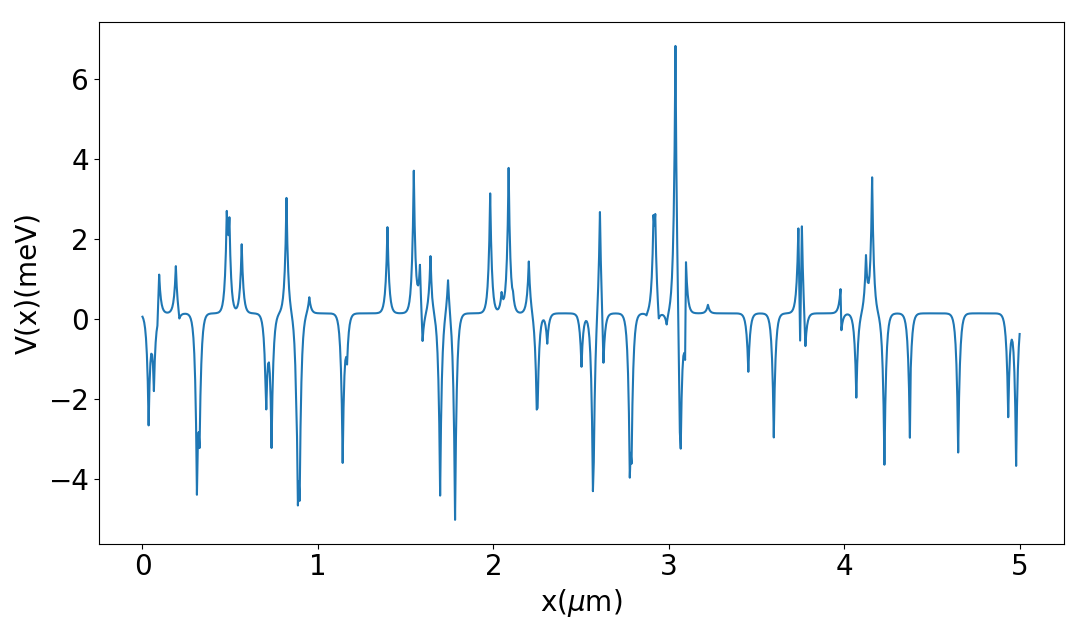}%
    }
    \hspace{0.6cm}
    \subfloat{%
    \includegraphics[clip,width=\columnwidth]{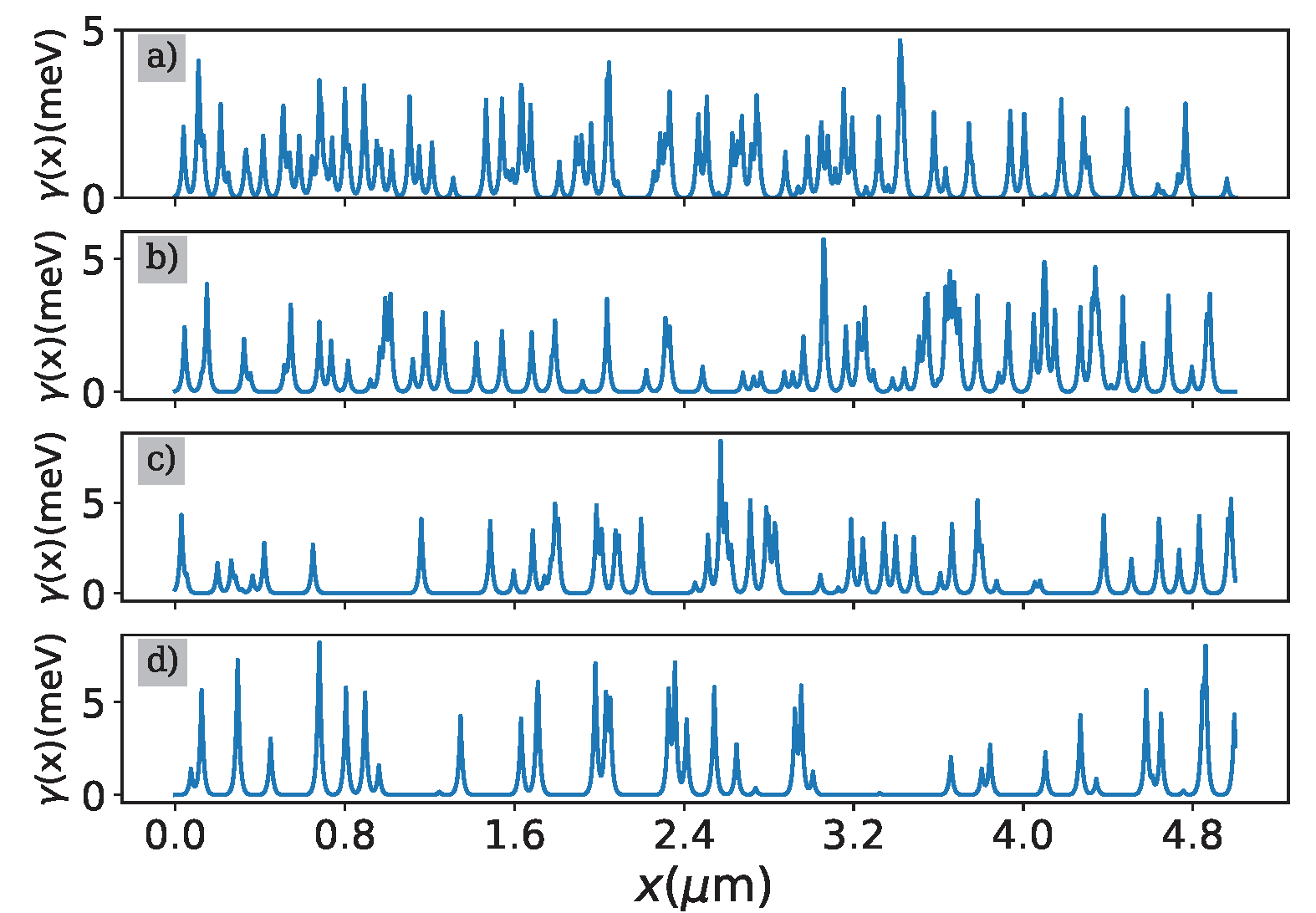}%
    }
\caption{Disorder profiles used in the calculation. {\em Top}: Normalized semiconductor disorder potential with amplitude $\sqrt{\langle V(x_i)^2 \rangle} = 1~$meV. The disorder amplitude is controlled by multiplying the disorder profile by an overall factor $V_0$. {\em Bottom}: Position dependent effective SM-SC coupling, $\gamma(x_i) = \Lambda\Tilde{g}(0,i)$, generated by disorder in the parent superconductor  (see Eq. \ref{eq:self_energy}). The disorder profiles are characterized by the same value of the average coupling, $\langle \gamma(x_i) \rangle = 0.6$ meV, but correspond to different values of $\langle \gamma(x_i)^2\rangle$: 0.92, 1.16, 1.42 and 1.86 meV$^2$ (from top to bottom).}   \label{fig:SM_disorder_profile}
\end{figure}

We model the spin-orbit coupled semiconductor-superconductor hybrid system in the presence of short-range correlated disorder (inside the SM and/or the SC) and parallel Zeeman field using a one-dimensional (1D) effective Hamiltonian of the form
\begin{equation}\label{eq:Hamiltonian}
    \mathcal{H} = H_{SM} + H_{SC} + H_{SM-SC}, 
\end{equation}
where $H_{SM}$ and $H_{SC}$ are terms describing the semiconductor wire and superconductor, respectively and $H_{SM-SC}$ describes the coupling between the semiconductor wire and the parent superconductor. The 1D tight-binding model describing the semiconductor wire in the presence of disorder is given by the second quantized Hamiltonian
\begin{flalign}\label{eq:secondQuant}
   H_{SM} = & \nonumber \sum_{i,\sigma} \left[-t_{SM}(\hat{a}_{i\sigma}^\dagger\hat{a}_{i+1\sigma} + h.c.) + (V(x_i) - \mu)\hat{a}_{i\sigma}^\dagger\hat{a}_{i\sigma} \right] & \\ 
    & \nonumber + \dfrac{\alpha}{2} \sum_i\left[(\hat{a}_{i\uparrow}^\dagger\hat{a}_{i+1\downarrow} - \hat{a}_{i\downarrow}^\dagger\hat{a}_{i+1\uparrow} + h.c. \right] & \\ 
    & + \Gamma\sum_i(\hat{a}_{i\uparrow}^\dagger\hat{a}_{i\downarrow} + h.c.)
\end{flalign}
where $i$ labels the sites of the 1D lattice with lattice constant $a$, $\sigma$ denotes the electron spin, $t_{SM} = \hbar^2/2m^* a^2$ is the amplitude of the nearest-neighbor hopping, $V(x_i)$ is the disorder potential, $\mu$ is the chemical potential, $\alpha$ is the Rashba spin orbit coupling, and $\Gamma$ is the Zeeman (half) splitting generated by an external field applied along the direction of the wire. 
The random disorder potential is generated using the ansatz $V(x_i) = \sum_n A_n e^{{-\lvert x_i-\xi_n \rvert}/{\lambda_0}}$, where $\xi_n$ gives the (random) position along the wire of an impurity that produces a contribution to the disorder potential of (random) amplitude $A_n$ that decays exponentially with a decay length $\lambda_0$. After generating a disorder profile, we subtract an overall constant, so that $\langle V(x_i) \rangle = 0$, and normalize it so that ${\langle V(x_i)^2\rangle}=1$. To control the amplitude of the disorder potential, in the numerical calculations we multiply the normalized disorder profile by an overall amplitude $V_0$. Hence, the disorder strength is $V_0= \sqrt{\langle V(x_i)^2 \rangle}$. An example of normalized disorder profile is shown in the top panel of Fig. \ref{fig:SM_disorder_profile}. 
The parameter values used in our numerical calculations are: lattice constant $a = 5.0$ nm with $N=1000$ lattice points (resulting in a 5 $\mu$m long wire), effective mass $m^* = 0.023m_0$, hopping $t_{SM}$ = 66.26 meV and spin-orbit coupling strength $\alpha \cdot a$ = 250 meV $\cdot \textup{~\AA}$.

The superconducting component of the hybrid structure is described at the mean-field level by the tight binding Hamiltonian 
\begin{flalign}
    H_{SC} = \sum_{\langle i,j \rangle, \sigma} \left(t_{ij} - \mu\delta_{ij}\right)\hat{c}_{i\sigma}^\dagger\hat{c}_{j\sigma} + \Delta_0\sum_i\left(\hat{c}_{i\uparrow}^\dagger\hat{c}_{i\downarrow}^\dagger + \hat{c}_{i\uparrow}\hat{c}_{i\downarrow}\right),
\end{flalign}
where $i$ and $j$ label nearest neighbor SC lattice sites, $\hat{c}_{i\sigma}^\dagger$ is the creation operator for a fermionic state with spin $\sigma$, and $\Delta_0$ is the pairing potential of the parent superconductor. Finally, the SM-SC coupling Hamiltonian is given by
\begin{equation}
    H_{SM-SC} = \sum_{\langle i,j \rangle}\sum_{\Tilde{\sigma}\sigma} \left[ t_{i,j}^{\Tilde{\sigma}\sigma}\hat{a}_{i\Tilde{\sigma}}^\dagger\hat{c}_{j\sigma} + h.c. \right],
\end{equation}
where $i$ and $j$ label lattice sites near the SM-SC interface, $\Tilde{\sigma}$ and $\sigma$ are spin quantum numbers associated with the SM and the SC states, respectively, and $t_{i,j}^{\Tilde{\sigma}\sigma}$ is the effective hopping matrix element between the lattice sites of the semiconductor and the superconductor on the two sides of the SM-SC interface.

We are interested in the low energy physics of the SM nanowire, taking into account the proximity effects induced by the superconductor. Consequently, we integrate out the SC degrees of freedom and derive an effective Green's function for the semiconductor \cite{PhysRevB.106.085429}. The proximity effect of the superconductor is incorporated as a self-energy contribution, $\Sigma_{SC}$:
\begin{equation}\label{eq:Greensfunc}
    G_{SM}(\omega) = \left[\omega~\!I_{4N} - H_{SM} - \Sigma_{SC}(\omega)\right]^{-1}.
\end{equation}
The self-energy is given by,
\begin{equation}
\Sigma_{\sigma\sigma'}(r,r',\omega) = \sum_{r_1,r_2}\Tilde{t}(r,r_1)G_{\sigma\sigma'}(r_1,r_2,\omega)\Tilde{t}^\dagger(r_2,r'),
\end{equation}
where $\Tilde{t}(r,r_1)$ is a matrix describing the tunneling between the SM and the SC and $G_{\sigma\sigma'}(r_1,r_2,\omega)$ is the Green's function of the superconductor at the SM-SC interface. 
Note that no approximations are made in deriving Eq. \ref{eq:Greensfunc}; it models the system at the same level as the initial full Hamiltonian in Eq. \ref{eq:Hamiltonian}.
We use the simplest form of the tunneling matrix elements $\Tilde{t}(r,r_1) = \Tilde{t}\delta(z)\delta(z_1)\delta(r^{||} - r^{||}_1)$, with $r^{||}$ and $z$ denoting in-plane and out-of-plane coordinates. Explicitly, we have \cite{PhysRevB.84.144522, PhysRevB.82.214509}
\begin{equation}\label{eq:self_energy}
    \Sigma_{SC}(i) = -\Lambda \dfrac{\Tilde{g}(\omega;i)}{\sqrt{\Delta_0^2 - \omega^2}}(\omega\sigma_0\tau_0 + \Delta_0\sigma_y\tau_y)
\end{equation}
where $\sigma_{\mu}$ and $\tau_{\mu}$ ($\mu =  x,y,z$) are Pauli matrices acting in the spin and particle-hole spaces, respectively. Here, $\Lambda = \pi\nu_f^{\infty}\Tilde{t}^2$ where $\nu_f^{\infty}$ is the interface density of states at the Fermi energy for an infinitely thick superconductor. We define the effective SM-SC coupling as 
$\gamma(i) = \Lambda\Tilde{g}(\omega=0,i)$. Note that the superconductor disorder generates a position-dependent Green's function, which, in turn, induces a position-dependent effective SM-SC coupling. Also note that in the presence of disorder the interface Green's function becomes quasi-local
\cite{PhysRevB.106.085429}. Here, we work within a local approximation that neglects the short-range contributions to the effective SM-SC coupling, i.e., we assume $\gamma(i,j) = \delta_{ij}\gamma(i)$. 
We model the effective SM-SC coupling induced by the disordered superconductor as a position-dependent function with random fluctuations.
The strength of the SC disorder is controlled by amplitude of these fluctuations, for a given average value of $\gamma(i)$. More specifically, we consider different spatial profiles characterized by a constant 
$\langle \gamma(x_i) \rangle$ and different values of $\sqrt{\langle\gamma^2\rangle}$, as shown in the lower panel of Fig. \ref{fig:SM_disorder_profile}. We characterize the effective strength of the superconducting disorder using the dimensionless parameter $\gamma_s = \sqrt{\langle\gamma^2\rangle}/\langle\gamma\rangle$. 

In Eq. (\ref{eq:self_energy}), the parameter $\Delta_0=0.3~$meV represents the pairing potential of the parent SC. The ratio between $\gamma = \langle \gamma(i) \rangle$ and $\Delta_0$ determines the strength of the SM-SC coupling, with $\gamma/\Delta_0 > 1$  representing the strong coupling regime and $\gamma/\Delta_0 < 1$ corresponding to weak coupling regime. We note that in Eq. \ref{eq:self_energy} the diagonal term proportional to $\omega$ is responsible for energy renormalization, while the off-diagonal contribution proportional to $\Delta_0$ is responsible for the induced pairing \cite{sau2010non, PhysRevB.82.094522, PhysRevB.83.094525, PhysRevB.81.241310, PhysRevB.84.075142, Grein_2012, PhysRevB.85.064512}. In this paper we study in detail how the parameter $\gamma$ (i.e., the effective SM-SC coupling) affects the low-energy physics of the hybrid wire. 

To calculate the SM Green's function $G_{SM}(\omega)$, one can directly invert the $4N \times 4N$ matrix from Eq. \ref{eq:Greensfunc}. However, for large systems this brute force method is numerically expensive. To address this issue, we use a recursive Green's function approach \cite{1985ZPhyB..59..385M, lewenkopf2013recursive} that involves 3N inversions of $4 \times 4$ matrices. We divide the system into nearest-neighbor-coupled slices and define the auxiliary left, $G^{L}$, and right, $G^{R}$, Green's functions with on-slice values $G^{L(R)}_i$
\begin{equation}\label{eq:recur_green}
    G^{L(R)}_i = \left[g^{-1}(i) - \Sigma^{L(R)}_i\right]^{-1},
\end{equation}
\begin{equation}\label{eq:recur_left}
    \Sigma^L_i = \mathcal{T}^TG^L_{i-1}\mathcal{T},
\end{equation}
\begin{equation}\label{eq:recur_right}
    \Sigma^R_i = \mathcal{T}G^R_{i+1}\mathcal{T}^T.
\end{equation}
For a system with open boundary condition, the inter slice coupling, $\mathcal{T}$ is given by
\begin{equation}
    \mathcal{T} = -t\sigma_0\tau_z + \dfrac{i\alpha}{2}\sigma_y\tau_z.
\end{equation}
The local inverse Green's function $g^{-1}(i)$ in Eq. \ref{eq:recur_green} can be split up into position independent and position dependent contributions,
\begin{equation}\label{eq:local_inverse}
    g^{-1}_i = g^{-1}_0(\omega) - V_{dis}(i)\sigma_0\tau_z.
\end{equation}
Explicitly, the position independent contribution  is given by
\begin{flalign}
    g^{-1}_0(\omega) = & \nonumber\omega\left(1 + \dfrac{\gamma}{\sqrt{\Delta^2_0 - \omega^2}}\right)\sigma_0\tau_0 - (\epsilon_0 - \mu)\sigma_0\tau_z & \\ 
    & - \Gamma\sigma_x\tau_z + \dfrac{\gamma\Delta_0}{\sqrt{\Delta^2_0 - \omega^2}}\sigma_y\tau_y,
\end{flalign}
where $\epsilon_0 = 2t\cos{(\pi/(N+1))}$ is used to define the chemical potential with respect to the bottom of the SM band. The auxiliary left Green's function is calculated using Eqs. \ref{eq:recur_green} and \ref{eq:recur_left} starting with the slice $i = 1 $ and $\Sigma^L_1 = 0$. Similarly, we calculate the auxiliary right Green's function using Eqs. \ref{eq:recur_green} and \ref{eq:recur_right}, starting with $i = N$ and $\Sigma^R_N = 0$. Finally, we use the local inverse Green's function in Eq. \ref{eq:local_inverse} and the auxiliary left and right self-energies to calculate the diagonal elements of the SM Green's function. Specifically, for lattice site $i$ we have  
\begin{equation}\label{eq:onsite_green}
    [G_{SM}]_{ii} = [g^{-1}(i) - \Sigma^L_i - \Sigma^R_i]^{-1}. 
\end{equation}

Consider now an infinitely long,  clean system. The corresponding effective SM Green's function can be written in terms of a wave vector $k$, which is a good quantum number in this case. Specifically, after we Fourier transform the (real space) Green's function we have
\begin{flalign}\label{eq:kspace_green}
    G^{-1}_{SM}(\omega,k) = & \omega\left(1 + \dfrac{\gamma}{\sqrt{\Delta^2_0 - \omega^2}}\right) + \dfrac{\gamma\Delta_0}{\sqrt{\Delta^2_0 - \omega^2}}\sigma_y\tau_y & \\
    & \nonumber - \left[\xi(k) + \Gamma\sigma_x + \alpha\sin{(k)}\sigma_y\right]\tau_z, 
\end{flalign}
where $\xi(k)=2t_{SM}(1-\cos k)-\mu$ is the bare SM energy band in the absence of spin-orbit coupling and Zeeman field.
The low energy states of the hybrid system are given by the poles of the effective Green's function in Eq. \ref{eq:kspace_green} and can be obtained by solving the equation det$[G_{eff}(\omega,k)] = 0$. For $k=0$, we have \cite{PhysRevB.96.014510}
\begin{equation}\label{eq:lowest_energy}
    \omega\left(1+\dfrac{\gamma}{\sqrt{\Delta_0^2 - \omega^2}}\right) = \left(\pm\right)\sqrt{{\mu}^2 + \dfrac{\gamma^2\Delta_0^2}{\Delta_0^2-\omega^2}} \pm \Gamma,
\end{equation}
where the chemical potential is measured from the bottom of the band (in absence of SOI). The solution to this equation at $\omega = 0$ gives the critical Zeeman field at the topological quantum phase transition (TQPT) \cite{PhysRevB.105.205122},
\begin{equation}\label{critical_zeeman}
    \Gamma_c = \sqrt{\gamma^2 + {\mu}^2}.
\end{equation}
In the presence of a (smooth) non-homogeneous potential $V(x)$, we can generalize Eq. \ref{critical_zeeman} and write the {\em local} topological condition,
\begin{equation}\label{critical_zeeman_local}
    \Gamma \ge \sqrt{\gamma^2 + ({\mu}-V(x_i))^2}.
\end{equation}
When this local condition is satisfied within a segment of length comparable to (or larger than) the coherence length, we expect the emergence of (partially overlapping) Majorana bound states with maxima near the ends of this segment. Re-arranging the terms, we can write the local topological condition as 
\begin{equation}\label{eq:TQPT_condn}
[V^+(x_i) - {\mu}][{\mu} - V^-(x_i)] \ge \gamma^2,
\end{equation}
where $V^\pm(x_i) = V(x_i) \pm \Gamma$. Qualitatively, this condition is also relevant in the presence of disorder, with $V(x_i)$ being a {\em smooth} disorder potential that includes only the low-$k$ components of the actual disorder potential. 

Note that in a minimal model that does not explicitly incorporate the proximity-induced renormalization, the topological condition is expressed in terms of the induced pairing potential, $\Delta_{ind}$, as $\Gamma_c = \sqrt{\Tilde{\mu}^2 + \Delta_{ind}^2}$. In our model, $\Delta_{ind}$ is given by the minimum with respect to $k$ of the lowest energy solution of the equation det$[G_{SM}(\omega,k)] = 0$ at $\Gamma = 0$. One obtains 
\begin{equation}
    \Delta_{ind}\sqrt{\Delta_0 + \Delta_{ind}} = \gamma\sqrt{\Delta_0 - \Delta_{ind}}.
\end{equation}
In the weak coupling regime, $\gamma < \Delta_0$,  we have $\Delta_{ind} \approx \gamma\Delta_0/(\gamma + \Delta_0)\approx \gamma$. Therefore, in the weak coupling regime the critical Zeeman field from Eq. \ref{critical_zeeman} reduces to the ``standard'' expression given by the minimal model.

One important effect of the proximity-induced energy renormalization is the reduction of the effective disorder strength with increasing SM-SC coupling. It has been established that strong-enough disorder destroys the topological phase and causes a quantum phase transition to a trivial localized (Anderson) phase \cite{Motrunich2001}. The onset of this transition is associated with the localization length (which is reduced by disorder) becoming  shorter than the SC coherence length (which increases with decreasing the gap that characterizes the clean system, i.e., $\Delta_{ind}$). In other words, the topological SC phase is associated with a large gap-to-effective disorder ratio, $\Delta_{ind}/V_0^{eff}$, while the trivial localized phase corresponds to small values of this ratio.
For the induced gap, the equation $\Delta_{ind} \approx \gamma\Delta_0/(\gamma + \Delta_0)$ gives a good approximation if $\gamma$ is not much larger than $\Delta_0$ \cite{PhysRevB.96.014510}. On the other hand, the disorder potential (and all other SM bare parameters) is renormalized (at low energies) by a factor $Z\approx\Delta_0/(\gamma+\Delta_0)$, i.e., $V_0^{eff} = ZV_0$. Consequently, the ratio that controls the effects of disorder becomes $\Delta_{ind}/V_0^{eff}= \gamma/V_0$. Increasing the effective SM-SC coupling generates a larger gap-to-effective disorder ratio, which favors the topological SC phase. This effects plays a critical role in understanding the numerical results described below.

Having determined the effective SM Green's function we can calculate the (low-energy) density of states (DOS), $\rho(\omega)$, and local density of states (LDOS), $\rho_{\Delta l}(\omega)$, using the relations
\begin{equation}\label{eq:TDOS}
    \rho(\omega) = -\dfrac{1}{\pi} \text{Im}\sum_i\text{Tr}[G_{SM}(\omega + i\eta)]_{ii},
\end{equation}
\begin{equation}\label{eq:local_DOS}
    \rho_{\Delta l}(\omega) = -\dfrac{1}{\pi} \text{Im}\sum_{i\in\Delta l}\text{Tr}[G_{SM}(\omega + i\eta)]_{ii},
\end{equation}
where $\eta$ is a finite broadening (typically $\sim$1 $\mu$eV) and Tr is the trace over spin and particle-hole spaces. 
Both $\rho$ and $\rho_{\Delta l}$ depend on the control parameters, $\mu$ and $\Gamma$, as well as the SM-SC coupling ($\gamma$).
Note that the LDOS is integrated over a short segment of the wire, $\Delta l$, which may include one or more sites. When $\Delta l$ includes a single site, the local density of states is given by
\begin{equation}\label{eq:local_site_DOS}
    \rho_{i}(\omega) = -\dfrac{1}{\pi} \text{Im}\text{Tr}[G_{SM}(\omega + i\eta)]_{ii}.
\end{equation}

The final tool used in our study of the SM-SC hybrid system is an end-to-end correlation function defined as follows. Consider the local density of states defined at the ends of the wire by integrating over segments of length $\Delta l = 100~$nm, $\rho_{l}(\omega)$ (left end) and $\rho_{r}(\omega)$ (right end). The corresponding summations in Eq. \ref{eq:local_DOS} are done over the leftmost (rightmost) 20 sites. Next, we determine the lowest energy local maxima, $\rho_{l}^*=\rho_l(\omega_l)$  and $\rho_{r}^*=\rho_r(\omega_r)$, characterizing the left and right LDOS  within an energy range $0 \leq \omega\leq \omega_{\rm max}$. Assume that the two maxima occur at energies $\omega_l$ and $\omega_r$, respectively.
Basically, if left and right local maxima with energies less than $\omega_{\rm max}$ exist and if they occur at (almost) the same energy, we assign a value $C=1$ to the end-to-end correlation function; otherwise, $C=0$. More specifically, we consider two LDOS peaks to occur at effectively the same energy if the difference $|\omega_l-\omega_r|$ is less than a certain value $\Delta \omega$. In addition, we impose a threshold $\rho_{\rm min}$ on the size on the LDOS peaks; LDOS peaks smaller than $\rho_{\rm min}$ are disregarded. This is to eliminate contributions from (exponentially small) ``tails'' of low-energy modes localized away from the ends of the wire. To summarize, the end-to-end correlation function is $C=1$ if there exist left and right LDOS local maxima with the properties
\begin{equation}
\omega_l, \omega_r \leq \omega_{\rm max}, ~~~~|\omega_l-\omega_r| \leq \Delta\omega, ~~~~\rho_l^*, \rho_r^* >\rho_{\rm min};
\end{equation}
otherwise, $C=0$. Of course, the correlation function depends on the ``filter'' parameters $\omega_{\rm max}$, $\Delta\omega$, and $\rho_{\rm min}$, as well as the broadening parameter $\eta$. In the calculations, we use $\omega_{\rm max}= 20~\mu$eV, $\Delta\omega=1~\mu$eV, and 
\begin{equation}
\rho_{\rm min}= \frac{\Delta l}{2L}\frac{\Delta_0}{\pi \eta(\Delta_0+\gamma)}.
\end{equation}
Note that $\rho_{\rm min}$ corresponds to half of the end-of-wire LDOS generated by a zero-energy state uniformly distributed throughout the wire. We have verified that changing these parameters does not modify the qualitative behavior of the correlation function, which in this work represents only an additional tool to be used together with the DOS and LDOS. A more detailed analysis of the correlation function, for example determining the information that could be obtained by varying the filter parameters or establishing the connection with experiment (e.g., obtaining correlation maps based on differential conduction measurements),  is beyond the scope of this study.

\section{Results}

We investigate the low-energy physics of the hybrid SM-SC system by calculating the DOS, LDOS, and end-to-end correlation function, as described above, for different parameter regimes. We consider four different cases: (a) clean system (which is used to benchmark the disorder results), (b) hybrid system with potential disorder in the SM, (c) hybrid system with disorder in the parent SC, and (d) system with disorder in both the SM wire and the parent SC. For each case we consider two different SM-SC coupling regimes: a weak coupling regime characterized by $\langle\gamma\rangle/\Delta_0 =0.5$ and a strong coupling regime with $\langle\gamma\rangle/\Delta_0 =2$. We note that in the weak coupling regime a fraction $\Delta_0/(\Delta_0+\langle\gamma\rangle) \approx 66.7\%$ of the spectral weight of a (nearly) zero-energy mode is distributed inside the SM wire, while $33.3\%$ is in the SC. The percentages are reversed in the strong coupling case.

\begin{figure}[t]
\centering
\includegraphics[width=0.5\textwidth]{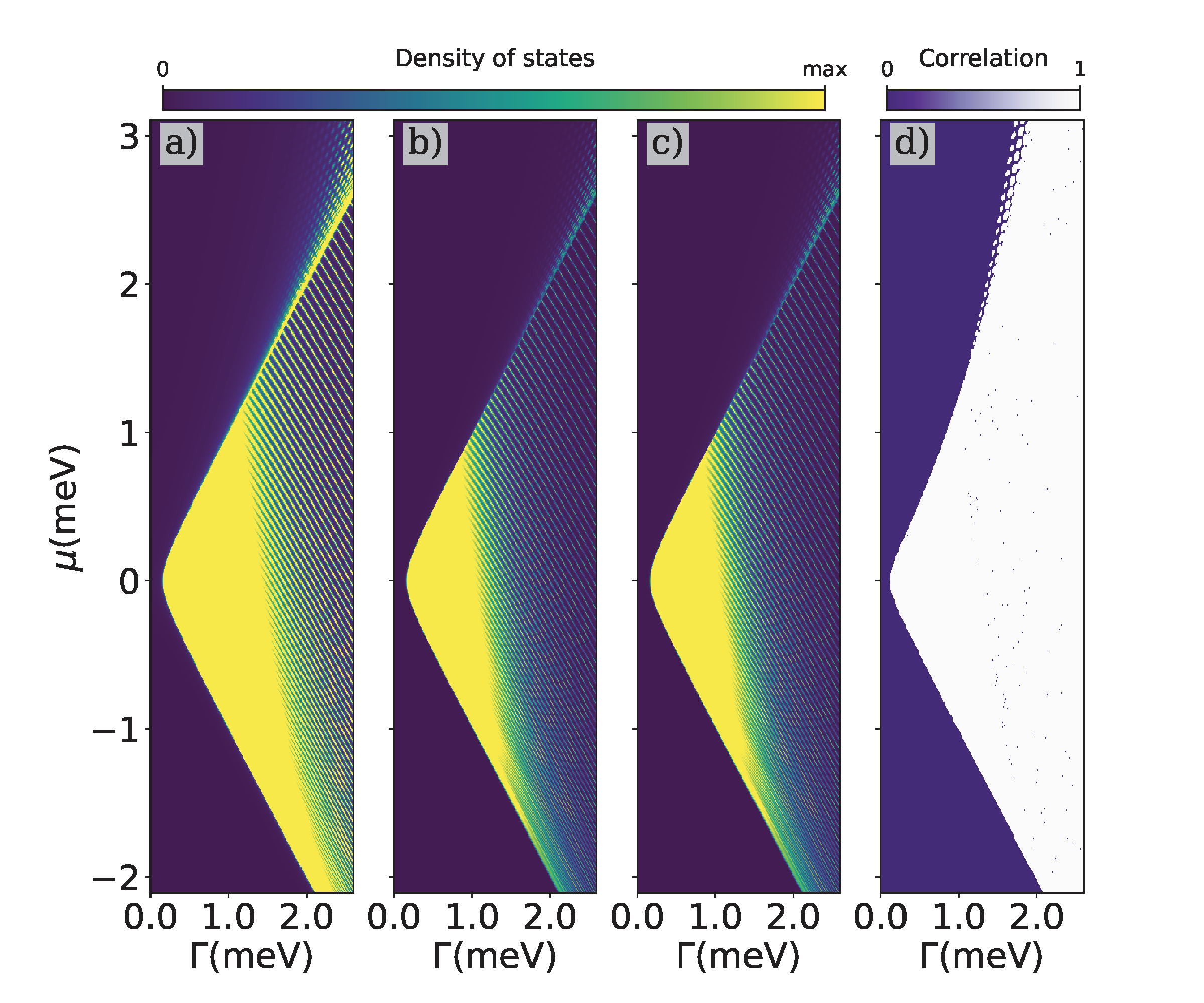}
\caption{Clean system of length $L=5~\mu$m with weak SM-SC coupling ($\gamma=0.15~$meV; $\gamma/\Delta_0 = 0.5$). (a) Zero-energy DOS as function of Zeeman field and chemical potential. The zero-energy DOS is non-zero inside the topological region $\Gamma^2 \geq {\mu}^2 + \gamma^2$, indicating the presence of midgap MZMs. With increasing Zeeman field the zero-energy DOS decreases as a result of larger Majorana splitting oscillations (note the faint ``stripe'' structure). 
(b) Zero-energy LDOS at the left end of the wire, $\rho_l(0)$, as function of $\Gamma$ and $\mu$. (c) Zero-energy LDOS at the right end of the wire, $\rho_r(0)$, as function of $\Gamma$ and $\mu$.
Note that the left and right LDOS signatures are identical, being generated by symmetric MZMs localized at the opposite ends of the system. The zero-energy LDOS also decreases with increasing Zeeman field (see main text).
(d) End-to-end correlation as a function of Zeeman field and chemical potential. Note the non-zero correlations outside the topological region (at large $\mu$ values) generated by the presence of ``intrinsic'' Andreev bound states at the ends of the system (see main text). Also note the absence of a ``stripe'' structure, as a result of the correlation function being finite in the presence of finite energy correlated peaks.}    \label{fig:clean}
\end{figure}

\subsection{Clean System} \label{sec:clean}

\begin{figure}[t]
\centering
\includegraphics[width=0.5\textwidth]{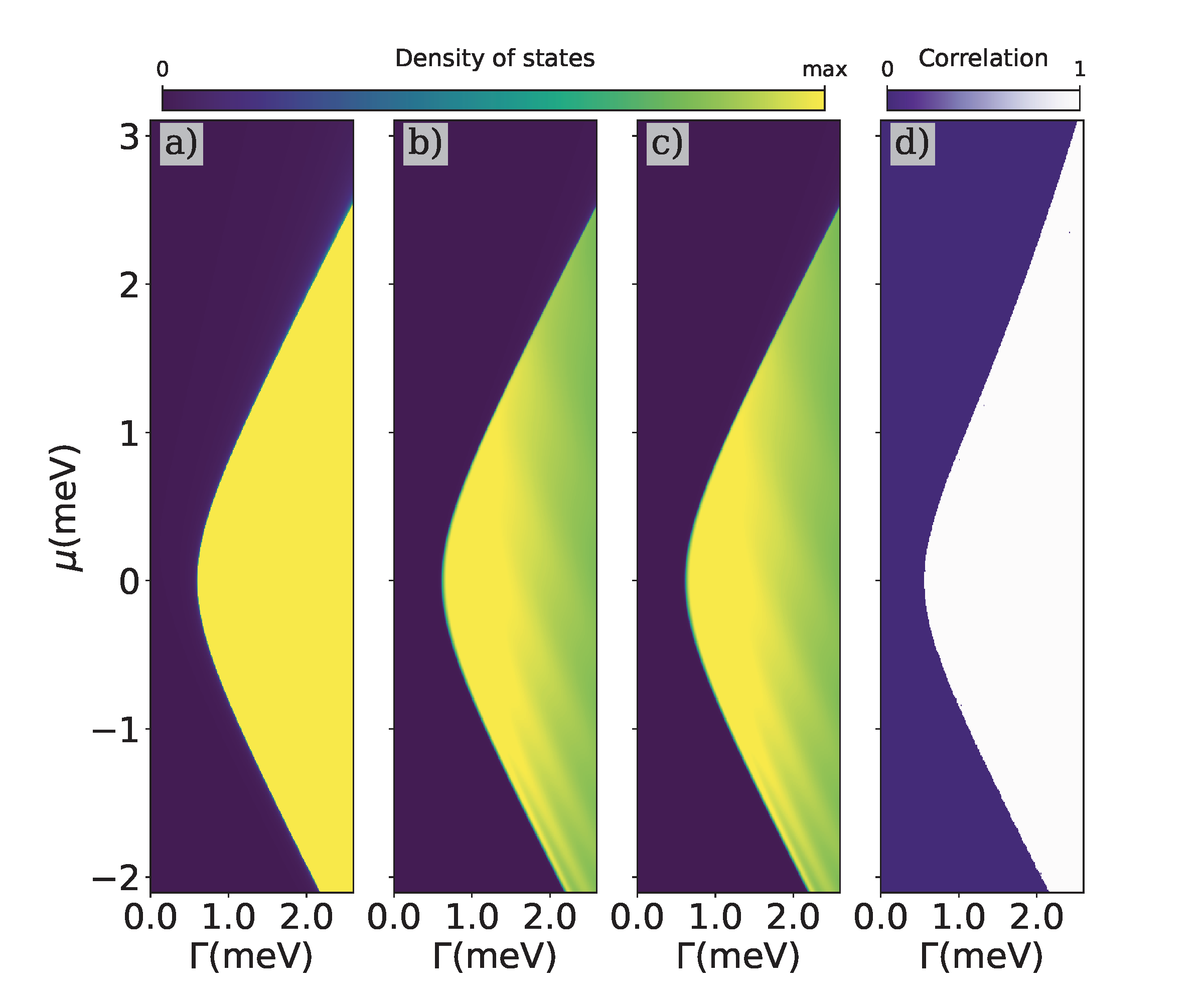}
\caption{Clean system with strong SM-SC coupling ($\gamma=0.6~$mev; $\gamma/\Delta_0 = 2$), showing (a) the zero-energy DOS, (b) the left end zero-energy LDOS, (c) the right end zero-energy LDOS, and (d) the end-to-end correlation as functions of chemical potential $\mu$ and Zeeman field $\Gamma$. 
Note the negligible dependence of the DOS and the weak dependence of the LDOS on the Zeeman field (as compared to Fig.  \ref{fig:clean}), as well as the absence of any  ``stripe'' structure. This is a consequence of the MZMs being more localized, which results in smaller energy splitting oscillations. The effect of the spectral weight near the ends of the wire decreasing with $\Gamma$ is weaker than that in Fig.  \ref{fig:clean}, but  is still visible in panels (b) and (c). Also, note the presence of trivial non-zero end-to-end correlations at high values of the chemical potential.}        \label{fig:clean_strong}
\end{figure}

We begin by considering the clean case. The weak coupling results are shown in  Fig. \ref{fig:clean}, while the corresponding diagrams for the strongly coupled system are shown in Fig. \ref{fig:clean_strong}. The zero-energy total density of states as function of Zeeman field ($\Gamma$) and chemical potential ($\mu$) for a weakly coupled system ($\gamma/\Delta_0 = 0.5$) is shown in Fig. \ref{fig:clean}(a). Non-zero values of the DOS obtain in the topological SC region, where MZMs are present at the wire ends. Indeed, the light green/yellow region in Fig. \ref{fig:clean}(a) corresponds to the topological condition $\Gamma^2 > \mu^2 + \gamma^2$. Note that the value of the zero-energy DOS decreases with increasing magnetic field. This is (in part) the effect of having small energy splitting oscillations with an amplitude that increases with increasing $\Gamma$ as a result of an enhancement of the characteristic length of the Majorana modes (which, in turn, leads to the enhancement of the exponentially small overlap of the Majorana wave functions). 
These so-called Majorana oscillations are visible in Fig. \ref{fig:clean}(a) as (faint) stripes characterizing the DOS at larger values of Zeeman field. 

In Fig. \ref{fig:clean}(b) and (c) we show the zero-energy LDOS at the left and the right ends, respectively. The quantities $\rho_l(0)$ and $\rho_r(0)$ are calculated using Eq. \ref{eq:local_DOS} with $\Delta l=100~$nm, as described above. Note that the system is characterized by the same LDOS at both ends, which indicates the presence of (identical) MZMs at the edges of the wire. Again, the decrease of the zero energy LDOS with 
increasing Zeeman field is due in part to an enhancement of the amplitude of the Majorana splitting oscillations with $\Gamma$. Indeed, when the splitting energy is larger than the value of the broadening used in the calculation ($\eta=1~\mu$eV), the split modes do not contribute to the zero energy LDOS (or DOS).
Additionally, since the LDOS at each end is integrated over a short segment ($100~$nm), the enhancement of the characteristic length of the Majorana wave functions with the Zeeman field further reduces $\rho_l(0)$ and $\rho_r(0)$ (as spectral weight becomes distributed over larger segments of the wire, instead of being strongly localized near the ends). 

The end-to-end correlation  is shown in  Fig. \ref{fig:clean}(d) as a function of chemical potential $\mu$ and Zeeman field $\Gamma$. Unlike the DOS and LDOS maps, the correlation map does not have a ``stripe'' structure associated with the Majorana energy splitting oscillations. This is because the correlation function includes finite energy information; a pair of split Majorana modes does not contribute to the zero-energy DOS and LDOS, but generates correlated LDOS peaks at finite energy, which results in a correlation function $C=1$ (as long as the peak energy is less than $\omega_{\rm max}$).
An important feature in  Fig. \ref{fig:clean}(d) is the presence of non-zero end-to-end correlations outside the topological region (at high values of the chemical potential). In this region, close-enough to the phase boundary, the bulk gap is comparable to (or smaller than) the  energy window ($\omega_{\rm max} = 20~\mu$eV) used to calculate the correlation function. In addition, the system supports subgap ``intrinsic'' Andreev bound states (ABSs) localized near the ends of the wire \cite{Huang2018}, which generate (correlated) finite energy peaks in the LDOS.  This is an early warning regarding the possibility of having (accidental) non-zero end-to-end correlations associated with trivial states, particularly in the large chemical potential regime where the bulk gap is relatively small. 

\begin{figure}[t]
    \subfloat{%
    \includegraphics[clip,width=\columnwidth]{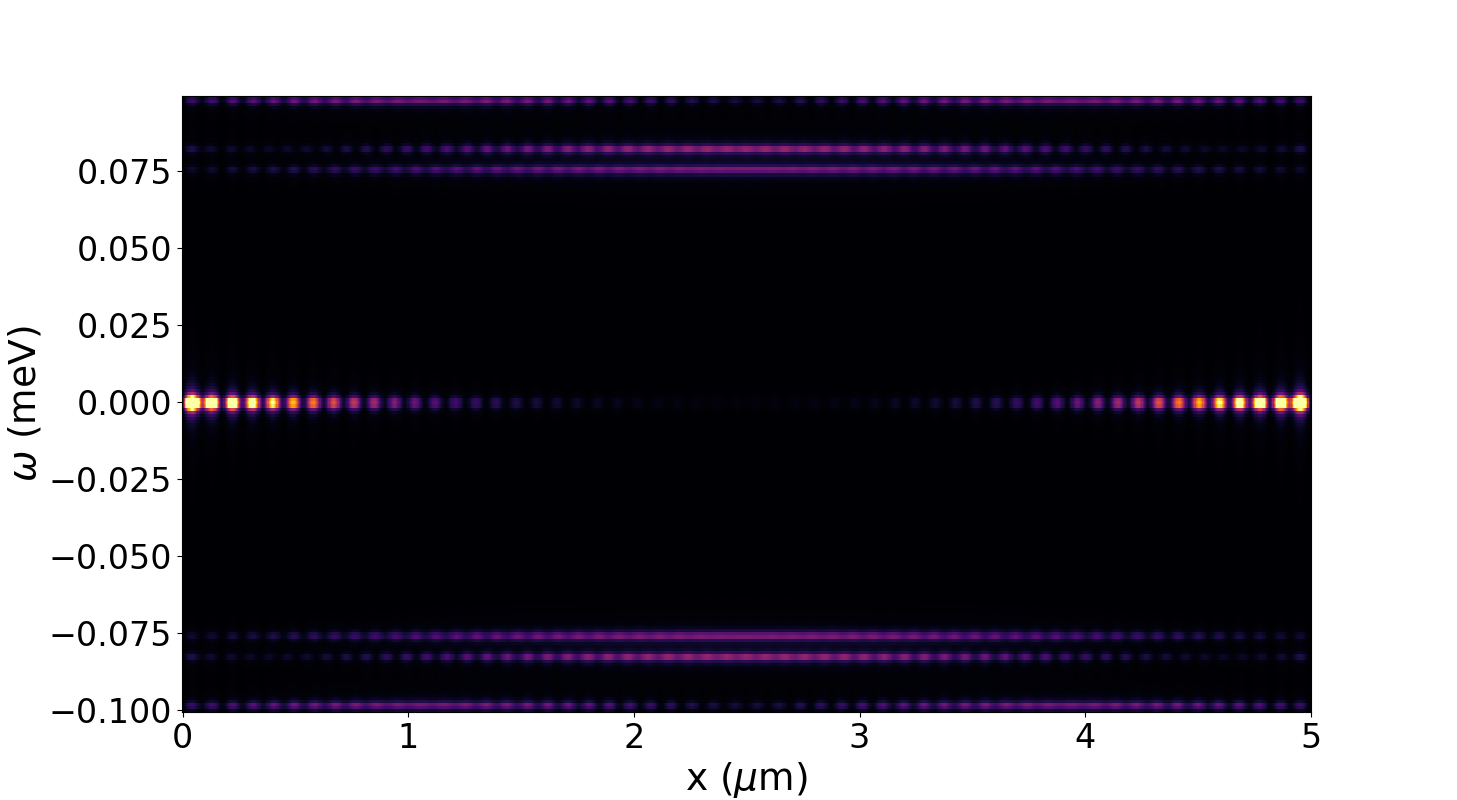}%
    }
    \hspace{0.6cm}
    \subfloat{%
    \includegraphics[clip,width=\columnwidth]{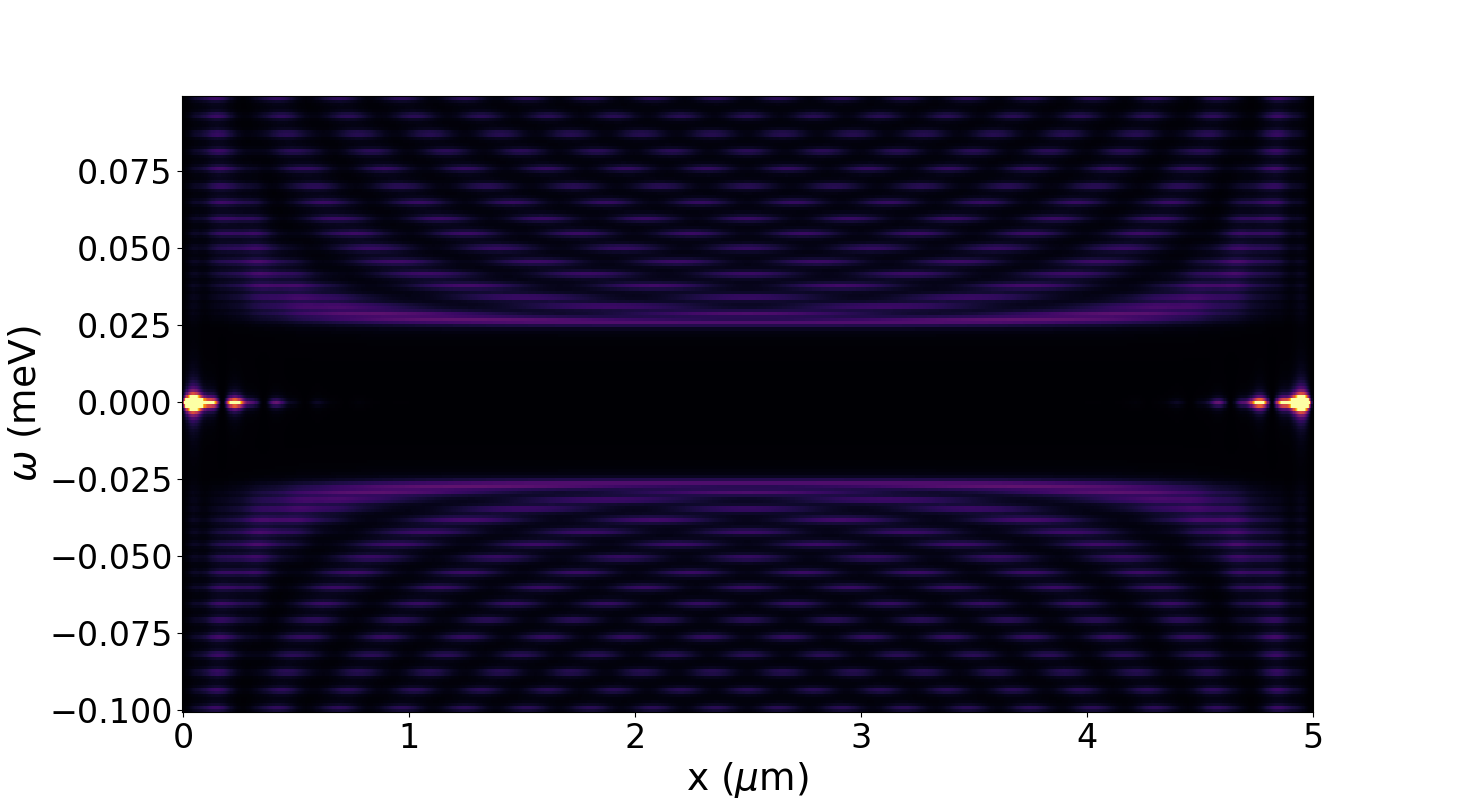}%
    }
\caption{LDOS as a function of energy and position for a clean system with weak SM-SC coupling (top) and strong SM-SC coupling (bottom) and control parameter values  $\mu = 0.7~$meV and $\Gamma = 1.0~$meV (which correspond to a topological SC phase). The strong LDOS features near the ends of the wire are generated by a pair of MZMs. Note that the characteristic length scale of the MZMs in the strong-coupling regime (bottom) is much shorter than its weak-coupling correspondent (top), although the (bulk) topological gap  of the strongly-coupled system  is smaller than its weak coupling counterpart ($\sim\!25~\mu$eV in the bottom panel and $\sim\!75~\mu$eV in the top panel).}   \label{fig:LDOS_x}
\end{figure}

We repeat the analysis described above for a clean system with strong SM-SC coupling,  $\gamma/\Delta_0 = 0.6/0.3 =2$. The results are shown in  Fig. \ref{fig:clean_strong}. In panel (a) we represent the zero-energy DOS as a function of $\mu$ and $\Gamma$. Comparison with Fig. \ref{fig:clean}(a) 
reveals a much weaker (practically negligible) dependence on the Zeeman field. This is because in the strongly coupled system the characteristic length of the Majorana wavefunction is significantly shorter that the corresponding length scale in a weakly coupled system (also see Fig. \ref{fig:LDOS_x}), which results in Majorana energy splitting oscillations having a smaller amplitude. For the parameters used in Fig. \ref{fig:clean_strong}, this amplitude does not exceed the broadening $\eta$ and, consequently, the Majorana contribution to the zero-energy DOS is practically the same over the whole parameter range corresponding to the topological phase,  $\Gamma^2 > \mu^2 + \gamma^2$. Of course, the ``stripe'' features associated with the Majorana oscillations also disappear.
In Fig. \ref{fig:clean_strong}(b) and (c) we show the zero-energy LDOS at the left and right ends of the wire, respectively. Note that the two  LDOS maps are identical, which is a results of the MZMs that generate them having symmetric real space profiles. Of course, this symmetry can be broken by applying local potentials; this will affect the (non-vanishing) values of the LDOS, but not the parameter region characterized by a finite LDOS. Also note that the LDOS at zero energy decreases with increasing Zeeman field,  but not as much as it does in the weak coupling case. Again, this is a result of the Majoranas having shorter characteristic lengths, comparable to $\Delta l$ (the length of the segment over which the LDOS is integrated), for all control parameter values; the increase of this length scale with $\Gamma$  accounts for the decrease of the LDOS at large Zeeman field values. 
Finally, in Fig. \ref{fig:clean_strong}(d) we show the end-to-end correlation as a function of $\mu$ and $\Gamma$.  Again, we emphasize the presence of (significant) end-to-end correlations associated with intrinsic Andreev bound states inside the topologically-trivial region, near the phase boundary corresponding to $\mu>\gamma$. We emphasize that these ``accidental'' correlations can be eliminated by breaking the (real space) symmetry of the system (e.g., by applying a local potential at one end of the wire). In a broken symmetry system, the left and right  Andreev-induced LDOS peaks will generically occur at different energies (hence, they will not generate end-to-end correlations). By contrast, the Majorana-induced correlations are immune    to (weak-enough) local potentials.  
 
We conclude that the main differences between the weak-coupling and the strong-coupling behaviors stem from the characteristic length scales of the low-energy modes being significantly shorter in the strong-coupling case. To clarify this point, we calculate the LDOS as a function of energy (within an energy window $-0.1$--$0.1$ meV) and position $x$ along the wire for a set of parameters,  $\mu = 0.7$ meV, $\Gamma = 1.0$ meV, corresponding to the topological phase (for both $\gamma= 0.15~$meV and $\gamma=0.6~$meV). The results shown in Fig. \ref{fig:LDOS_x} clearly demonstrate the presence of MZMs localized near the ends of the wire and the fact that the localization length of these modes is much shorter in the strongly coupled system (bottom panel) as compared with the weakly coupled system (top panel).  Also note that this property is not associated with a larger value of the topological (bulk) gap in the strongly coupled system.

\begin{figure}[t]
\centering
\includegraphics[width=0.5\textwidth]{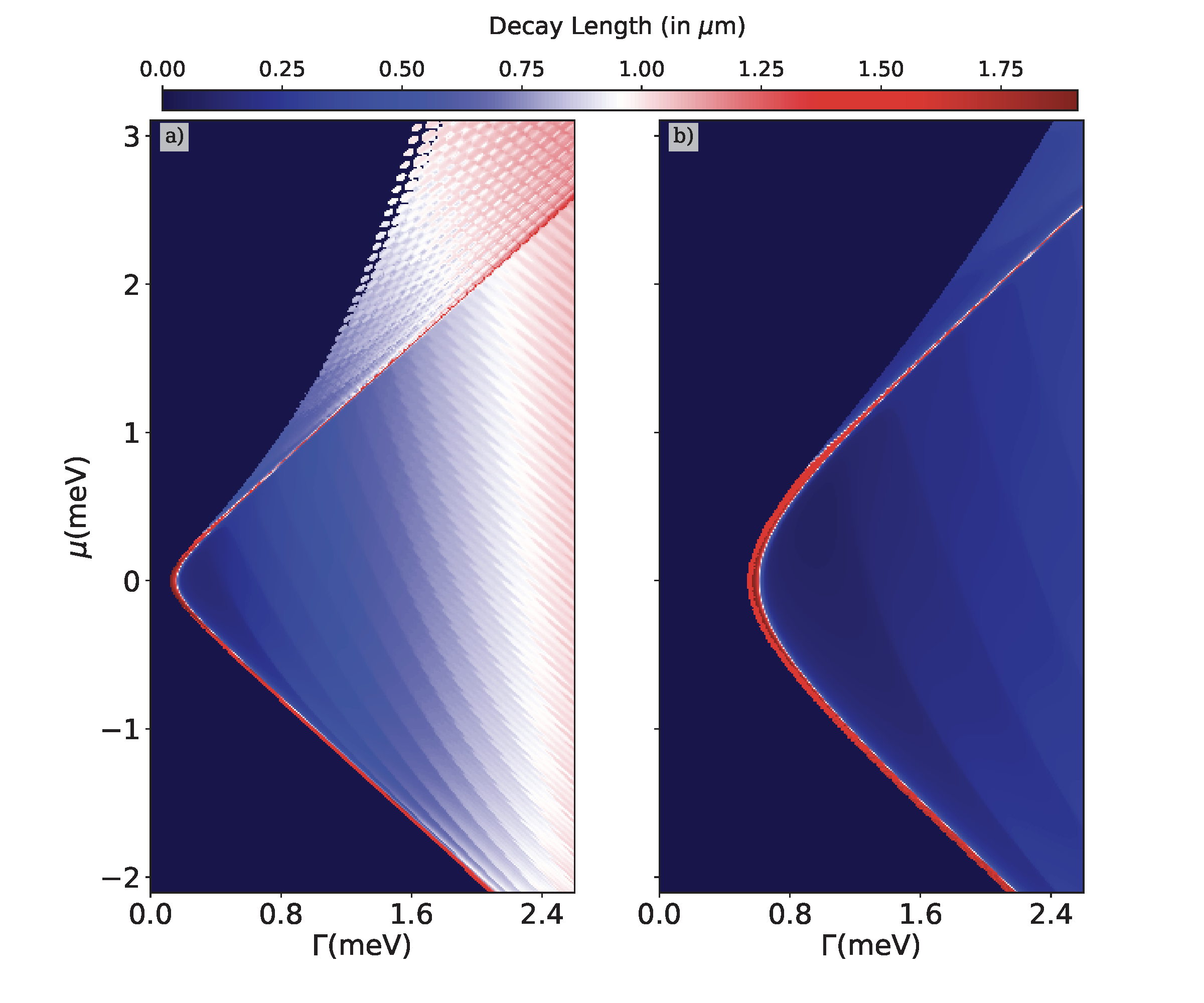}
\caption{Dependence of the decay length of the low-energy modes localized near the ends of the wire on the control parameters ($\Gamma$ and $\mu$)  for a weakly-coupled system (left, $\gamma =0.15~$meV) and a strongly-coupled system (right, $\gamma=0.6~$meV). The decay length $l_D$ is defined as the distance from the (left or right) end of the wire corresponding to $1-1/e \approx 0.63$ of the spectral weight of the low-energy mode --- see Eq. (\ref{Eq_ld}). Note the large $l_d$ contribution at the phase boundary (red ``line''), which is generated by the zero-energy delocalized  state associated with the vanishing of the bulk gap at the TQPT. The finite contributions outside the topological phase (at high $\mu$ values) are generated by low-energy intrinsic Andreev bound states. The key point is that strong SM-SC coupling significantly reduces the characteristic length scale of the low-energy modes, as clearly revealed by the comparison between the two panels. The black regions correspond to gapped systems with no low-energy modes.} \label{fig:length_clean_g_var}
\end{figure}

To fully characterize the dependence of the length scale of low-energy modes localized near the ends of the wire (particularly of Majorana modes) on the system parameters, we define the decay length $l_D$ as the length (measured from the end of the wire) corresponding to $1-1/e \approx 0.63$ of the total spectral weight of the localized mode. For an exponentially localized mode with energy $\omega$, this would correspond to reducing the LDOS (as compared with the maximum value at the boundary)  by a factor of $e$. Specifically, we have
\begin{equation}
 \rho_{l_D}(\omega) = 0.63~\!\rho_{L/2}(\omega), \label{Eq_ld}
\end{equation}
where $\rho_{l_D}$ is the LDOS integrated over a segment $l_D$ near one end of the wire (e.g., the left end) and $\rho_{L/2}$ is the LDOS integrated over half of the wire length, which, by symmetry, represents the total weight of the localized mode (even when there is a non-zero overlap with the mode localized near the opposite end).  The results are shown in Fig.  \ref{fig:length_clean_g_var}, with the left and right panels corresponding to weak and strong coupling, respectively. In practice, we first determine $\rho_{L/2}$, then we integrate the LDOS starting from the leftmost site ($i=1$), adding contributions until we reach  a value (approximately) equal to $0.63~\!\rho_{L/2}$ at site $i=l_D/a$. 
Note that the calculation is done for the  parameter space region characterized by the presence of low-energy (sub-gap) modes, which generate a non-zero end-to-end correlation; in the regions with no low-energy modes $l_D$ is set to zero (by convention). 
First, one can clearly see that (i) $l_D$ increases with increasing the Zeeman field and (ii) the values of $l_D$ in the weakly coupled system are significantly larger (by a factor 2-3) as compared to the corresponding values in the strongly coupled system. An interesting feature is the presence of a large $l_D$ contribution at the topological phase boundary, which appears as a red ``line'' in Fig. \ref{fig:length_clean_g_var}. This contribution is generated by the (near) zero-energy delocalized state associated with the closing of the bulk gap at the TQPT. Finally, note the non-zero contributions outside the topological region (at high $\mu$ values). As discussed above, these contributions are generated by low-energy intrinsic Andreev bound states (having energies less than $\omega_{\rm max}$). Again, the characteristic length scale of these ABSs in the strong coupling regime (right panel) is much shorter than the length scale characterizing ABSs in the weakly coupled system (left panel).   

\subsection{Disorder in the semiconductor}     \label{sec:Dis_semi} 

In this section we investigate the effects of disorder on the low-energy physics of the hybrid wire by introducing the random potential shown in Fig. \ref{fig:SM_disorder_profile} for three different values of the disorder amplitude: $V_0 = 0.5$ meV, $1.0$ meV, and $2.0$ meV. We note that the effective disorder profile, which corresponds to a random potential generated by charge impurities located inside the semiconductor nanowire,  was constructed based on the microscopic analysis in Ref. \cite{Woods2021}. 

\begin{figure}[t]
\centering
\includegraphics[clip,width=\columnwidth]{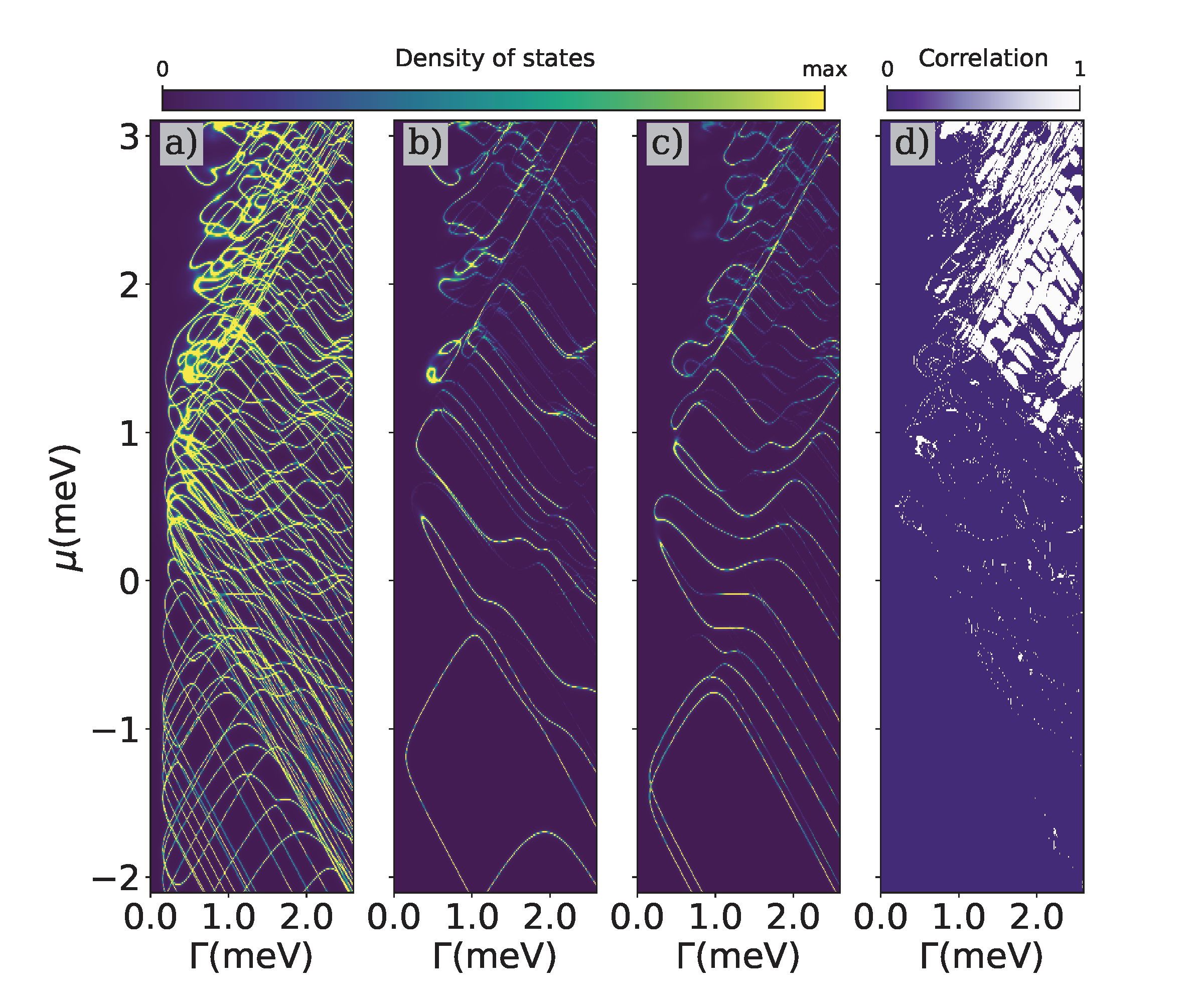}
\caption{Disordered system of length $L=5~\mu$m with weak SM-SC coupling ($\gamma=0.15~$meV; $\gamma/\Delta_0 = 0.5$) and strong SM disorder ($V_0 = 2.0$ meV).  (a) Zero-energy DOS as a function of Zeeman field, $\Gamma$, and chemical potential, $\mu$. (b) Map of the zero-energy LDOS at the left end of the wire. (c) Map of the zero-energy LDOS at the right end of the wire. (d) End-to-end correlation map. 
Strong SM disorder reduces the zero-energy DOS and LDOS to a mesh of ``lines'' associated with accidental, disorder-induced zero-energy ABSs evolving in the $\Gamma$--$\mu$ plane. The zero-energy LDOS features in (b) and (c) are uncorrelated. Accidental correlations from finite-energy modes generate small ``islands,'' as illustrated in (d).}
    \label{fig:SM_disorder_weak_coupl}
\end{figure}

First, we consider a hybrid system with weak SM-SC coupling, $\gamma= 0.15~$meV, and strong disorder, $V_0 = 2.0~$meV. The corresponding DOS, LDOS, and correlation maps are shown in Fig. \ref{fig:SM_disorder_weak_coupl}.  Panel (a) shows the zero-energy DOS as a function of $\mu$ and $\Gamma$. By comparison with the corresponding panel in Fig. \ref{fig:clean}, we notice two important features: (i) The quasi-continuous region characterized by a non-zero DOS in Fig. \ref{fig:clean}(a)  has evolved into a mesh of  ``lines.'' (ii) The parameter region where these ``lines'' occur extends outside the clean topological phase, covering the area with $\Gamma >\gamma$ for a wide range of chemical potential values. The low-energy states that generate the zero-energy DOS ``mesh'' are disorder-induced states \cite{sarma2023spectral}, each ``line'' tracing the evolution of one of those states in the $\Gamma-\mu$ plane. Some of the low-energy disorder-induced states have non-zero weight near the ends of the wire, generating a finite zero-energy LDOS. The corresponding features for the  left and right ends  are shown in Fig. \ref{fig:SM_disorder_weak_coupl}(b) and (c), respectively. Note that the density of ``lines'' in the LDOS maps is significantly lower than that in panel (a), revealing the fact that most of the disorder-induced states are localized away from the ends of the system (i.e., in the bulk).  By comparison,  in the clean system  (Fig. \ref{fig:clean}) practically all contributions to the zero-energy DOS come from MZMs localized near the ends of the system, which generate similar contributions to the LDOS. Also, a direct comparison between panels (b) and (c) in Fig. \ref{fig:SM_disorder_weak_coupl} reveals that the zero-energy LDOS features at the left and right ends of the wire are not correlated, which confirms that they are generated by accidental, disorder-induced zero energy Andreev bound states (ABSs).
On the other hand, there are finite energy (accidental) correlations occurring within small ``islands'' in the parameter space, as illustrated in Fig. \ref{fig:SM_disorder_weak_coupl}(d). We point out that reducing the energy window $\omega_{\rm max} = 20~\mu$eV within which we calculate the correlation function results in these correlation ``islands'' shrinking (and, eventually, most of them disappearing). 

We emphasize that the zero-energy DOS and LDOS maps shown in  Fig. \ref{fig:SM_disorder_weak_coupl} are calculated with relatively high energy resolution, $\eta = 1~\mu$eV. Lowering the resolution results in thicker ``lines'' that eventually merge and cover most of the parameter space, as the system supports a multitude of disorder-induced low-energy modes (being effectively gapless). Thus, the experimental observation of zero-bias conductance peaks (over relatively large parameter ranges) with energy resolution representing a significant fraction of the (clean) topological gap does not demonstrate the presence of MZMs.  On the other hand, high-resolution LDOS (or differential conductance) maps could clearly distinguish between Majorana-induced ``stripy'' features and ``lines'' associated with disorder-induced ABSs. 

\begin{figure}[t]
\centering
\includegraphics[width=0.5\textwidth]{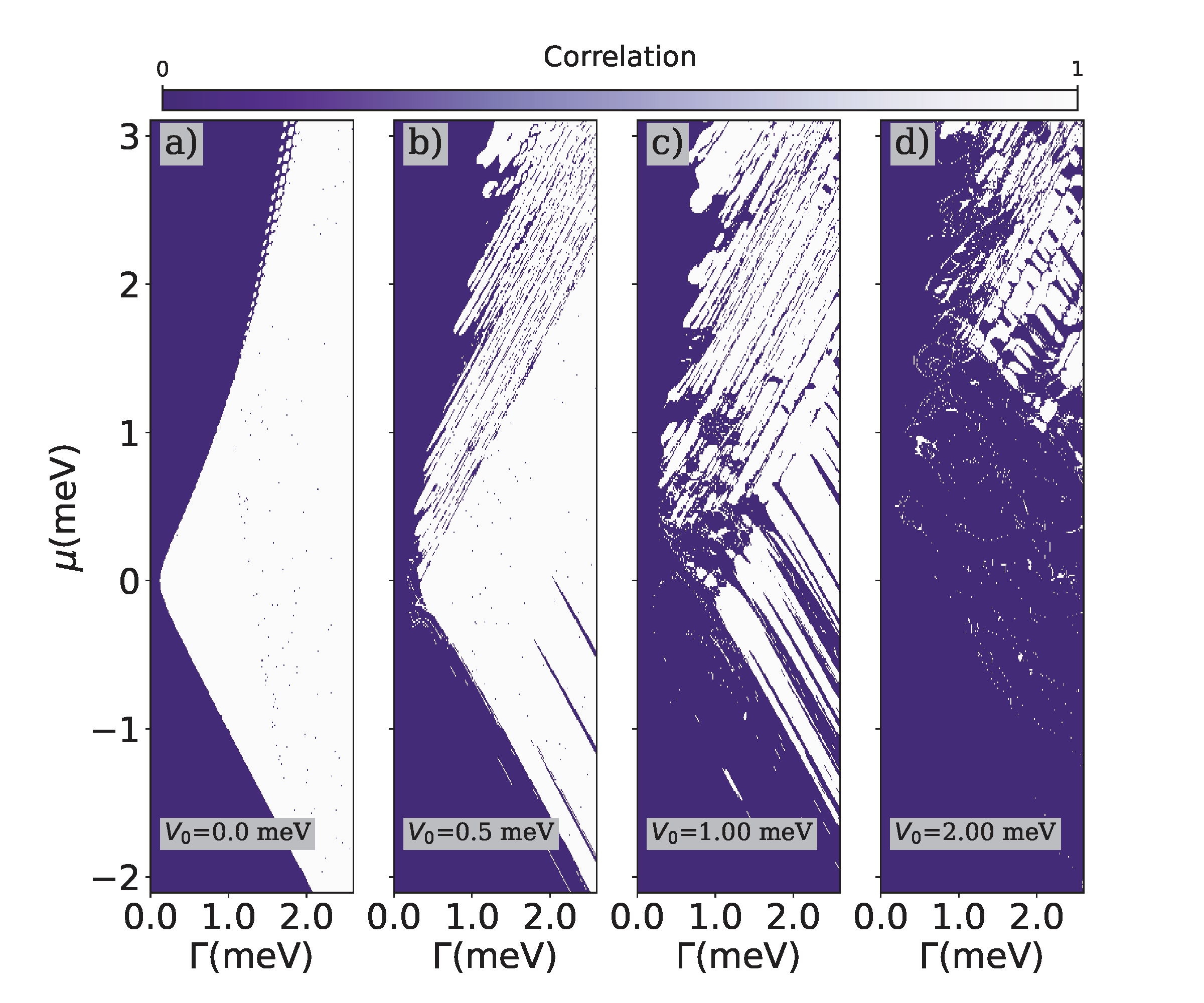}
\caption{Evolution of the end-to-end correlation map with disorder for a weakly coupled hybrid system. The values of the disorder potential amplitude, are (going from left to right): (a) $V_0=0$, (b) $V_0=0.5~$meV, (c) $V_0=1.0~$meV and (d) $V_0=2.0~$meV. With increasing disorder, the correlated region recedes away from low chemical potential values towards the top right corner of the phase diagram and, eventually, gets fragmented into small islands.}
    \label{fig:dis_var_g_weak}
\end{figure}

\begin{figure}[t]
\centering
\includegraphics[clip,width=\columnwidth]{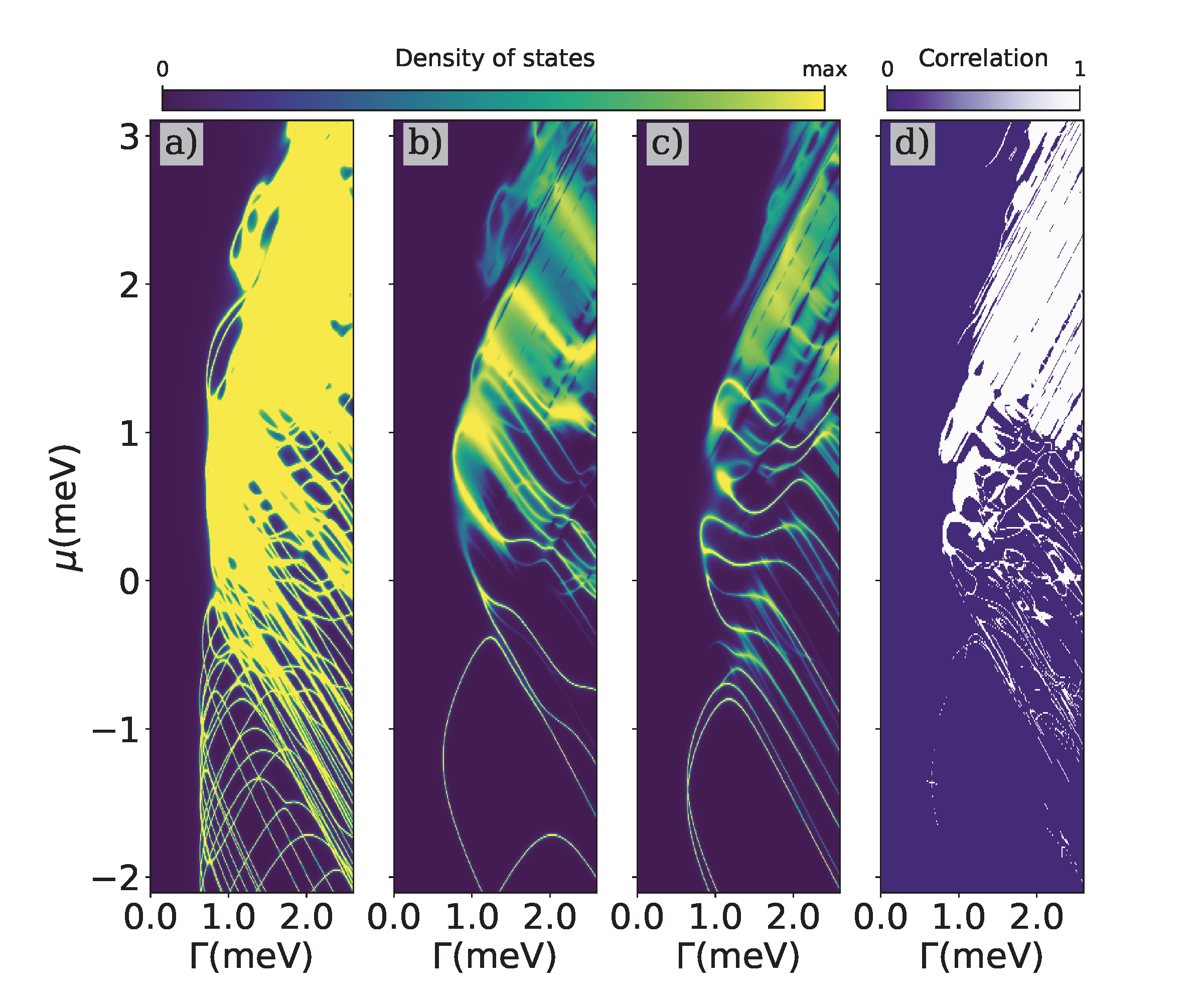}
\caption{Disordered system with strong SM-SC coupling, $\gamma =0.6~$meV, and disorder potential amplitude $V_0=2~$meV. All other parameters are the same as in Fig. \ref{fig:SM_disorder_weak_coupl}. (a) Zero-energy DOS as a function of Zeeman field and chemical potential. (b) Map of the zero-energy LDOS at the left end of the wire. (c) Map of the zero-energy LDOS at the right end of the wire. (d) End-to-end correlation map. 
 Note the relatively large areas characterized by a finite zero-energy DOS (or LDOS) and the robust end-to-end correlated ``phase'' [panel (d)]. Strong SM-SC coupling results in a weaker ``effective disorder'' (as compared to  Fig. \ref{fig:SM_disorder_weak_coupl}).}  \label{fig:SM_disorder_strong_coupl}
\end{figure}

To better understand the significance of the of the correlation map in Fig. \ref{fig:SM_disorder_weak_coupl}(d), we calculate the dependence of the correlation function on the disorder strength, starting with the clean case and considering three non-zero values of $V_0$. The corresponding correlation maps are shown in  Fig. \ref{fig:dis_var_g_weak}.  Upon increasing the disorder strength, the region characterized by low values of the chemical potential becomes uncorrelated. This is due to the fact that in the low-$\mu$ regime the low-energy states (e.g., the zero-energy states revealed by the DOS shown in Fig. \ref{fig:SM_disorder_weak_coupl}) become localized, typically within the bulk, and do not generated correlations. Nonetheless, for disorder strengths up to $V_0 \sim 1~$meV there is still a large, almost continuous region characterized by nonzero correlations. The low-energy states responsible for these correlations are Majorana modes localized near the opposite ends of the wire, as we checked explicitly (also see below, Fig.  \ref{fig:semi_dis_g_weak}). Further increasing the disorder strength results in the correlated region breaking into small islands, which correspond to accidental correlations generated by disorder-induced low-energy states.  Hence, the observation of large correlated areas in the control parameter space is a useful operational criterion for identifying topological regions that supports Majorana modes. Note that a natural energy (or magnetic field) scale that is experimentally accessible is given by the minimum Zeeman (or magnetic) field at which zero-energy features emerge, which is controlled by the effective SM-SC coupling $\gamma$ (see, e.g., Figs. \ref{fig:SM_disorder_weak_coupl} and \ref{fig:SM_disorder_strong_coupl}). Correlated regions with characteristic ``lengths'' significantly larger that this scale indicate the presence of a topological superconducting phase; correlated islands comparable to or smaller than this scale are generated by accidental disorder-induced low-energy modes. 

\begin{figure}[t]
\centering
\includegraphics[width=0.5\textwidth]{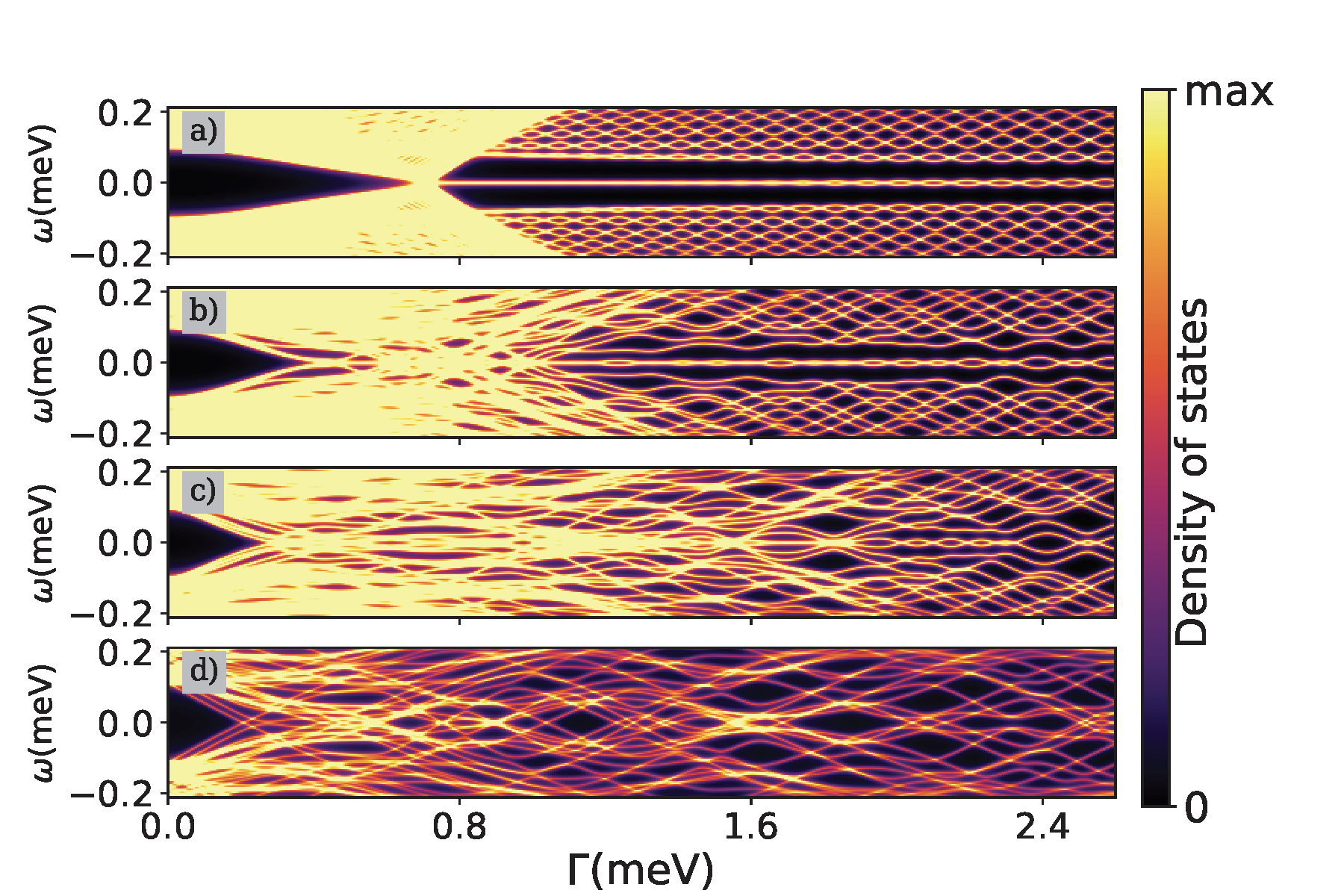}
\caption{DOS as a function of Zeeman field and energy of a hybrid system with weak SM-SC coupling, fixed chemical potential, $\mu = 0.7~$meV, and different disorder strengths. The values of the disorder potential amplitude, are (going from left to right): (a) $V_0=0$, (b) $V_0=0.5~$meV, (c) $V_0=1.0~$meV and (d) $V_0=2.0~$meV. In (a) and (b) the (weakly oscillating) Majorana mode is separated from the rest of the spectrum by a finite energy gap. For $V_0\sim1~$ the gap collapses, but the Majorana modes still survives at large-enough values of the Zeeman field. In the strong disorder regime --- panel (d) --- there are no Majorana modes.}         \label{fig:semi_dis_g_weak}
\end{figure}

The identification of the low-energy modes responsible for various correlation features was done by taking constant chemical potential cuts through the ``phase diagrams'' and calculating the DOS as a function of Zeeman field and energy. For example, in Fig. \ref{fig:semi_dis_g_weak} we show the DOS spectra corresponding to $\mu=0.7~$meV and different disorder strengths. For the clean system [panel (a)], we clearly see the closing and re-opening of the bulk gap at $\Gamma \approx 0.7~$meV and the emergence of a (weakly oscillating) Majorana mode at larger Zeeman field values. The spatial profile of the Majorana mode (or any other low-energy mode) can be obtained by fixing the values of $\omega$ and $\Gamma$ (to correspond to the desired DOS feature) and calculating the LDOS as a function of position. Note that this does not involve any numerical cost, since the DOS data itself is obtained by integrating the LDOS over the entire length of the wire. Weak disorder [panel (b)] results in the emergence of disorder-induced zero-energy states in the vicinity  of the clean phase boundary ($0.5 \lesssim \Gamma \lesssim 1.1~$meV). However, at larger values of $\Gamma$ the Majorana mode is still separated from the rest of the spectrum by a finite energy gap. In the intermediate disorder regime [panel (c)], the gap collapses, although the Majorana mode is still present. This mode is responsible for the (large) correlated region in Fig. \ref{fig:dis_var_g_weak}. Further increasing the disorder strength results in the system becoming trivial, with no Majorana modes at the ends of the wire. This strong disorder regime corresponds to small (accidental) correlation islands.  

Next, we consider the strong SM-SC coupling regime and repeat the above analysis for a system with $\gamma=0.6~$mev (i.e., $\gamma/\Delta_0 = 2$), all other parameters being the same. The DOS, LDOS, and correlation map corresponding to $V_0=2~$meV  are shown in Fig. \ref{fig:SM_disorder_strong_coupl}. Unlike the weak coupling case (see Fig. \ref{fig:SM_disorder_weak_coupl}), there are finite, relatively large (two-dimensional) regions of the parameter space characterized by finite zero-energy DOS and LDOS. Consequently, the correlated region --- see panel (d) --- is also large and practically continuous, although the correlations are almost completely suppressed at low $\mu$ values, similar to the weak coupling case.    
The effective disorder in Fig. \ref{fig:SM_disorder_strong_coupl} appears to be weaker than the (nominally identical) disorder in Fig. \ref{fig:SM_disorder_weak_coupl}. 
The key reason for this behavior is the low-energy proximity-induced renormalization \cite{PhysRevB.96.014510}, which reduces the effective strength of the disorder potential by a factor $\sim\Delta_0/(\Delta_0+\gamma$). Another way of thinking about this effect is in terms of the spectral weight distribution between the SM wire and the SC. In the strong coupling case, only approximately $33\%$  of the spectral weight of a low-energy mode is inside the SM, versus $\sim 67\%$ in the weak coupling case, hence these modes ``feel'' a weaker disorder potential. In addition, the strongly coupled system has low-energy modes characterized by shorter length scales (see Fig.  \ref{fig:length_clean_g_var}), which reduces the finite size effects, including the Majorana energy splitting oscillations. The net result is the robust end-to-end correlated ``phase'' shown in Fig. \ref{fig:SM_disorder_strong_coupl}(d). We note that differences between the left and right LDOS maps in panels (b) and (c) are associated with (i) the presence of accidental, uncorrelated, disorder-induced  zero-energy ABSs and (ii) the difference in the ``visibility'' of the left and right MZMs. Also note that the correlation map in panel (d) includes correlations between finite energy modes, up to $\omega_{\rm max} = 20~\mu$eV. As mentioned above, reducing this window  will reduce the region characterized by a finite end-to-end correlation function. However, in the strong coupling case the correlated region remains large (as compared to the ``natural'' energy scale $\gamma$) for $\omega_{\rm max}$ values down to a few $\mu$eV. We should also point out that the (accidental) disorder-induced ABSs can generate false positive signatures in the correlation function, when such ABSs accidentally emerge at both ends of the system for certain parameter values. However, these accidental correlations correspond to small ``islands'' in the parameter space, since the dependence of the left and right ABSs on the control parameters is different. In addition, the difference in the  ``visibility'' of the left and right MZMs can result in false negative correlation signatures, if (at least) one of the MZMs is ``pushed away'' from the end of the system by the disorder potential (so that its contribution to the LDOS becomes smaller than $\rho_{\rm min}$). Nonetheless, our ``operational criterion'' for identifying a topological SC phase --- having a correlated region significantly larger that the ``natural'' energy (or magnetic field) scale controlled by $\gamma$ --- holds in the presence of both false positives and false negatives, as they affect a relatively small fraction of the correlated region.  

\begin{figure}[t]
\centering
\includegraphics[width=0.5\textwidth]{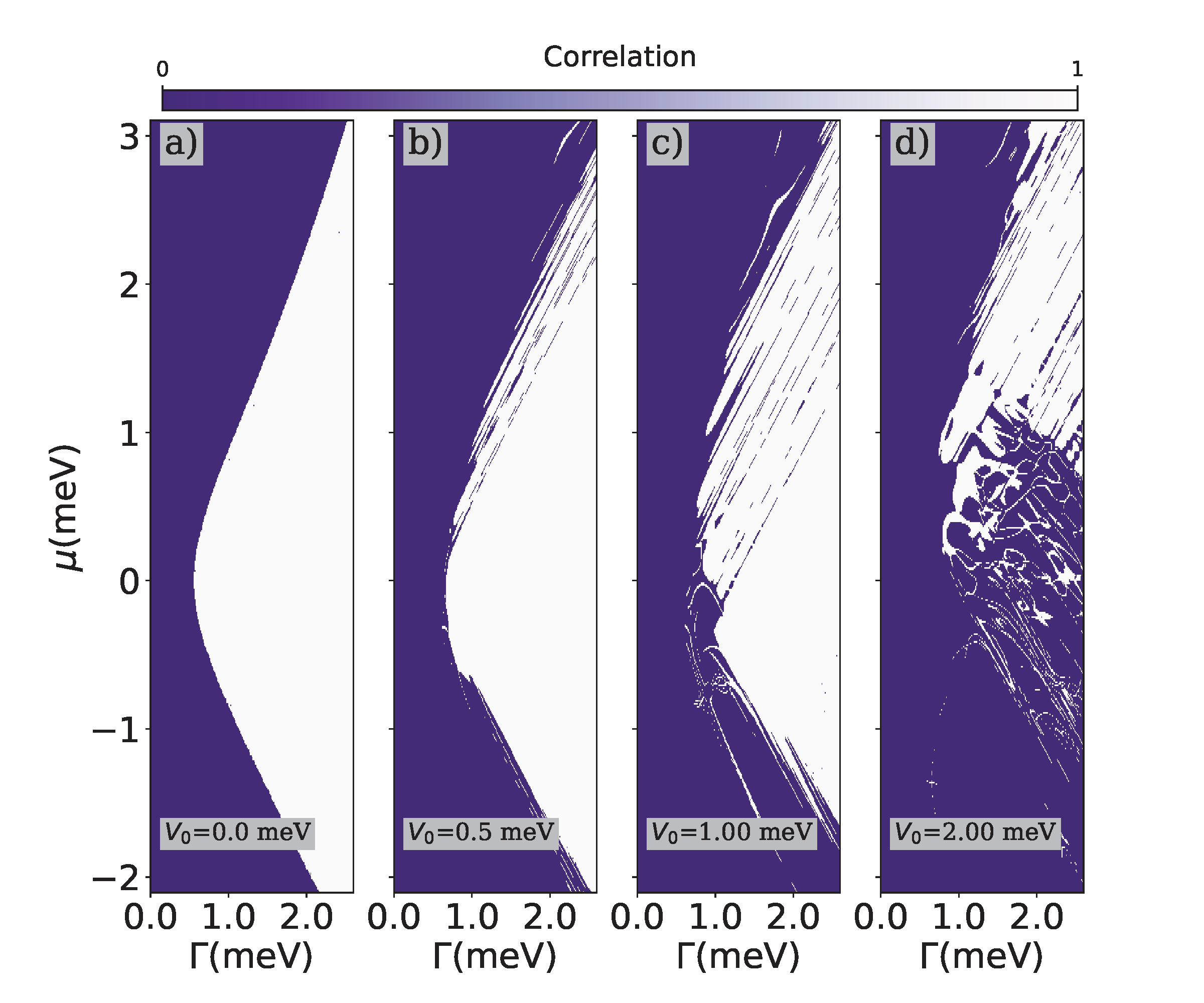}
\caption{Evolution of the end-to-end correlation map with the disorder strength for a system with strong SM-SC coupling ($\gamma=0.6~$meV). The values of the disorder potential amplitude, are (going from left to right): (a) $V_0=0$, (b) $V_0=0.5~$meV, (c) $V_0=1.0~$meV and (d) $V_0=2.0~$meV. Comparison with Fig. \ref{fig:dis_var_g_weak} shows that enhancing the SM-SC coupling strength alleviates the effects of disorder. In particular, for $V_0=2~$meV the strongly coupled system supports a topological SC phase over a large control parameter area [panel (d)], while the weakly coupled system is topologically trivial within the relevant parameter range.}           \label{fig:dis_var_g_strong}
\end{figure}

To further support our claim that stronger SM-SC coupling reduces the effects of a disorder potential (inside the SM), we calculate the correlation maps for different disorder amplitudes and compare the results, which are shown in Fig. \ref{fig:dis_var_g_strong}, with the corresponding panels in Fig. \ref{fig:dis_var_g_weak}. As in the weak coupling case, the presence of disorder reduces the correlated region, starting with the low-chemical potential regime. However, in the strong coupling case there is a large correlated region (with an area about half the area corresponding to a clean system) even at $V_0=2~$meV; by contrast, the corresponding weak coupling correlation map consists of small islands. We also note that the stable correlated region tends to ``migrate'' tovards the top right corner of the ``phase diagram,'' i.e., toward larger values of the chemical potential and Zeeman field. 

\begin{figure}[t]
    \centering
    \includegraphics[width=0.5\textwidth]{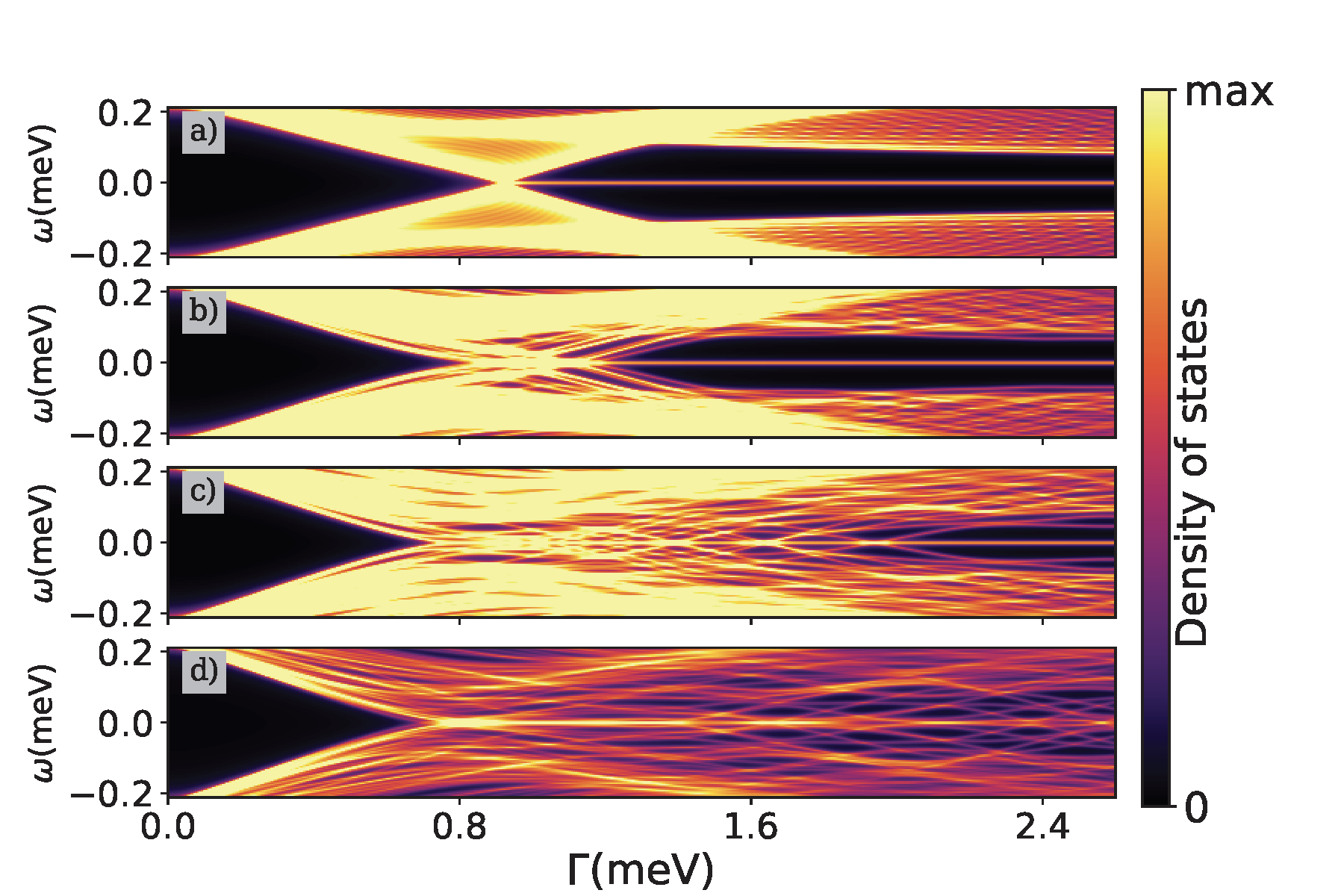}
    \caption{DOS as a function of Zeeman field and energy for a hybrid system with strong SM-SC coupling, $\gamma=0.6~$meV,  fixed chemical potential, $\mu = 0.7~$meV, and different disorder strengths. The values of the disorder potential amplitude, are (going from left to right): (a) $V_0=0$, (b) $V_0=0.5~$meV, (c) $V_0=1.0~$meV and (d) $V_0=2.0~$meV. The quasiparticle gap separating the Majorana mode from the rest of the spectrum (at high-enough Zeeman field values) is reduced by disorder, but ``survives'' up to disorder strengths $V_0>1~$meV, while the Majorana mode can even  be identified in the lower panel ($V_0=2~$meV.) These robust MZMs are responsible for the large correlated regions in Fig. \ref{fig:dis_var_g_strong}.}
	\label{fig:semi_dis_g_strong}
\end{figure}

Finally, to identify the modes responsible for the end-to-end correlations, we calculate the DOS as function of energy and Zeeman field  along representative horizontal cuts (i.e., at constant chemical potential), as well as the LDOS as a function of position for parameters corresponding to various low-energy features (e.g., ner zero-energy modes). As an example, in Fig. \ref{fig:semi_dis_g_strong} we show the DOS spectra corresponding to $\mu=0.7~$meV and different disorder strengths. Comparison with the weak coupling spectra in Fig. \ref{fig:semi_dis_g_weak} reveals several important features. As expected, the closing and reopening of the bulk gap of the clean system [Fig. \ref{fig:semi_dis_g_strong}(a)] occurs at a higher value of the Zeeman field [as compared to Fig. \ref{fig:semi_dis_g_weak}(a)], $\Gamma_c=\sqrt{\gamma^2+\Delta_0^2}\approx 0.92~$meV. The weight within the SM wire of the Majorana mode emerging at $\Gamma > \Gamma_c$  is smaller than that corresponding to the weakly coupled system, consistent with the expected fraction, $\Delta_0/(\gamma + \Delta_0)$, and the Majorana energy splitting oscillations are strongly suppresses as a result of the corresponding wave functions being more localized near the ends of the system (also, see Fig. \ref{fig:length_clean_g_var}). Upon increasing the disorder strength, low-energy modes proliferate, starting with the parameter region near the topological phase boundary, while the quasiparticle gap at larger Zeeman field values gets reduced and, eventually, collapses. While this behavior is qualitatively similar to that of weak coupled systems, there are significant quantitative differences: the disorder strength associated with the collapse of quasiparticle gap is larger (above $1~$meV) and, most importantly, the Majorana modes survive up to disorder strengths $V_0>2~$meV, generating strong end-to-end correlation signatures over large areas in parameter space, as illustrated in Fig. \ref{fig:dis_var_g_strong}. 

The main conclusions of our analysis of the effects generated by charge impurity disorder in the semiconductor wire are: (a) The presence of MZMs localized near the ends of the wire generates relatively large regions in parameter space characterized by a finite end-to-end correlation function.  (b) Upon increasing the disorder strength, the correlated regions ``shrink'' toward larger values of the chemical potential and Zeeman field (i.e., the top right corner of the ``phase diagram'') and  eventually break down into small ``islands'' associated with accidental, disorder-induced low-energy modes. (c) Increasing the strength of the effective SM-SC coupling enhances the robustness of the topological SC phase (and MZMs) against disorder and reduces the characteristic length scale of the low-energy modes.   

\subsection{Disorder in the superconductor}             \label{sec:Dis_super}

In this section we investigate the low-energy effects due to the presence of disorder in the parent superconductor. We emphasize that  this type of disorder is {\em required} to ensure a robust superconducting proximity effect \cite{PhysRevB.106.085429}. On the other hand, the presence of disorder in the parent SC (due to, e.g., surface roughness) is expected to generate ``induced disorder'' in the SM wire. To understand the effects of this ``induced  disorder,'' we use a phenomenological disorder model based on the microscopic calculation from Ref. \onlinecite{PhysRevB.106.085429}. More specifically, we model the proximity-effect generated by a thin SC film with surface disorder as a self-energy contribution with position-dependent effective SM-SC coupling, as given by Eq. (\ref{eq:self_energy}). Examples of position-dependent SM-SC couplings are shown in Fig. \ref{fig:SM_disorder_profile} (bottom panels). The strength of the SM-SC coupling is characterized by the average value, $\langle \gamma\rangle$, while the strength of the effective disorder is characterized by the values of $\langle \gamma^2\rangle$, where $\langle \dots \rangle$ means averaging over position. Note that, for a given value of $\langle \gamma\rangle$, a larger  $\langle \gamma^2\rangle$ implies more disorder. As in the previous section, we consider both weak coupling, $\langle \gamma\rangle=0.15~$meV, and strong coupling, $\langle \gamma\rangle=0.6~$meV, as well as different disorder ``strengths,'' which we characterize by the dimensionless ratio $\gamma_s=\sqrt{\langle \gamma^2\rangle}/\langle \gamma\rangle \geq 1$.

\begin{figure}[t]
\centering
\includegraphics[width=0.5\textwidth]{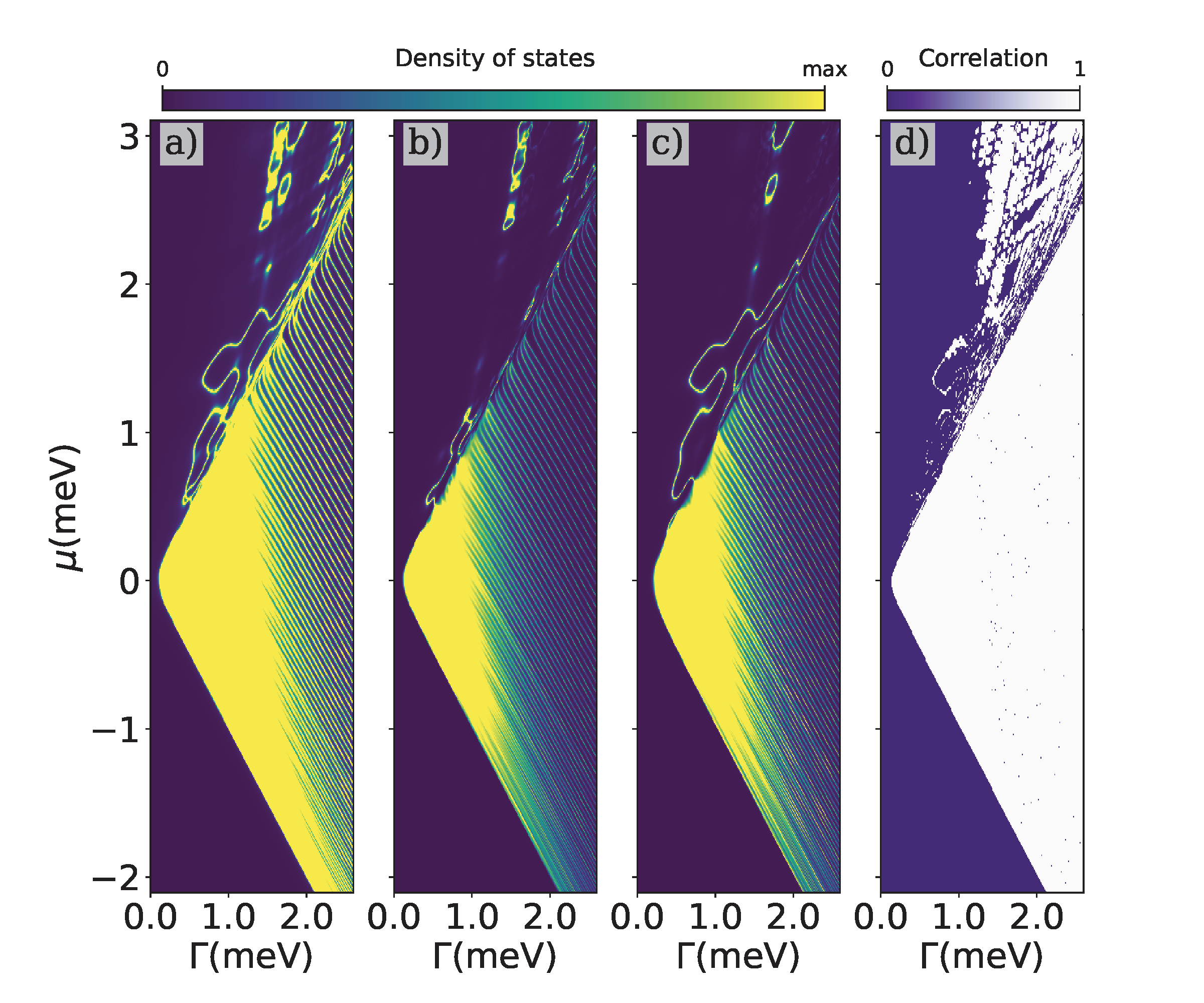}
\caption{Disordered system of length $L=5~\mu$m with weak SM-SC coupling, $\langle\gamma\rangle=0.15~$meV, and SC disorder of strength $\gamma_s=1.76$. (a) Zero-energy DOS as a function of Zeeman field, $\Gamma$, and chemical potential, $\mu$. (b) Map of the zero-energy LDOS at the left end of the wire. (c) Map of the zero-energy LDOS at the right end of the wire. (d) End-to-end correlation map. Note the similarity with the corresponding clean maps in Fig. \ref{fig:clean}, which suggests that in the weak coupling regime the effects of SC disorder are minimal. While the topological region is basically unaffected, disorder-induced low-energy modes emerge at large values of the chemical potential, in the topologically-trivial region.}  \label{fig:super_dis_g_weak}
\end{figure}

First, we consider a hybrid system with weak SM-SC coupling,  $\langle \gamma\rangle=0.15~$meV, and dimensionless disorder ``strength'' $\gamma_s= 1.76$. 
The corresponding DOS, LDOS, and correlation maps are shown in Fig. \ref{fig:super_dis_g_weak}.  Panel (a) shows the zero-energy DOS as a function of $\mu$ and $\Gamma$, (b) and (c) are the left and right LDOS maps, respectively, and (d) shows the end-to-end correlation map. We point out the negligible impact of the SC disorder on the ``phase diagrams,'' which are similar to those corresponding to a clean system (Fig. \ref{fig:clean}). In particular the low chemical potential regime is practically unaffected by the presence of SC disorder, in sharp contrast with the SM disorder case discussed in the previous section. 
\begin{figure}[t]
    \centering
    \includegraphics[width=0.5\textwidth]{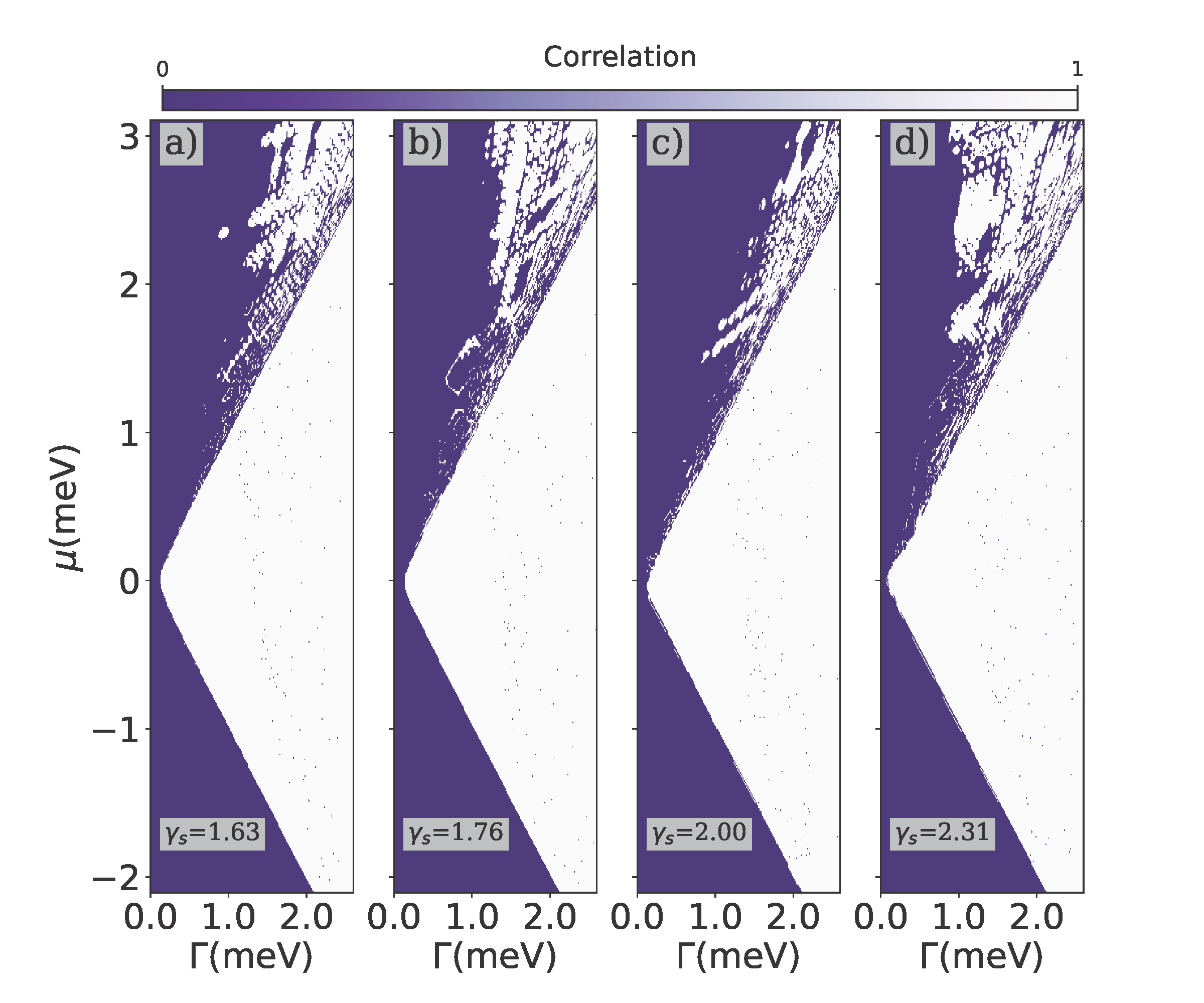}
    \caption{End-to-end correlation maps for a hybrid system with weak SM-SC coupling in the presence of SC disorder. The value of the average SM-SC coupling is $\langle \gamma \rangle = 0.15~$meV, while  the dimensionless disorder strength, $\gamma_s$, is: (a) 1.63, (b) 1.76, (c) 2.00 and (d) 2.31. Notice that the topological region (see Fig. \ref{fig:clean}) is practically unaffected by the presence of disorder in the parent SC, while the trivial correlated region (at larger values of the chemical potential) is highly fragmented.}
    \label{fig:super_dis_g_015}
\end{figure}
On the other hand, one can notice some disorder effects at large $\mu$ values, inside the topologically-trivial region. These low-energy features are associated with disorder-induced ABSs localized near the ends of the system. Indeed, almost every zero-energy DOS feature occurring outside (i.e., above) the topological ``triangle'' in  Fig. \ref{fig:super_dis_g_weak}(a) can be identified in panels (b) or (c), i.e., in the LDOS at the ends of the wire. This implies that (almost all) zero-energy modes have significant weight near one of the ends of the system.
On the other hand, the large trivial correlated area outside the phase boundary characterizing the clean system --- see Fig. \ref{fig:clean}(d) --- is broken into disconnected ``islands.'' This is because the presence of SC disorder breaks the symmetry of the clean system and, as a consequence, the low-energy ABSs at the left and right ends of the wire have, in general, different energies. 

To strengthen these observations, we calculate the correlation maps for a weakly coupled system, $\langle \gamma\rangle=0.15~$meV, with SC disorder having different ``strength'' values, $\gamma_s$. We should point out that, while in the case of SM disorder we consider a single disorder profile  (multiplied by the overall amplitude $V_0$), for SC disorder different $\gamma_s$ values correspond to different disorder realizations (see Fig. \ref{fig:SM_disorder_profile}), since multiplication by an overall factor would affect the average coupling, $\langle \gamma\rangle$, which in this calculation has a fixed value.  The results are shown in Fig. \ref{fig:super_dis_g_015}. Clearly, in the weak coupling regime disorder inside the superconductor has a small effect on the low-energy physics of the hybrid structure, even for large $\gamma_s$ values, when the position-dependent fluctuations of the effective SM-SC coupling are huge (see Fig. \ref{fig:SM_disorder_profile}). Intuitively, this is not surprising, as we are considering proximity-induced disorder in a system characterized by a weak proximity effect. The low-energy modes (i.e., MZMs for control parameter values inside the ``topological triangle'' and trivial ABSs outside this region, in the large chemical potential regime) have most of their spectral weigth inside the SM wire and, consequently, are weakly affected by the presence of disorder in the parent superconductor. The most notable effect of SC disorder occurs in the trivial region (within the large chemical potential regime). In contrast with the clean case shown in Fig. \ref{fig:clean} (d), the trivial correlated area is highly fragmented. Also, for the disorder realizations corresponding to panels (a), (b), and (d) (see Fig. 
\ref{fig:super_dis_g_015}), this correlated region typically extends to lower values of the Zeeman field, as compared to the clean case. This behavior has two main sources. On the one hand, the presence of disorder can lower the energy of the ABSs, which can emerge below the correlation threshold, $\omega_{\rm max}$, at lower Zeeman field values (as compared to the clean case). On the other hand, disorder breaks the symmetry of the wire and, consequently, the ABSs localized at the opposite ends can have different energies (which results in the absence of end-to-end correlations). Note that the breaking of the left-right symmetry can also be done in a controlled manner, by applying local electrostatic potentials. We emphasize that the trivial region characterized by nonzero correlations shrinks upon reducing the energy ``window'' $\omega_{\rm max}$, while the characteristic size of the correlated ``islands'' decreases upon lowering the threshold LDOS peak height $\rho_{\rm min}$. By contrast, the correlation function within topological region is not affected by these parameter changes.

\begin{figure}[t]
    \centering
    \includegraphics[width=0.5\textwidth]{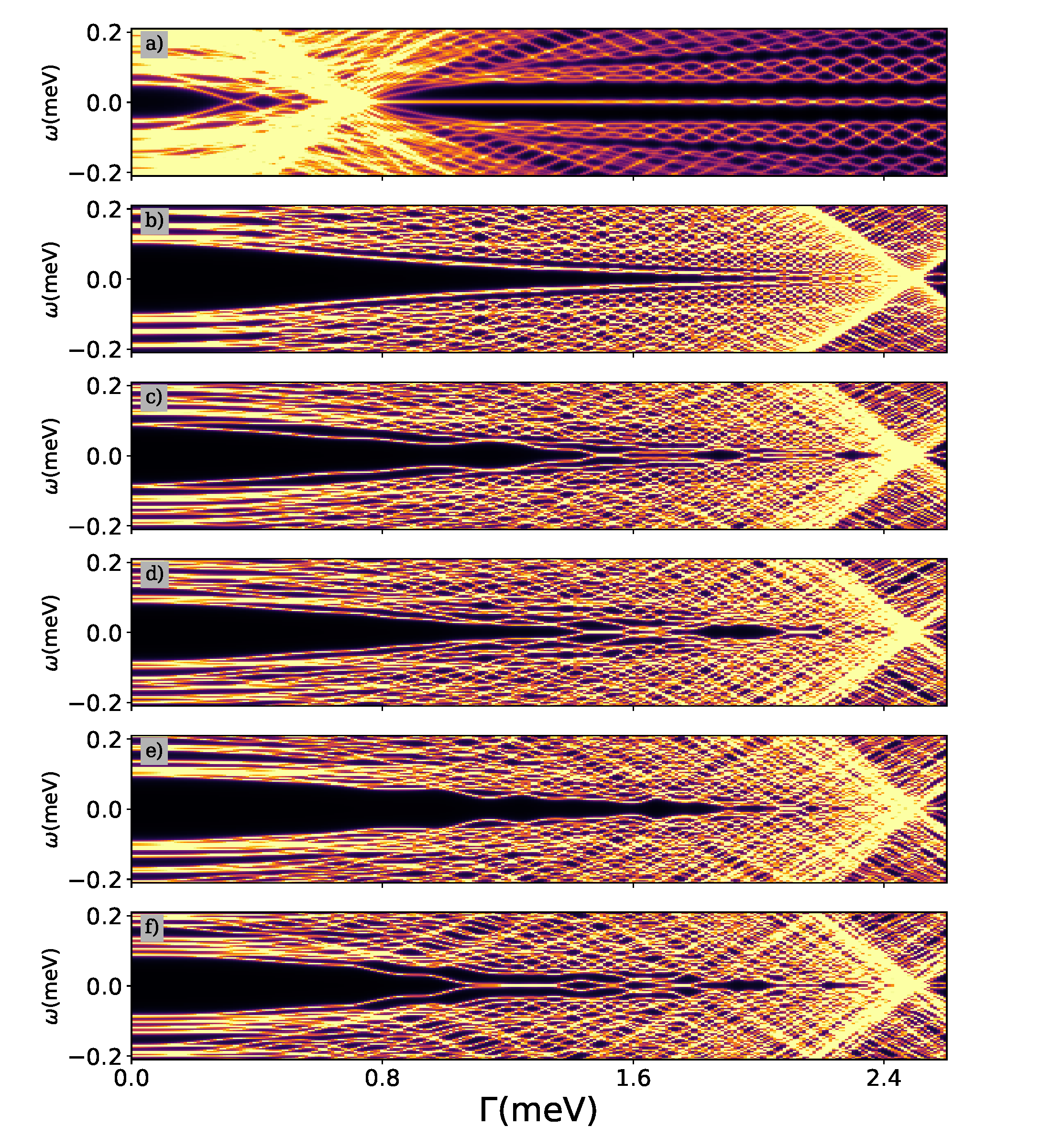}
    \caption{DOS as a function of Zeeman field and energy for a hybrid system with weak SM-SC coupling ($\gamma=0.15~$meV) and disorder in the parent SC. The top panel (a) corresponds to $\mu=0.7~$meV and dimensionless disorder strength $\gamma_s=2.31$, while the rest of the panels represent horizontal cuts at  $\mu=2.5~$meV through ``phase diagrams'' corresponding to different disorder strengths values (also see Fig. \ref{fig:super_dis_g_015}): (b) $\gamma_s=1$ (clean system),  (c) $\gamma_s=1.63$, (d) $\gamma_s=1.76$, (e) $\gamma_s=2.00$ and (f) $\gamma_s=2.31$.  The Majorana modes are weakly affected by the SC disorder and generate end-to-end correlations  throughout the whole topological region (see Fig. \ref{fig:super_dis_g_015}). The correlations outside the topological  region (at large chemical potential values) are generated by low-energy ABSs, which are affected by the presence of disorder and can become gapless --- see, e.g., panel (f) for $1.1 \lesssim \Gamma \lesssim 1.4~$meV.}
\label{fig:super_dis_g_var_weak}
\end{figure}

The topological or trivial nature of the correlations shown in Fig. \ref{fig:super_dis_g_015} can be easily identified by calculating the dependence of the DOS on energy and control parameters and can be further confirmed by determining the position dependence of the LDOS. As an example, in  Fig. \ref{fig:super_dis_g_var_weak} we show the dependence of the DOS on energy and Zeeman field along constant chemical potential cuts through the correlation maps. The top panel, which corresponds to $\mu=0.7~$meV and $\gamma_s=2.31$, illustrates the robustness of the Majorana modes against (strong) SC disorder in the weak SM-SC coupling limit. Note that the quasiparticle gap separating the Majorana mode from the finite energy spectrum is practically identical with the clean gap shown in Fig. \ref{fig:semi_dis_g_weak} (a). The only notable effect of disorder is to induce low-energy states in the vicinity of the topological quantum phase transition (i.e., $\Gamma \approx 0.7~$meV), including trivial zero-energy modes. To clarify the nature of the end-to-end correlations emerging outside the topological region, we consider horizontal cuts through the correlation maps in Fig. \ref{fig:super_dis_g_015} corresponding to $\mu=2.5~$meV. For comparison, in \ref{fig:super_dis_g_var_weak} (b) we show the corresponding clean spectrum. Note the presence of low-energy (intrinsic) ABSs, which generate finite end-to-end correlations outside the topological region (also see Fig. \ref{fig:clean}). The presence of disorder can further reduce the energy of these trivial modes, which can ``stick'' near zero energy over finite Zeeman field ranges. The most striking example can be seen in panel (f) for Zeeman field values $1.1 \lesssim \Gamma \lesssim 1.4~$meV. Note that this near-zero energy trivial mode is responsible for the relatively large correlated island that emerges in this parameter range in Fig. \ref{fig:super_dis_g_015}. We conclude that, in the weak SM-SC coupling limit, SC disorder has a minimal effect on the low-energy physics of the hybrid system  (most notably on the trivial low-energy ABSs that emerge at large chemical potential) and, practically, does not modify the topological phase diagram. 

\begin{figure}[t]
    \centering
    \includegraphics[width=0.5\textwidth]{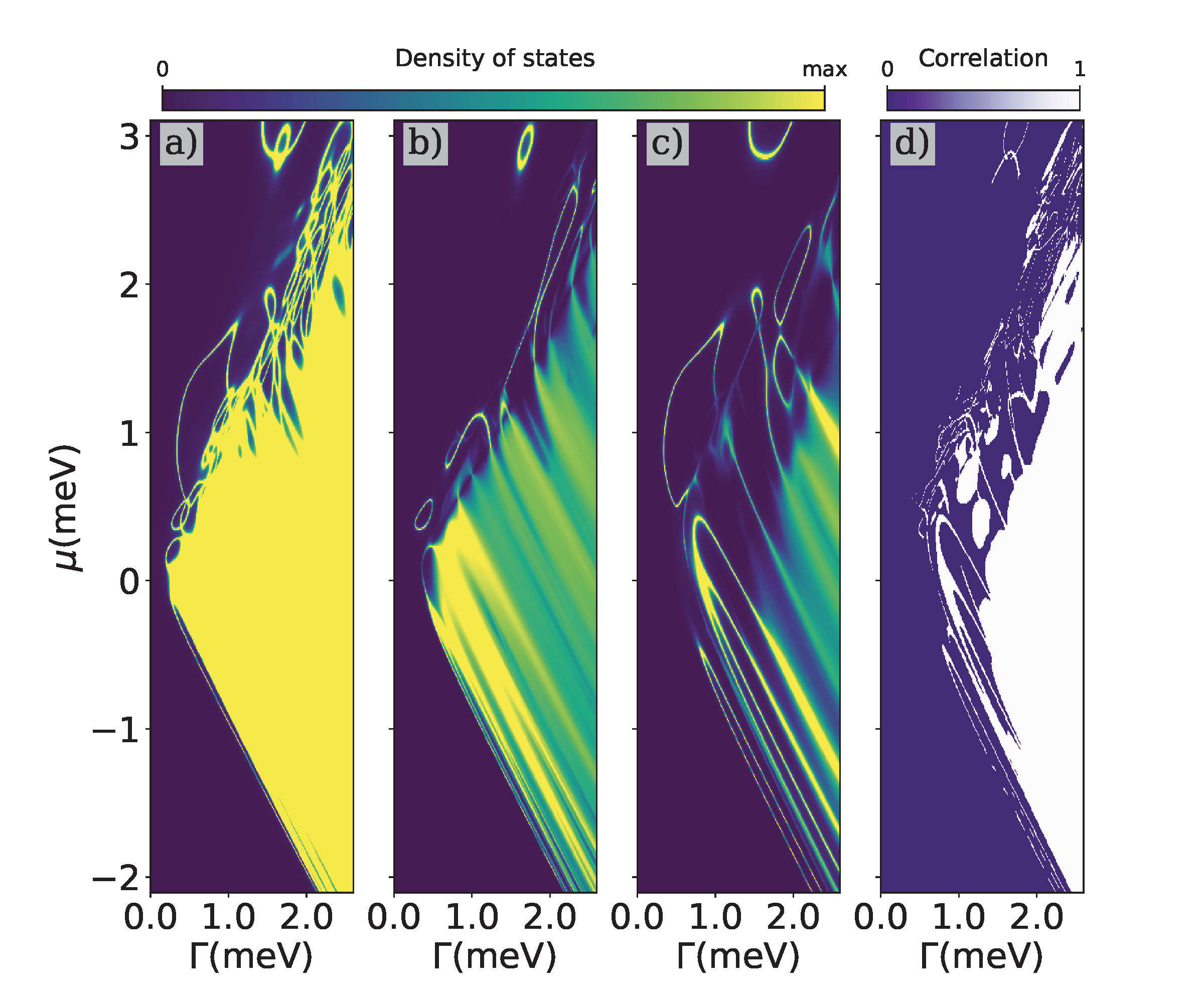}
    \caption{Hybrid system with SC disorder and strong SM-SC coupling,  $\langle \gamma\rangle = 0.6~$meV. The dimensionless disorder strength is  $\gamma_s=1.8$. All other parameters are  the same as in Fig. \ref{fig:super_dis_g_weak}. (a) Zero-energy DOS as a function of Zeeman field and chemical potential. (b-c) Maps of the LDOS at the left and right ends of the wire, respectively. (d) End-to-end correlation map. Note that, as compared to the weakly-coupled results shown in Fig. \ref{fig:super_dis_g_weak}, the SC disorder effects are significantly stronger, particularly near the phase boundary with $\mu > 0$. Howewer, a large topological region survives at higher values of the Zeeman field, $\Gamma \gtrsim 1.5~$meV.}
	\label{fig:super_dis_g_strong}
\end{figure}

We repeat the above analysis for a system in the strongly coupled regime with $\langle \gamma \rangle = 0.6~$meV. The DOS, LDOS, and correlation map corresponding to a dimensionless disorder ``strength'' $\gamma_s = 1.8$ are shown in Fig. \ref{fig:super_dis_g_strong}. A straightforward comparison with Fig. \ref{fig:super_dis_g_weak} suggests that the disorder effects are substantially stronger in this regime, but a large topological region (characterized by the presence of zero energy states and finite end-to end correlations) still survives. Furthermore, comparison with the clean case (see Fig. \ref{fig:clean_strong}) shows the presence of zero energy states outside of the nominally-topological region --- yellow area in Fig. \ref{fig:clean_strong} (a) --- including near $\mu=0$ where the finite energy DOS (yellow) extends to lower values of the Zeeman field, i.e., below $\Gamma = \langle \gamma \rangle = 0.6~$meV. This is consistent with our earlier observation that disorder reduces the quasiparticle gap, which can vanish for certain parameter values outside the topological region. However, in clean wires and in systems with SM disorder there are no zero energy states below the minimum Zeeman field value $\Gamma = \gamma$ (see Figs. \ref{fig:clean}, \ref{fig:clean_strong},  \ref{fig:SM_disorder_weak_coupl}, and \ref{fig:SM_disorder_strong_coupl}).  One can understand this lowering of the minimum Zeeman field associated with the emergence of zero energy modes as a consequence of the inhomogeneous SM-SC coupling. More specifically, if a (sufficiently long) segment of the wire has an average SM-SC coupling significantly lower than $\langle \gamma \rangle$, it can host zero energy modes at Zeeman field values lower than $\langle \gamma \rangle$. Of course, the corresponding modes will (typically) not be localized near the ends of the system, so they will not contribute to the end-to-end correlation, which is illustrated by the result shown in panel (d).  
\begin{figure}[t]
    \centering
    \includegraphics[width=0.5\textwidth]{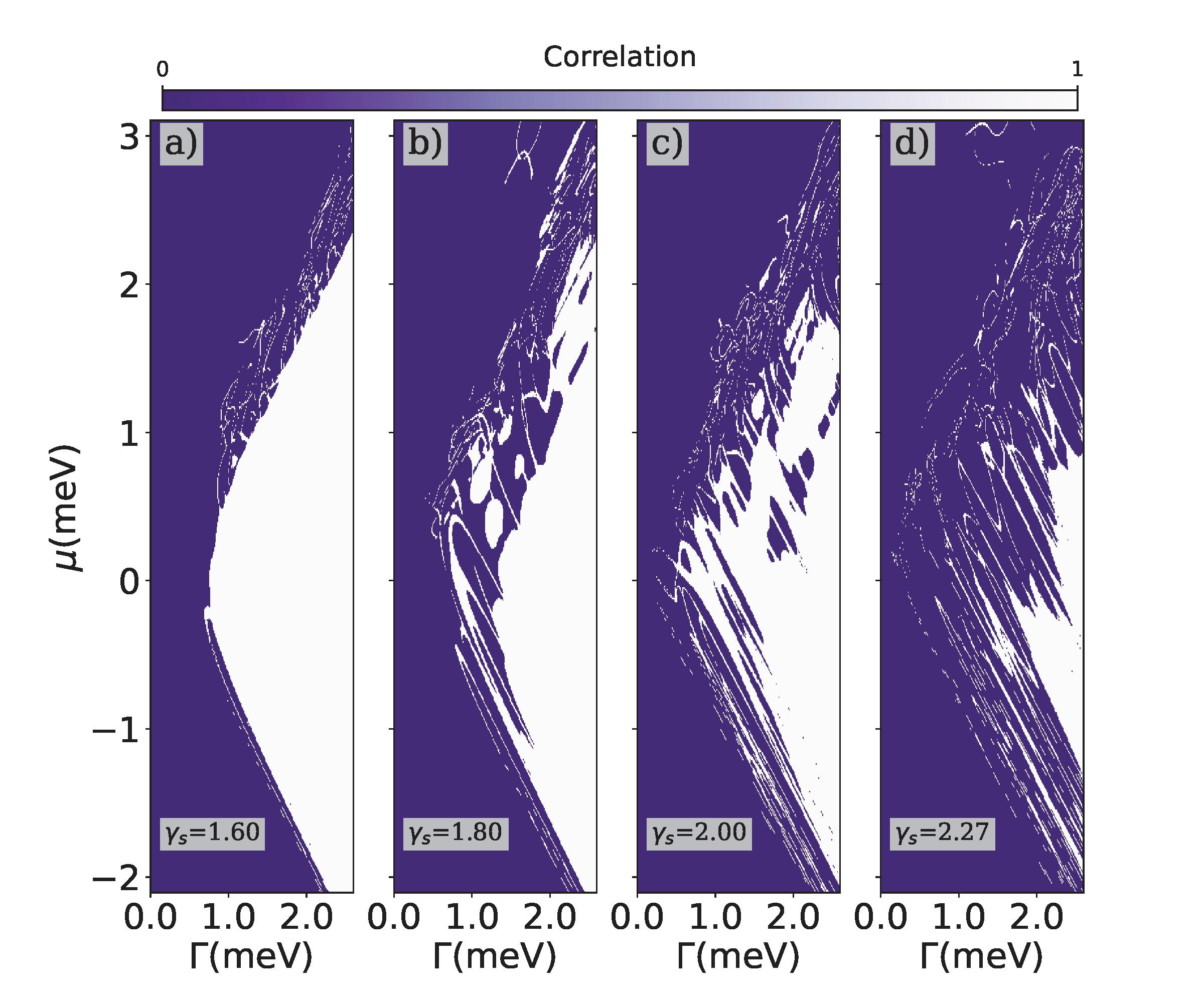}
    \caption{End-to-end correlation maps for a hybrid system with strong SM-SC coupling, $\langle \gamma \rangle = 0.6~$meV, and different values of the dimensionless disorder strength:  (a) $\gamma_s=1.6$, (b) $\gamma_s=1.8$, (c) $\gamma=2$, and (d) $\gamma_s=2.27$. Note that, upon increasing the disorder strength, the correlated region fragments near the phase boundary and shrinks toward larger Zeeman field values.}
	\label{fig:super_dis_g_06}
\end{figure} 
The LDOS at the left and right ends of the wire is shown in Fig. \ref{fig:super_dis_g_strong} panels (b) and (c), respectively.  We note a strong suppression of the LDOS, as compared with the clean case, particularly at the right end of the system. We emphasize that a low value of the zero-energy LDOS [e.g., the green/dark green region in panel (c) with $\Gamma \gtrsim 1.5~$meV] does not necessarily imply the absence of MZMs, or the fact that they acquire a finite energy gap. It may simply mean that the visibility of the Majorana modes was reduced in the presence of disorder due to spectral weight transfer away from the end of the system. In other words, the Majorana mode was ``pushed'' away from the edge of the system. Of course, this may lead to false negative signatures of Majorana physics in the LDOS or the end-to end correlation. By contrast, the opening of an energy gap (due to a small overlap of the Majorana wave functions) does not affect the end to end correlation, as long as the modes remain localized near the ends of the system. 
Nonetheless, the correlation map shown in Fig. \ref{fig:super_dis_g_strong} (d) clearly reveals that in the strong coupling regime the topological phase becomes unstable near the (clean) phase boundary, particularly for $\mu > 0$. For $\Gamma_s=1.8$, a substantial topological region survives at large values of the Zeeman field, $\Gamma \gtrsim 1.5~$meV.

\begin{figure}[t]
    \centering	
    \includegraphics[width=0.5\textwidth]{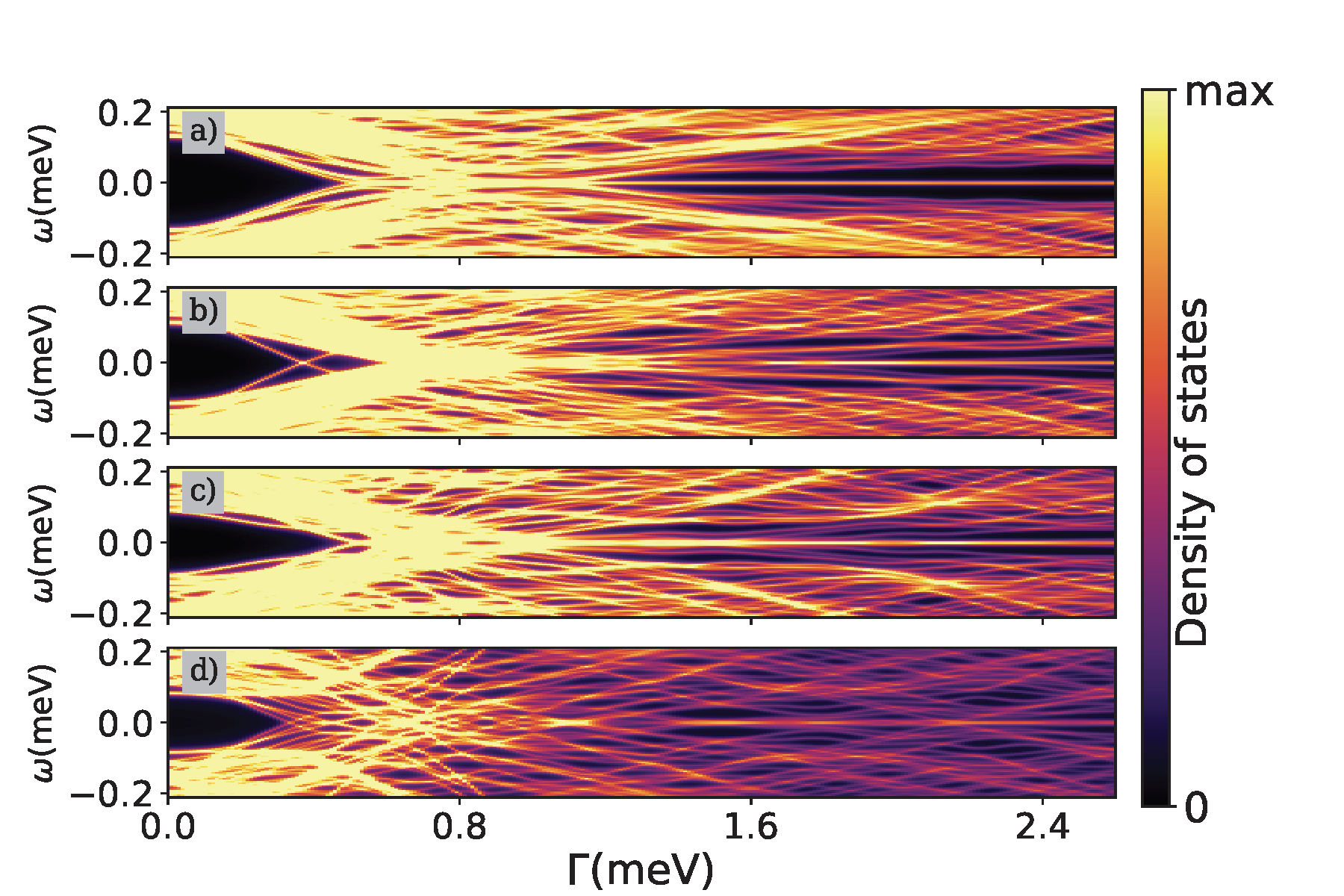}
    \caption{DOS as a function of Zeeman field and energy for a hybrid system with SC disorder,  strong SM-SC coupling, $\langle\gamma\rangle = 0.6~$meV, fixed chemical potential, $\mu = 0.7~$meV, and different values of the dimensionless disorder strength $\gamma_s$: (a) 1.6, (b) 1.8, (c) 2.0, and (d) 2.27. As compared to the weak coupling case (see, e.g.,  Fig. \ref{fig:super_dis_g_var_weak}), the effects of SC disorder are significantly stronger and result in low-energy subgap states emerging in the topological regime. These disorder-induced low-energy states hybridize with the Majorana modes, which can affect the visibility of the MZMs and, eventually, their stability.}
\label{fig:super_dis_g_var_strong}
\end{figure}

To better understand the dependence on the disorder strength, we calculate the end-to-end correlation maps for a system with strong SM-SC coupling (correspobnding to an average effective coupling strength $\langle \gamma \rangle = 0.6~$meV) and different values $\gamma_s$ of the dimensionless disorder strength. The results of the numerical calculation are shown in Fig. \ref{fig:super_dis_g_06}. Upon increasing $\gamma_s$ (i.e., the spatial inhomogeneity of the effective SM-SC coupling), the correlated reagion shrinks from a large area that practically coicides with the topological region of the clean system [panel (a) and, more generally, the regime $\gamma_s \lesssim 1.5$; for comparison, see Fig. \ref{fig:clean_strong}], to a relatively small region with $\Gamma \gtrsim 2~$meV and $|\mu|\lesssim 1~$meV for $\gamma_s=2.27$ [panel (d)]. Note that the main (connected) correlated region is surrounded by a highly fragmented area. While the correlation maps do not provide precise information about the location of the topological phase boundary in the $\Gamma-\mu$ plane, there is strong overlap between the topological region and the main (connected) correlated area. By contrast, small correlated ``islands'' and ``filamentary peninsulas'' are typically associated with accidental correlations generated by topologically trivial low-energy states. Hence, in the strong coupling regime (e.g., $\langle \gamma \rangle = 0.6~$meV), the topological SC phase can become practically  inaccesible if the Zeeman fields that can be realized in the laboratory (before the collapse of the parent SC gap) are only 2-3 times larger than the effective SM-SC coupling. For a more quantitative caracterization of this limitation, one would need specific information about the value of the SM-SC coupling and about the SC disorder (e.g., characteristic lengthscale, dimensionless disorder strength, etc.).

The reduction of the correlated area  by SC disorder (in the strong coupling limit) is the result of a proliferation of disorder-induced low-energy states. To illustrate this point, we calculate the DOS as a function of energy and Zeeman field for $\mu=0.7~$meV and different values of the dimensionless disorder $\gamma_s$. The results in Fig. \ref{fig:super_dis_g_var_strong} clearly show the presence of disorder induced subgap modes, which collapse toward zero energy upon increasing the disorder strength. Nonetheless, a robust (practically unsplit) near-zero Majorana mode can still be identified for $\gamma_s \lesssim 2.2$. We emphasize again that in the strong coupling limit  the characteristic length scales of the low-energy modes are significantly smaller than the size of the system ($L=5~\mu$m) and, consequently, the disorder-induced low-energy modes can overlap significantly with only one Majorana mode (at least for moderate and low values of the chemical potential). As a result, a Majorana mode can be ``pushed'' away from the edge of the wire, which affects  its visibility and may result in the vanishing of end-to-end correlation (i.e., a ``false negative''). 

The key conclusions of our analysis of SC disorder effects are: (a) In the weak SM-SC coupling limit the presence of disorder in the parent SC has negligible effects on the low-energy physics of the hybrid structure and practically does not affect the stability of the Majorana modes. (b) In the strong coupling regime, the presence of SC disorder results in the emergence of subgap states within the topological regime, but the Majorana modes are robust and clearly visible (i.e., localized near the end of the wire) for disorder strengths $\gamma_s = \sqrt{\langle\gamma^2\rangle}/\langle\gamma\rangle \lesssim 2$ (for $ \langle\gamma\rangle=0.6~$meV). Higher disorder strengths lead to a proliferation of disorder-induced low-energy modes. 

\begin{figure}[t]
    \subfloat{%
    \includegraphics[width=\columnwidth]{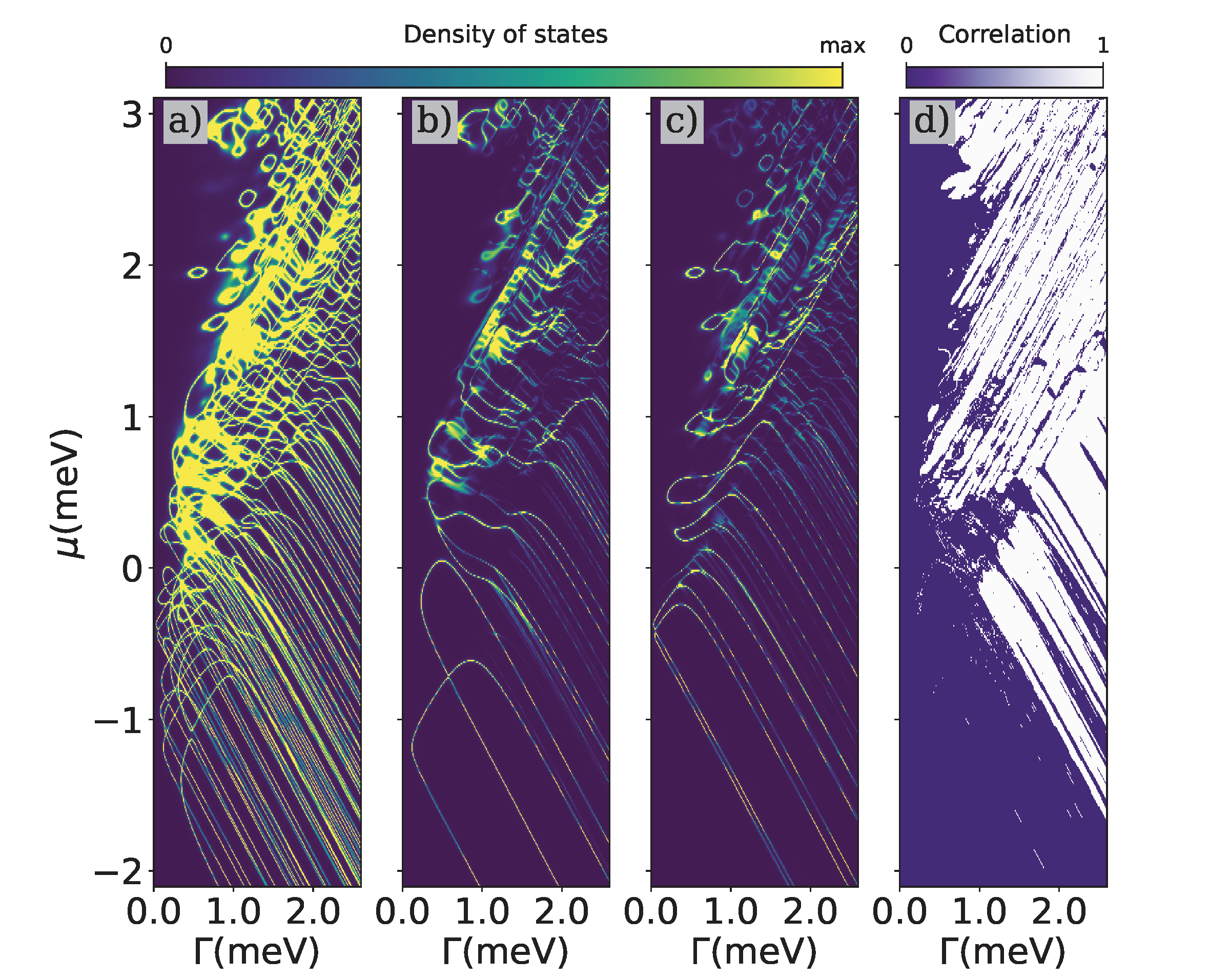}%
    }
    \hspace{0.4cm}
    \subfloat{%
    \includegraphics[width=\columnwidth]{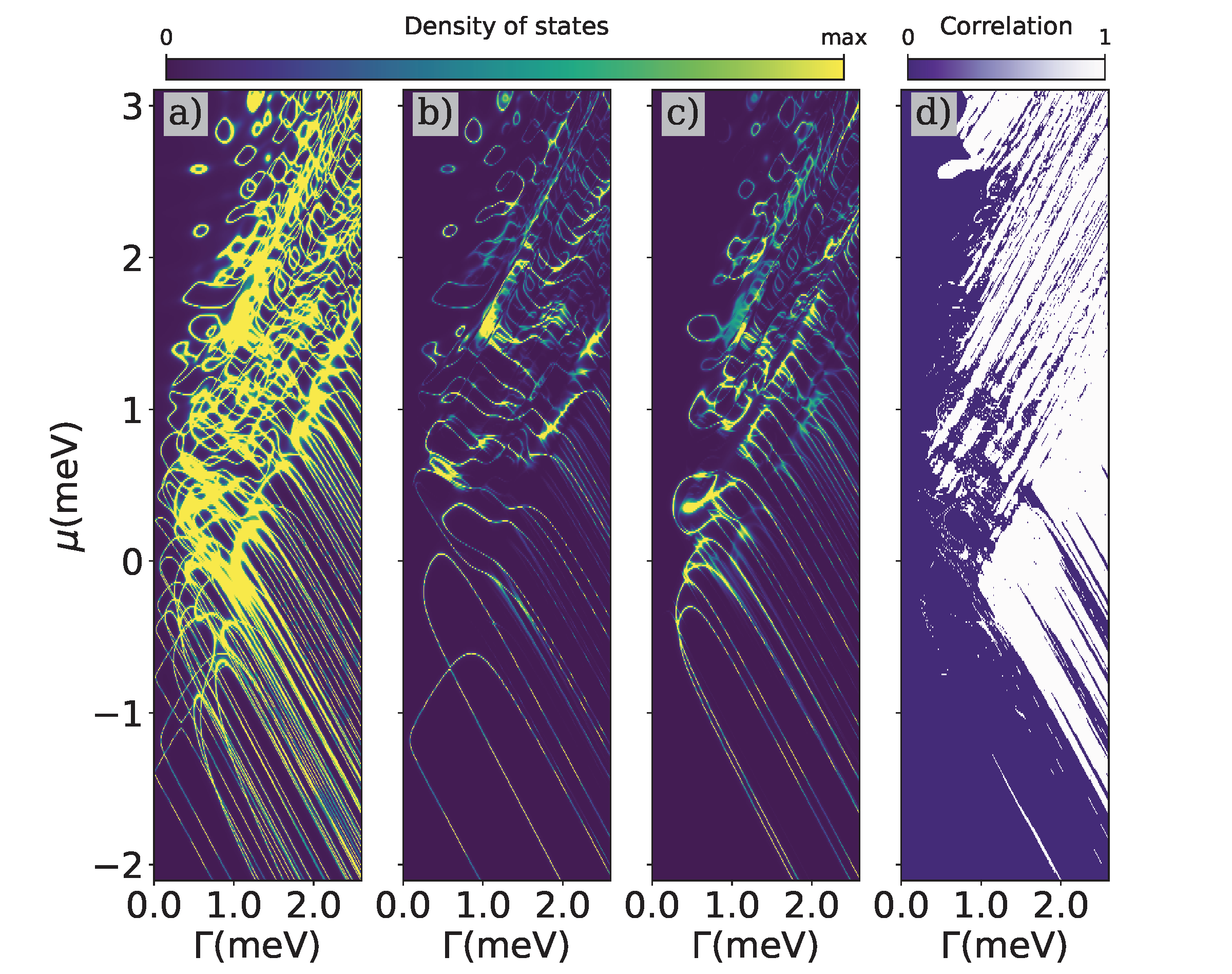}%
    }
    \caption{Disordered hybrid system of length $L=5~\mu$m with weak SM-SC coupling, $\langle\gamma\rangle = 0.15~$meV, in the presence of a SM disorder potential of amplitude $V_0=1~$meV and SC disorder of strength  $\gamma_s = 1.63$ (top panels) and $\gamma_s = 2.0$ (bottom panels). The Zeeman field--chemical potential maps correspond to: (a) the zero-energy DOS, (b) and (c) the zero-energy LDOS at the left ends of the wire, respectively, and (d) the end-to-end correlation function.}
    \label{fig:weak_sc_weak_sm}
\end{figure}

\subsection{Hybrid system with disorder in both the semiconductor and the superconductor}

In this section we investigate the combined effects of disorder in the SM and the SC, focusing on a few representative parameter regimes. First, we consider a weakly coupled system  with SM-SC coupling $\langle\gamma\rangle = 0.15~$meV in the presence of an  intermediate strength  SM disorder potential of amplitude $V_0=1~$meV. The maps showing the dependence of  the DOS, left and right ends LDOS, and end-to-end correlation on the Zeeman field and chemical potential are presented in  Fig. \ref{fig:weak_sc_weak_sm}, with the top panels corresponding to a dimensionless SC disorder strength $\gamma_s = 1.63$, while the bottom panels correspond to  $\gamma_s = 2$. The strong similarities between the corresponding top and bottom panels confirm that, in the weak coupling regime, the SC disorder has a negligible effect on the low-energy physics of the hybrid system. Further comparison with the correlation map in Fig. \ref{fig:dis_var_g_weak} (c) also indicates that adding SC disorder does not change significantly the correlated area characterizing the system with (only) SM disorder. On the other hand,  the LDOS maps [panels (b) and (c)] consist of a mesh of ``lines'', similar to the features shown in Fig. \ref{fig:SM_disorder_weak_coupl}, which are indicative of large energy splittings and, implicitly, of large characteristic length scales associated with the low-energy modes. Furthermore, a direct comparison of the  LDOS and DOS maps reveals the presence of a significant number of zero-energy modes in the bulk (with no signatures in the LDOS at the ends of the wire). This proliferation of disorder-induced low-energy modes, combined with the relatively large characteristic length scales associated with some of these states lead to an instability of the MZMs in weakly coupled finite systems with length $L$ on the order of a few microns (or less), which is the typical size of the systems available experimentally.  Finally, we point out that zero-energy modes can emerge at nearly zero Zeeman fields,  i.e., at $\Gamma < \langle\gamma\rangle$. As discussed in Section \ref{sec:Dis_super}, this is a direct consequence of the SM-SC coupling being nonuniform;  states localized in certain regions experience a weaker effective coupling.  By contrast, large (not fragmented) correlated regions only emerge at relatively large Zeeman field values, $\Gamma \gtrsim 1~$meV$~\gg \langle\gamma\rangle$. 

\begin{figure}[t]
    \subfloat{%
    \includegraphics[width=\columnwidth]{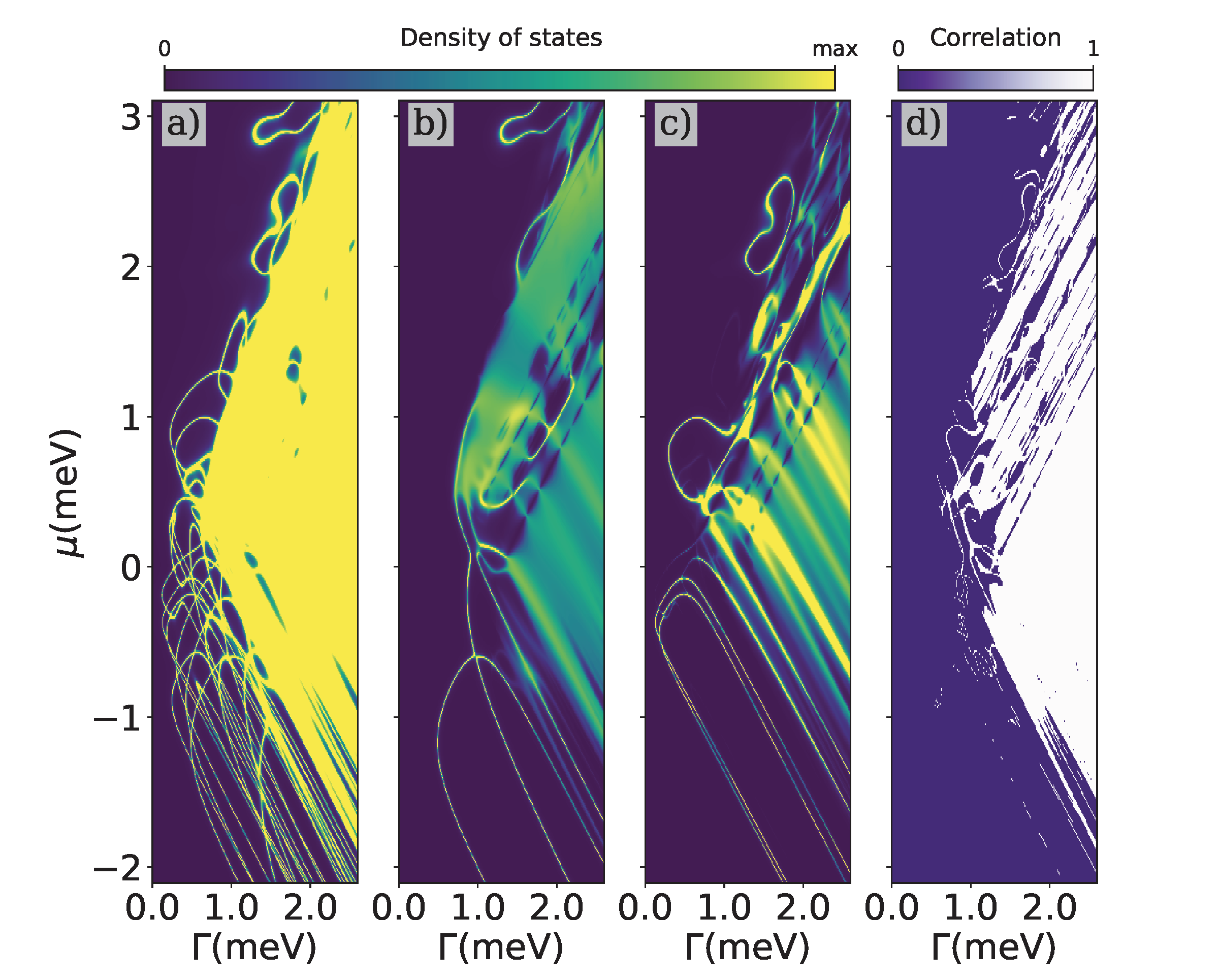}
    }
    \hspace{0.4cm}
    \subfloat{%
    \includegraphics[width=\columnwidth]{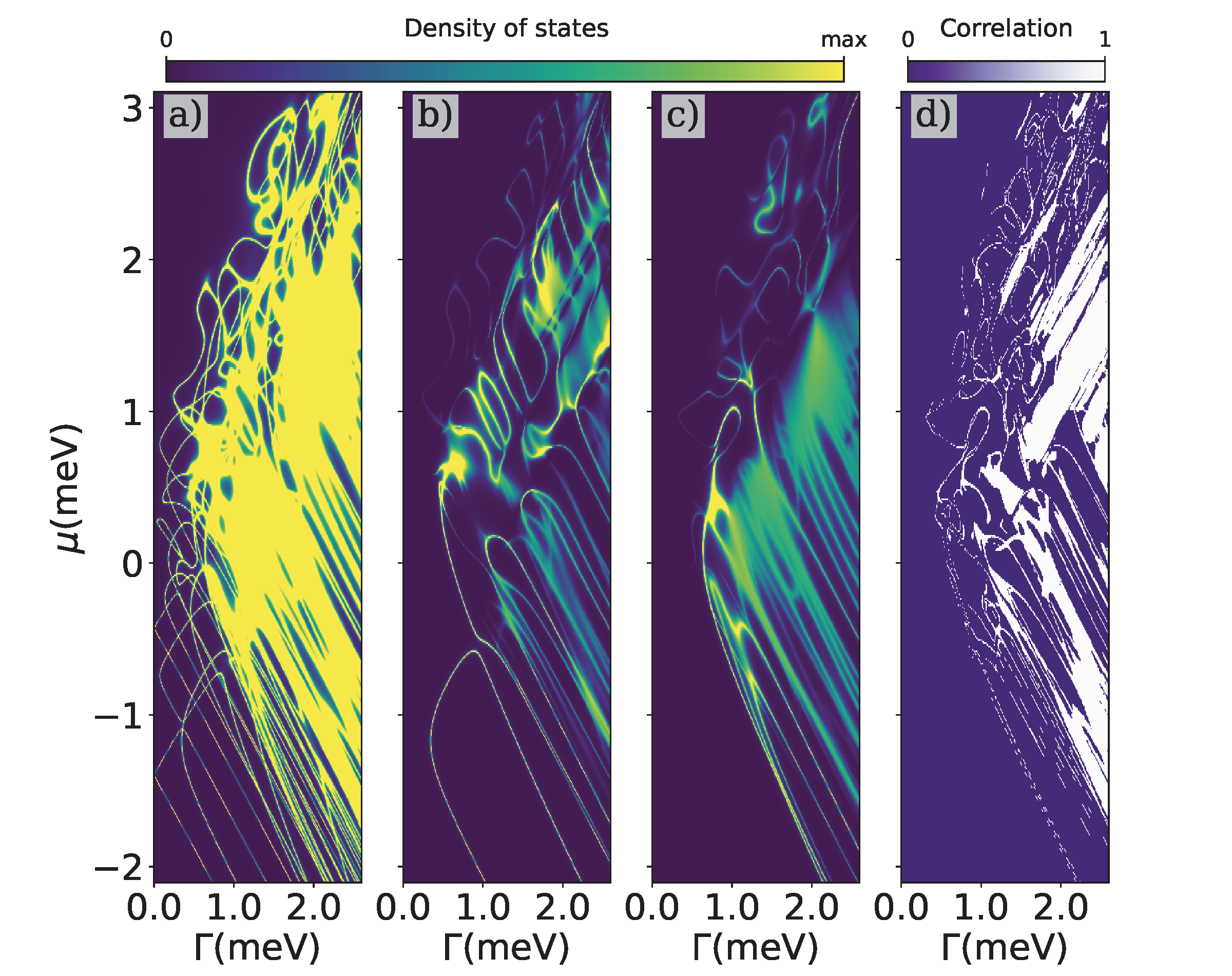}
    }
\caption{Zeeman field--chemical potential maps for a  disordered  hybrid system with strong SM-SC coupling, $\langle\gamma\rangle = 0.6~$meV, in the presence of a SM disorder potential of amplitude $V_0=1~$meV and SC disorder of strength  $\gamma_s = 1.63$ (top panels) and $\gamma_s = 2.0$ (bottom panels). The maps correspond to: (a) the zero-energy DOS, (b) and (c) the zero-energy LDOS at the left ends of the wire, respectively, and (d) the end-to-end correlation function. }
    \label{fig:weak_sc_weak_sm_strong_coupl}
\end{figure}

Next, we turn our attention to the strong coupling regime. In the preceding sections we have determined that increasing the effective SC-SM coupling (i) reduces the characteristic length scale of the low-energy modes and (ii) enhances the stability of the MZMs against SM disorder, but (iii) generates disorder-induced low-energy states and reduces the stability of MZMs in the presence of SC disorder. To characterize the net effect of these opposing trends, we consider a  strongly coupled system  with SM-SC coupling $\langle\gamma\rangle = 0.6~$meV in the presence of an  intermediate strength  SM disorder potential of amplitude $V_0=1~$meV, as well as SC disorder with dimensionless strength $\gamma_s = 1.63$ -- top panels of Fig. \ref{fig:weak_sc_weak_sm_strong_coupl} -- and  $\gamma_s = 2.0$ -- bottom panels of Fig. \ref{fig:weak_sc_weak_sm_strong_coupl}.The quantities shown in panels (a-d) are the same as in the corresponding panels of Fig. \ref{fig:weak_sc_weak_sm}. Several key features deserve some discussion. First,  comparing the (a) panels in Fig. \ref{fig:weak_sc_weak_sm_strong_coupl} with the (a) panel in Fig. \ref{fig:SM_disorder_strong_coupl} reveals a feature that we already discussed, but is better illustrated by this strong coupling example: in the presence of SC disorder the minimum field associated with the emergence of zero-energy states can be arbitrarily low, while in the absence of SC disorder all zero-energy features emerge above $\Gamma_{\rm min} = \gamma$. In practice, the low-field zero-energy features can reveal critical information regarding the disorder present in the hybrid structure. If the system is nearly clean (i.e., for weak disorder), zero-energy features (associated with MZMs) emerge above the critical Zeeman field $\Gamma_c =\sqrt{\gamma^2+\mu^2}$ that reaches the minimum value $\Gamma_{\rm min}=\gamma$ at a well-defined value of the chemical potential, $\mu=0$ (see, e.g., Figs. \ref{fig:clean} and \ref{fig:clean_strong}). If the system has intermediate/strong SM disorder, but no SC disorder, zero-energy features (typically associated with topologically trivial states) emerge above $\Gamma_{\rm min}=\gamma$ over a finite (and, possibly, large) range of chemical potential values; there are no zero-energy features below $\Gamma_{\rm min}$ (see, e.g., Figs. \ref{fig:SM_disorder_weak_coupl} and \ref{fig:SM_disorder_strong_coupl}). If, on the other hand, the system also has significant SC disorder, the zero-energy features can emerge at arbitrarily low Zeeman field  values (see Figs.  \ref{fig:weak_sc_weak_sm} and \ref{fig:weak_sc_weak_sm_strong_coupl}).

Second, we point out that the DOS and LDOS maps of the strongly coupled system exhibit continuous two-dimensional areas, particularly for $\gamma_s = 1.63$, in contrast with the ``filamentary'' structure of the maps characterizing the weak coupled system. This signals that robust MZMs are present within relatively large regions of the parameter space.  This indicates that the finite size effects are small and that the disorder-induced low-energy states hybridize with at most one MZM (rather than both); these properties are directly connected to the reduction of the characteristic length scales associated with strong SM-SC coupling. Finally, our third point is that, although SC disorder affects the visibility of the MZMs and, consequently, reduces the area of the regions characterized by finite LDOS and end-to-end correlations (particularly for $\gamma_s=2.0$), in the presence of moderate/low SC disorder (e.g., $\gamma_s=1.63$) the strongly coupled system is characterized by a topological phase and associated MZMs significantly more robust that their counterparts hosted by a weakly coupled system. More specifically, the (continuous) correlated regions in Fig. \ref{fig:weak_sc_weak_sm_strong_coupl} (d) (top panel), which emerge above a Zeeman field comparable to that characterizing the (continuous) correlated regions in Fig.\ref{fig:weak_sc_weak_sm}, are associated with MZMs that ``stick'' near zero energy  over large areas in the parameter space [see panels (b) and (c)]. Testing the non-Abelian properties of these MZMs would be meaningful, while a similar test using a weakly coupled system (with the same size and disorder strength) is pointless due to the large finite size effects.

\section{Summary and Discussion}

We have investigated numerically the low-energy effects of disorder in hybrid semiconductor-superconductor nanowires focusing on identifying parameter regimes that support topological superconductivity and Majorana zero modes in the presence of significant disorder levels in the semiconductor wire and in the parent superconductor. In addition to the ``standard'' control parameters -- chemical potential $\mu$ and Zeeman field $\Gamma$ -- we considered the dependence of the low-energy properties on the effective semiconductor-superconductor (SM-SC) coupling, $\gamma$, for systems with different types of disorder and different disorder strengths. Our analysis is based on numerical solutions for a one-dimensional tight-binding model, which is solved using a Green's function approach.  The superconducting proximity effect is incorporated through a self-energy term, while the efficiency of the numerical procedure is optimized by using a recursive Green's function method. Disorder in the semiconductor is introduced via a random potential with vanishing average and controllable overall amplitude, while disorder in the superconductor is modeled by a position-dependent effective SM-SC coupling, $\gamma(x)$, which enters as a prefactor in the self-energy.  The low-energy physics of the hybrid system is investigated by calculating the density of states (DOS), the local density of states (LDOS) at the ends of the wire, as well as a newly defined correlation function ($C$) that indicates the presence of low-energy modes (having the same energy) at the opposite edges. 

Our primary objective is to identify physical regimes of moderately disordered systems that would enable the realization of robust Majorana zero modes (MZMs) to be used in a future meaningful braiding experiment. From this perspective, being in a ``topological'' superconducting phase is necessary, but not sufficient. Of course, the finite size effects have to be small enough, i.e., the Majorana splitting energy should be less than a certain threshold. In addition, however, the Majorana modes should be localized close-enough to the ends of the wire. This condition may not necessarily be realized in the presence of intermediate/strong disorder, when (at least) one MZM is ``pushed away'' from the edge of the system and becomes ``invisible''. The system may be in a robust topological phase, but coupling to ``invisible'' MZMs would be practically difficult. Finally, these conditions should hold over a sufficiently large parameter range, rather than requiring extreme fine tuning.  

The overall picture revealed by our analysis is rather complex and characterized by several key requirements involving antithetical conditions. We find that the finite size effects, as well as the effects generated by the presence of disorder in the SM wire can be mitigated by enhancing the effective SM-SC coupling. More specifically, increasing $\gamma$ reduces the characteristic length scale of the MZMs and of the disorder-induced low-energy states. In turn, this reduces the energy splitting due to the overlap of the Majorana wave functions and alleviates the issue of having effectively delocalized disorder-induced low-energy states, i.e., disorder-induced subgap modes with characteristic length scales comparable with the length of the system; these modes can hybridize with a pair of Majoranas, destroying the stability of the topological phase. This length scale issue makes weakly coupled hybrid systems with lengths on the order of a few microns inappropriate for braiding experiments. In addition, as a result of the proximity-induced energy renormalization, enhancing the SM-SC coupling corresponds to a weaker effective disorder potential. However,  increasing the SM-SC coupling leads to higher requirements regarding the Zeeman field necessary to access the topological phase in systems with low SM disorder. Furthermore, disorder in the parent superconductor has a negligible effect in the weak coupling limit, but can affect the stability of the MZMs in the strongly coupled regime. In addition, the SM disorder has a stronger effect on the stability of MZMs in the low chemical potential regime, while SC disorder is most effective in the large chemical potential region of the phase diagram. 

Thus, the optimal regime for realizing robust, visible MZMs depends not only on some overall disorder strength, but also on the relative contributions from SM disorder and SC disorder. Our study provides useful practical guidelines for estimating these types of disorder based on the structure of the zero-energy  
LDOS maps (or zero-energy differential conductance maps)  calculated as a function of the control parameters (e.g., Zeeman field and chemical potential, or applied gate  potential), or measured experimentally. Assuming single band occupancy, low SM disorder is necessarily associated with low-field zero-energy features occurring {\em only} within a narrow chemical potential window (near $\mu=0$). By contrast, in the presence of moderate/strong SM disorder, the   low-field zero-energy features emerge within a chemical potential window of the order of the disorder potential amplitude. These features are typically associated with disorder-induced modes and do not generate end-to-end  correlations (i.e., they occur only at one end of the system). If SM disorder is low (i.e., if the characteristic amplitude of the disorder potential is significantly smaller than about $1~$meV), robust topological superconductivity can be realized in the intermediate coupling regime (with $\gamma$ comparable to the parent SC gap, $\Delta_0$) at relatively low values of the Zeeman field (i.e., $\Gamma$ larger than but  comparable to $\gamma$). The weak coupling regime is problematic (even in the clean limit) because of strong finite size effects (unless one considers very long wires). Unfortunately, the current experimental situation does not appear to be consistent with low SM disorder. Assuming intermediate/strong SM disorder, the next task is to evaluate the impact of the disorder (necessarily) present in the parent SC. We emphasize that SC disorder is required to ensure a robust proximity effect. However,  if the strength of SC disorder is low, the emergence of zero-energy features requires a minimum Zeeman field of the order $\gamma$ (the effective SM-SC coupling) over the entire chemical potential window consistent with the presence of these features. By contrast, in systems with strong SC disorder zero-energy features can emerge at arbitrarily low Zeeman fields.  In systems with weak SC disorder (and strong SM disorder), the optimal  SM-SC coupling corresponds to the strong coupling regime, $\gamma > \Delta_0$, which alleviates the effect of SM disorder. Again, stable MZMs can be accessed at Zeeman fields of order $\gamma$, but, of course, the corresponding values can be significantly larger than those corresponding to weak SM disorder (because of the stronger SM-SC coupling). Finally, if both SM and SC disorder are strong,  robust MZMs can only be accessed in strongly coupled systems at large values of the Zeeman field ($\Gamma$ much larger than $\gamma$). Note that systems with intermediate/weak SM-SC coupling  are characterized by huge finite size effects in the large $\Gamma$ limit. In practice, accessing this regime may  require enhancing the gyromagnetic ratio of the SM wire and/or optimizing the parent superconductor, to enable the application of larger magnetic fields. 

Finally, we point out the correlation function that we introduced as an auxiliary tool in our analysis. We believe that the full potential of this tool can be realized by performing a detailed  study of the dependence of the correlation maps on the ``filter parameters.'' A machine learning approach may be relevant in this context. In addition, defining the correlation function in terms of measurable quantities (e.g., differential conductance) could provide a valuable tool for analyzing experimental data.

\section{Acknowledgement}
ST acknowledges support from grant Grant No. NSF 2014157 and ONR Grant No: N00014-23-1-2061. TDS was supported by NSF-2014156 and ONR-N000142312061.

\bibliography{references}

\begin{thebibliography}{63}%
\makeatletter
\providecommand \@ifxundefined [1]{%
 \@ifx{#1\undefined}
}%
\providecommand \@ifnum [1]{%
 \ifnum #1\expandafter \@firstoftwo
 \else \expandafter \@secondoftwo
 \fi
}%
\providecommand \@ifx [1]{%
 \ifx #1\expandafter \@firstoftwo
 \else \expandafter \@secondoftwo
 \fi
}%
\providecommand \natexlab [1]{#1}%
\providecommand \enquote  [1]{``#1''}%
\providecommand \bibnamefont  [1]{#1}%
\providecommand \bibfnamefont [1]{#1}%
\providecommand \citenamefont [1]{#1}%
\providecommand \href@noop [0]{\@secondoftwo}%
\providecommand \href [0]{\begingroup \@sanitize@url \@href}%
\providecommand \@href[1]{\@@startlink{#1}\@@href}%
\providecommand \@@href[1]{\endgroup#1\@@endlink}%
\providecommand \@sanitize@url [0]{\catcode `\\12\catcode `\$12\catcode `\&12\catcode `\#12\catcode `\^12\catcode `\_12\catcode `\%12\relax}%
\providecommand \@@startlink[1]{}%
\providecommand \@@endlink[0]{}%
\providecommand \url  [0]{\begingroup\@sanitize@url \@url }%
\providecommand \@url [1]{\endgroup\@href {#1}{\urlprefix }}%
\providecommand \urlprefix  [0]{URL }%
\providecommand \Eprint [0]{\href }%
\providecommand \doibase [0]{https://doi.org/}%
\providecommand \selectlanguage [0]{\@gobble}%
\providecommand \bibinfo  [0]{\@secondoftwo}%
\providecommand \bibfield  [0]{\@secondoftwo}%
\providecommand \translation [1]{[#1]}%
\providecommand \BibitemOpen [0]{}%
\providecommand \bibitemStop [0]{}%
\providecommand \bibitemNoStop [0]{.\EOS\space}%
\providecommand \EOS [0]{\spacefactor3000\relax}%
\providecommand \BibitemShut  [1]{\csname bibitem#1\endcsname}%
\let\auto@bib@innerbib\@empty
\bibitem [{\citenamefont {Majorana}(1937)}]{Majorana1937}%
  \BibitemOpen
  \bibfield  {author} {\bibinfo {author} {\bibfnamefont {E.}~\bibnamefont {Majorana}},\ }\href@noop {} {\bibfield  {journal} {\bibinfo  {journal} {Nuovo Cimento}\ }\textbf {\bibinfo {volume} {14}},\ \bibinfo {pages} {171} (\bibinfo {year} {1937})}\BibitemShut {NoStop}%
\bibitem [{\citenamefont {Kitaev}(2003)}]{Kitaev2003}%
  \BibitemOpen
  \bibfield  {author} {\bibinfo {author} {\bibfnamefont {A.~Y.}\ \bibnamefont {Kitaev}},\ }\href@noop {} {\bibfield  {journal} {\bibinfo  {journal} {Annals. Phys.}\ }\textbf {\bibinfo {volume} {303}},\ \bibinfo {pages} {2} (\bibinfo {year} {2003})}\BibitemShut {NoStop}%
\bibitem [{\citenamefont {Nayak}\ \emph {et~al.}(2008)\citenamefont {Nayak}, \citenamefont {Simon}, \citenamefont {Stern}, \citenamefont {Freedman},\ and\ \citenamefont {Das~Sarma}}]{Nayak2008}%
  \BibitemOpen
  \bibfield  {author} {\bibinfo {author} {\bibfnamefont {C.}~\bibnamefont {Nayak}}, \bibinfo {author} {\bibfnamefont {S.~H.}\ \bibnamefont {Simon}}, \bibinfo {author} {\bibfnamefont {A.}~\bibnamefont {Stern}}, \bibinfo {author} {\bibfnamefont {M.}~\bibnamefont {Freedman}},\ and\ \bibinfo {author} {\bibfnamefont {S.}~\bibnamefont {Das~Sarma}},\ }\bibfield  {title} {\bibinfo {title} {Non-abelian anyons and topological quantum computation},\ }\href {https://doi.org/10.1103/RevModPhys.80.1083} {\bibfield  {journal} {\bibinfo  {journal} {Rev. Mod. Phys.}\ }\textbf {\bibinfo {volume} {80}},\ \bibinfo {pages} {1083} (\bibinfo {year} {2008})}\BibitemShut {NoStop}%
\bibitem [{\citenamefont {Wilczek}(1982)}]{wilczek1982quantum}%
  \BibitemOpen
  \bibfield  {author} {\bibinfo {author} {\bibfnamefont {F.}~\bibnamefont {Wilczek}},\ }\bibfield  {title} {\bibinfo {title} {Quantum mechanics of fractional-spin particles},\ }\href@noop {} {\bibfield  {journal} {\bibinfo  {journal} {Physical review letters}\ }\textbf {\bibinfo {volume} {49}},\ \bibinfo {pages} {957} (\bibinfo {year} {1982})}\BibitemShut {NoStop}%
\bibitem [{\citenamefont {Moore}\ and\ \citenamefont {Read}(1991)}]{Moore1991}%
  \BibitemOpen
  \bibfield  {author} {\bibinfo {author} {\bibfnamefont {G.}~\bibnamefont {Moore}}\ and\ \bibinfo {author} {\bibfnamefont {N.}~\bibnamefont {Read}},\ }\href@noop {} {\bibfield  {journal} {\bibinfo  {journal} {Nucl. Physics B}\ }\textbf {\bibinfo {volume} {360}},\ \bibinfo {pages} {362} (\bibinfo {year} {1991})}\BibitemShut {NoStop}%
\bibitem [{\citenamefont {Read}\ and\ \citenamefont {Green}(2000)}]{Read2000}%
  \BibitemOpen
  \bibfield  {author} {\bibinfo {author} {\bibfnamefont {N.}~\bibnamefont {Read}}\ and\ \bibinfo {author} {\bibfnamefont {D.}~\bibnamefont {Green}},\ }\bibfield  {title} {\bibinfo {title} {Paired states of fermions in two dimensions with breaking of parity and time-reversal symmetries and the fractional quantum hall effect},\ }\href {https://doi.org/10.1103/PhysRevB.61.10267} {\bibfield  {journal} {\bibinfo  {journal} {Phys. Rev. B}\ }\textbf {\bibinfo {volume} {61}},\ \bibinfo {pages} {10267} (\bibinfo {year} {2000})}\BibitemShut {NoStop}%
\bibitem [{\citenamefont {Nayak}\ and\ \citenamefont {Wilczek}(1996)}]{Nayak1996}%
  \BibitemOpen
  \bibfield  {author} {\bibinfo {author} {\bibfnamefont {C.}~\bibnamefont {Nayak}}\ and\ \bibinfo {author} {\bibfnamefont {F.}~\bibnamefont {Wilczek}},\ }\href@noop {} {\bibfield  {journal} {\bibinfo  {journal} {Nucl. Physics B}\ }\textbf {\bibinfo {volume} {479}},\ \bibinfo {pages} {529} (\bibinfo {year} {1996})}\BibitemShut {NoStop}%
\bibitem [{\citenamefont {Sau}\ \emph {et~al.}(2010{\natexlab{a}})\citenamefont {Sau}, \citenamefont {Lutchyn}, \citenamefont {Tewari},\ and\ \citenamefont {Sarma}}]{sau2010generic}%
  \BibitemOpen
  \bibfield  {author} {\bibinfo {author} {\bibfnamefont {J.~D.}\ \bibnamefont {Sau}}, \bibinfo {author} {\bibfnamefont {R.~M.}\ \bibnamefont {Lutchyn}}, \bibinfo {author} {\bibfnamefont {S.}~\bibnamefont {Tewari}},\ and\ \bibinfo {author} {\bibfnamefont {S.~D.}\ \bibnamefont {Sarma}},\ }\bibfield  {title} {\bibinfo {title} {Generic new platform for topological quantum computation using semiconductor heterostructures},\ }\href@noop {} {\bibfield  {journal} {\bibinfo  {journal} {Physical review letters}\ }\textbf {\bibinfo {volume} {104}},\ \bibinfo {pages} {040502} (\bibinfo {year} {2010}{\natexlab{a}})}\BibitemShut {NoStop}%
\bibitem [{\citenamefont {Sau}\ \emph {et~al.}(2010{\natexlab{b}})\citenamefont {Sau}, \citenamefont {Tewari}, \citenamefont {Lutchyn}, \citenamefont {Stanescu},\ and\ \citenamefont {Sarma}}]{sau2010non}%
  \BibitemOpen
  \bibfield  {author} {\bibinfo {author} {\bibfnamefont {J.~D.}\ \bibnamefont {Sau}}, \bibinfo {author} {\bibfnamefont {S.}~\bibnamefont {Tewari}}, \bibinfo {author} {\bibfnamefont {R.~M.}\ \bibnamefont {Lutchyn}}, \bibinfo {author} {\bibfnamefont {T.~D.}\ \bibnamefont {Stanescu}},\ and\ \bibinfo {author} {\bibfnamefont {S.~D.}\ \bibnamefont {Sarma}},\ }\bibfield  {title} {\bibinfo {title} {Non-abelian quantum order in spin-orbit-coupled semiconductors: Search for topological majorana particles in solid-state systems},\ }\href@noop {} {\bibfield  {journal} {\bibinfo  {journal} {Physical Review B}\ }\textbf {\bibinfo {volume} {82}},\ \bibinfo {pages} {214509} (\bibinfo {year} {2010}{\natexlab{b}})}\BibitemShut {NoStop}%
\bibitem [{\citenamefont {Oreg}\ \emph {et~al.}(2010)\citenamefont {Oreg}, \citenamefont {Refael},\ and\ \citenamefont {von Oppen}}]{oreg2010helical}%
  \BibitemOpen
  \bibfield  {author} {\bibinfo {author} {\bibfnamefont {Y.}~\bibnamefont {Oreg}}, \bibinfo {author} {\bibfnamefont {G.}~\bibnamefont {Refael}},\ and\ \bibinfo {author} {\bibfnamefont {F.}~\bibnamefont {von Oppen}},\ }\bibfield  {title} {\bibinfo {title} {Helical liquids and majorana bound states in quantum wires},\ }\href@noop {} {\bibfield  {journal} {\bibinfo  {journal} {Physical review letters}\ }\textbf {\bibinfo {volume} {105}},\ \bibinfo {pages} {177002} (\bibinfo {year} {2010})}\BibitemShut {NoStop}%
\bibitem [{\citenamefont {Lutchyn}\ \emph {et~al.}(2010)\citenamefont {Lutchyn}, \citenamefont {Sau},\ and\ \citenamefont {Sarma}}]{lutchyn2010majorana}%
  \BibitemOpen
  \bibfield  {author} {\bibinfo {author} {\bibfnamefont {R.~M.}\ \bibnamefont {Lutchyn}}, \bibinfo {author} {\bibfnamefont {J.~D.}\ \bibnamefont {Sau}},\ and\ \bibinfo {author} {\bibfnamefont {S.~D.}\ \bibnamefont {Sarma}},\ }\bibfield  {title} {\bibinfo {title} {Majorana fermions and a topological phase transition in semiconductor-superconductor heterostructures},\ }\href@noop {} {\bibfield  {journal} {\bibinfo  {journal} {Physical review letters}\ }\textbf {\bibinfo {volume} {105}},\ \bibinfo {pages} {077001} (\bibinfo {year} {2010})}\BibitemShut {NoStop}%
\bibitem [{\citenamefont {Mourik}\ \emph {et~al.}(2012)\citenamefont {Mourik}, \citenamefont {Zuo}, \citenamefont {Frolov}, \citenamefont {Plissard}, \citenamefont {Bakkers},\ and\ \citenamefont {Kouwenhoven}}]{mourik2012signatures}%
  \BibitemOpen
  \bibfield  {author} {\bibinfo {author} {\bibfnamefont {V.}~\bibnamefont {Mourik}}, \bibinfo {author} {\bibfnamefont {K.}~\bibnamefont {Zuo}}, \bibinfo {author} {\bibfnamefont {S.~M.}\ \bibnamefont {Frolov}}, \bibinfo {author} {\bibfnamefont {S.}~\bibnamefont {Plissard}}, \bibinfo {author} {\bibfnamefont {E.~P.}\ \bibnamefont {Bakkers}},\ and\ \bibinfo {author} {\bibfnamefont {L.~P.}\ \bibnamefont {Kouwenhoven}},\ }\bibfield  {title} {\bibinfo {title} {Signatures of majorana fermions in hybrid superconductor-semiconductor nanowire devices},\ }\href@noop {} {\bibfield  {journal} {\bibinfo  {journal} {Science}\ }\textbf {\bibinfo {volume} {336}},\ \bibinfo {pages} {1003} (\bibinfo {year} {2012})}\BibitemShut {NoStop}%
\bibitem [{\citenamefont {Deng}\ \emph {et~al.}(2012)\citenamefont {Deng}, \citenamefont {Yu}, \citenamefont {Huang}, \citenamefont {Larsson}, \citenamefont {Caroff},\ and\ \citenamefont {Xu}}]{Deng2012}%
  \BibitemOpen
  \bibfield  {author} {\bibinfo {author} {\bibfnamefont {M.~T.}\ \bibnamefont {Deng}}, \bibinfo {author} {\bibfnamefont {C.~L.}\ \bibnamefont {Yu}}, \bibinfo {author} {\bibfnamefont {G.~Y.}\ \bibnamefont {Huang}}, \bibinfo {author} {\bibfnamefont {M.}~\bibnamefont {Larsson}}, \bibinfo {author} {\bibfnamefont {P.}~\bibnamefont {Caroff}},\ and\ \bibinfo {author} {\bibfnamefont {H.~Q.}\ \bibnamefont {Xu}},\ }\bibfield  {title} {\bibinfo {title} {Anomalous zero-bias conductance peak in a nb--insb nanowire--nb hybrid device},\ }\href {https://doi.org/10.1021/nl303758w} {\bibfield  {journal} {\bibinfo  {journal} {Nano Letters}\ }\textbf {\bibinfo {volume} {12}},\ \bibinfo {pages} {6414} (\bibinfo {year} {2012})}\BibitemShut {NoStop}%
\bibitem [{\citenamefont {Das}\ \emph {et~al.}(2012)\citenamefont {Das}, \citenamefont {Ronen}, \citenamefont {Most}, \citenamefont {Oreg}, \citenamefont {Heiblum},\ and\ \citenamefont {Shtrikman}}]{Das2012}%
  \BibitemOpen
  \bibfield  {author} {\bibinfo {author} {\bibfnamefont {A.}~\bibnamefont {Das}}, \bibinfo {author} {\bibfnamefont {Y.}~\bibnamefont {Ronen}}, \bibinfo {author} {\bibfnamefont {Y.}~\bibnamefont {Most}}, \bibinfo {author} {\bibfnamefont {Y.}~\bibnamefont {Oreg}}, \bibinfo {author} {\bibfnamefont {M.}~\bibnamefont {Heiblum}},\ and\ \bibinfo {author} {\bibfnamefont {H.}~\bibnamefont {Shtrikman}},\ }\bibfield  {title} {\bibinfo {title} {Zero-bias peaks and splitting in an al--inas nanowire topological superconductor as a signature of majorana fermions},\ }\href {https://doi.org/10.1038/nphys2479} {\bibfield  {journal} {\bibinfo  {journal} {Nature Physics}\ }\textbf {\bibinfo {volume} {8}},\ \bibinfo {pages} {887} (\bibinfo {year} {2012})}\BibitemShut {NoStop}%
\bibitem [{\citenamefont {Rokhinson}\ \emph {et~al.}(2012)\citenamefont {Rokhinson}, \citenamefont {Liu},\ and\ \citenamefont {Furdyna}}]{rokhinson2012fractional}%
  \BibitemOpen
  \bibfield  {author} {\bibinfo {author} {\bibfnamefont {L.~P.}\ \bibnamefont {Rokhinson}}, \bibinfo {author} {\bibfnamefont {X.}~\bibnamefont {Liu}},\ and\ \bibinfo {author} {\bibfnamefont {J.~K.}\ \bibnamefont {Furdyna}},\ }\bibfield  {title} {\bibinfo {title} {The fractional ac josephson effect in a semiconductor--superconductor nanowire as a signature of majorana particles},\ }\href@noop {} {\bibfield  {journal} {\bibinfo  {journal} {Nature Physics}\ }\textbf {\bibinfo {volume} {8}},\ \bibinfo {pages} {795} (\bibinfo {year} {2012})}\BibitemShut {NoStop}%
\bibitem [{\citenamefont {Churchill}\ \emph {et~al.}(2013)\citenamefont {Churchill}, \citenamefont {Fatemi}, \citenamefont {Grove-Rasmussen}, \citenamefont {Deng}, \citenamefont {Caroff}, \citenamefont {Xu},\ and\ \citenamefont {Marcus}}]{churchill2013superconductor}%
  \BibitemOpen
  \bibfield  {author} {\bibinfo {author} {\bibfnamefont {H.}~\bibnamefont {Churchill}}, \bibinfo {author} {\bibfnamefont {V.}~\bibnamefont {Fatemi}}, \bibinfo {author} {\bibfnamefont {K.}~\bibnamefont {Grove-Rasmussen}}, \bibinfo {author} {\bibfnamefont {M.}~\bibnamefont {Deng}}, \bibinfo {author} {\bibfnamefont {P.}~\bibnamefont {Caroff}}, \bibinfo {author} {\bibfnamefont {H.}~\bibnamefont {Xu}},\ and\ \bibinfo {author} {\bibfnamefont {C.~M.}\ \bibnamefont {Marcus}},\ }\bibfield  {title} {\bibinfo {title} {Superconductor-nanowire devices from tunneling to the multichannel regime: Zero-bias oscillations and magnetoconductance crossover},\ }\href@noop {} {\bibfield  {journal} {\bibinfo  {journal} {Physical Review B}\ }\textbf {\bibinfo {volume} {87}},\ \bibinfo {pages} {241401} (\bibinfo {year} {2013})}\BibitemShut {NoStop}%
\bibitem [{\citenamefont {Finck}\ \emph {et~al.}(2013)\citenamefont {Finck}, \citenamefont {Van~Harlingen}, \citenamefont {Mohseni}, \citenamefont {Jung},\ and\ \citenamefont {Li}}]{finck2013anomalous}%
  \BibitemOpen
  \bibfield  {author} {\bibinfo {author} {\bibfnamefont {A.}~\bibnamefont {Finck}}, \bibinfo {author} {\bibfnamefont {D.~J.}\ \bibnamefont {Van~Harlingen}}, \bibinfo {author} {\bibfnamefont {P.}~\bibnamefont {Mohseni}}, \bibinfo {author} {\bibfnamefont {K.}~\bibnamefont {Jung}},\ and\ \bibinfo {author} {\bibfnamefont {X.}~\bibnamefont {Li}},\ }\bibfield  {title} {\bibinfo {title} {Anomalous modulation of a zero-bias peak in a hybrid nanowire-superconductor device},\ }\href@noop {} {\bibfield  {journal} {\bibinfo  {journal} {Physical review letters}\ }\textbf {\bibinfo {volume} {110}},\ \bibinfo {pages} {126406} (\bibinfo {year} {2013})}\BibitemShut {NoStop}%
\bibitem [{\citenamefont {Deng}\ \emph {et~al.}(2016)\citenamefont {Deng}, \citenamefont {Vaitiek{\.e}nas}, \citenamefont {Hansen}, \citenamefont {Danon}, \citenamefont {Leijnse}, \citenamefont {Flensberg}, \citenamefont {Nyg{\aa}rd}, \citenamefont {Krogstrup},\ and\ \citenamefont {Marcus}}]{deng2016majorana}%
  \BibitemOpen
  \bibfield  {author} {\bibinfo {author} {\bibfnamefont {M.}~\bibnamefont {Deng}}, \bibinfo {author} {\bibfnamefont {S.}~\bibnamefont {Vaitiek{\.e}nas}}, \bibinfo {author} {\bibfnamefont {E.~B.}\ \bibnamefont {Hansen}}, \bibinfo {author} {\bibfnamefont {J.}~\bibnamefont {Danon}}, \bibinfo {author} {\bibfnamefont {M.}~\bibnamefont {Leijnse}}, \bibinfo {author} {\bibfnamefont {K.}~\bibnamefont {Flensberg}}, \bibinfo {author} {\bibfnamefont {J.}~\bibnamefont {Nyg{\aa}rd}}, \bibinfo {author} {\bibfnamefont {P.}~\bibnamefont {Krogstrup}},\ and\ \bibinfo {author} {\bibfnamefont {C.~M.}\ \bibnamefont {Marcus}},\ }\bibfield  {title} {\bibinfo {title} {Majorana bound state in a coupled quantum-dot hybrid-nanowire system},\ }\href@noop {} {\bibfield  {journal} {\bibinfo  {journal} {Science}\ }\textbf {\bibinfo {volume} {354}},\ \bibinfo {pages} {1557} (\bibinfo {year} {2016})}\BibitemShut {NoStop}%
\bibitem [{\citenamefont {Zhang}\ \emph {et~al.}(2017)\citenamefont {Zhang}, \citenamefont {G{\"u}l}, \citenamefont {Conesa-Boj}, \citenamefont {Nowak}, \citenamefont {Wimmer}, \citenamefont {Zuo}, \citenamefont {Mourik}, \citenamefont {De~Vries}, \citenamefont {Van~Veen}, \citenamefont {De~Moor} \emph {et~al.}}]{zhang2017ballistic}%
  \BibitemOpen
  \bibfield  {author} {\bibinfo {author} {\bibfnamefont {H.}~\bibnamefont {Zhang}}, \bibinfo {author} {\bibfnamefont {{\"O}.}~\bibnamefont {G{\"u}l}}, \bibinfo {author} {\bibfnamefont {S.}~\bibnamefont {Conesa-Boj}}, \bibinfo {author} {\bibfnamefont {M.~P.}\ \bibnamefont {Nowak}}, \bibinfo {author} {\bibfnamefont {M.}~\bibnamefont {Wimmer}}, \bibinfo {author} {\bibfnamefont {K.}~\bibnamefont {Zuo}}, \bibinfo {author} {\bibfnamefont {V.}~\bibnamefont {Mourik}}, \bibinfo {author} {\bibfnamefont {F.~K.}\ \bibnamefont {De~Vries}}, \bibinfo {author} {\bibfnamefont {J.}~\bibnamefont {Van~Veen}}, \bibinfo {author} {\bibfnamefont {M.~W.}\ \bibnamefont {De~Moor}}, \emph {et~al.},\ }\bibfield  {title} {\bibinfo {title} {Ballistic superconductivity in semiconductor nanowires},\ }\href@noop {} {\bibfield  {journal} {\bibinfo  {journal} {Nature communications}\ }\textbf {\bibinfo {volume} {8}},\ \bibinfo {pages} {16025} (\bibinfo {year} {2017})}\BibitemShut {NoStop}%
\bibitem [{\citenamefont {Chen}\ \emph {et~al.}(2017)\citenamefont {Chen}, \citenamefont {Yu}, \citenamefont {Stenger}, \citenamefont {Hocevar}, \citenamefont {Car}, \citenamefont {Plissard}, \citenamefont {Bakkers}, \citenamefont {Stanescu},\ and\ \citenamefont {Frolov}}]{chen2017experimental}%
  \BibitemOpen
  \bibfield  {author} {\bibinfo {author} {\bibfnamefont {J.}~\bibnamefont {Chen}}, \bibinfo {author} {\bibfnamefont {P.}~\bibnamefont {Yu}}, \bibinfo {author} {\bibfnamefont {J.}~\bibnamefont {Stenger}}, \bibinfo {author} {\bibfnamefont {M.}~\bibnamefont {Hocevar}}, \bibinfo {author} {\bibfnamefont {D.}~\bibnamefont {Car}}, \bibinfo {author} {\bibfnamefont {S.~R.}\ \bibnamefont {Plissard}}, \bibinfo {author} {\bibfnamefont {E.~P.}\ \bibnamefont {Bakkers}}, \bibinfo {author} {\bibfnamefont {T.~D.}\ \bibnamefont {Stanescu}},\ and\ \bibinfo {author} {\bibfnamefont {S.~M.}\ \bibnamefont {Frolov}},\ }\bibfield  {title} {\bibinfo {title} {Experimental phase diagram of zero-bias conductance peaks in superconductor/semiconductor nanowire devices},\ }\href@noop {} {\bibfield  {journal} {\bibinfo  {journal} {Science advances}\ }\textbf {\bibinfo {volume} {3}},\ \bibinfo {pages} {e1701476} (\bibinfo {year} {2017})}\BibitemShut {NoStop}%
\bibitem [{\citenamefont {Nichele}\ \emph {et~al.}(2017)\citenamefont {Nichele}, \citenamefont {Drachmann}, \citenamefont {Whiticar}, \citenamefont {O’Farrell}, \citenamefont {Suominen}, \citenamefont {Fornieri}, \citenamefont {Wang}, \citenamefont {Gardner}, \citenamefont {Thomas}, \citenamefont {Hatke} \emph {et~al.}}]{nichele2017scaling}%
  \BibitemOpen
  \bibfield  {author} {\bibinfo {author} {\bibfnamefont {F.}~\bibnamefont {Nichele}}, \bibinfo {author} {\bibfnamefont {A.~C.}\ \bibnamefont {Drachmann}}, \bibinfo {author} {\bibfnamefont {A.~M.}\ \bibnamefont {Whiticar}}, \bibinfo {author} {\bibfnamefont {E.~C.}\ \bibnamefont {O’Farrell}}, \bibinfo {author} {\bibfnamefont {H.~J.}\ \bibnamefont {Suominen}}, \bibinfo {author} {\bibfnamefont {A.}~\bibnamefont {Fornieri}}, \bibinfo {author} {\bibfnamefont {T.}~\bibnamefont {Wang}}, \bibinfo {author} {\bibfnamefont {G.~C.}\ \bibnamefont {Gardner}}, \bibinfo {author} {\bibfnamefont {C.}~\bibnamefont {Thomas}}, \bibinfo {author} {\bibfnamefont {A.~T.}\ \bibnamefont {Hatke}}, \emph {et~al.},\ }\bibfield  {title} {\bibinfo {title} {Scaling of majorana zero-bias conductance peaks},\ }\href@noop {} {\bibfield  {journal} {\bibinfo  {journal} {Physical review letters}\ }\textbf {\bibinfo {volume} {119}},\ \bibinfo {pages} {136803} (\bibinfo {year} {2017})}\BibitemShut {NoStop}%
\bibitem [{\citenamefont {Albrecht}\ \emph {et~al.}(2017)\citenamefont {Albrecht}, \citenamefont {Hansen}, \citenamefont {Higginbotham}, \citenamefont {Kuemmeth}, \citenamefont {Jespersen}, \citenamefont {Nyg{\aa}rd}, \citenamefont {Krogstrup}, \citenamefont {Danon}, \citenamefont {Flensberg},\ and\ \citenamefont {Marcus}}]{albrecht2017transport}%
  \BibitemOpen
  \bibfield  {author} {\bibinfo {author} {\bibfnamefont {S.}~\bibnamefont {Albrecht}}, \bibinfo {author} {\bibfnamefont {E.}~\bibnamefont {Hansen}}, \bibinfo {author} {\bibfnamefont {A.~P.}\ \bibnamefont {Higginbotham}}, \bibinfo {author} {\bibfnamefont {F.}~\bibnamefont {Kuemmeth}}, \bibinfo {author} {\bibfnamefont {T.}~\bibnamefont {Jespersen}}, \bibinfo {author} {\bibfnamefont {J.}~\bibnamefont {Nyg{\aa}rd}}, \bibinfo {author} {\bibfnamefont {P.}~\bibnamefont {Krogstrup}}, \bibinfo {author} {\bibfnamefont {J.}~\bibnamefont {Danon}}, \bibinfo {author} {\bibfnamefont {K.}~\bibnamefont {Flensberg}},\ and\ \bibinfo {author} {\bibfnamefont {C.}~\bibnamefont {Marcus}},\ }\bibfield  {title} {\bibinfo {title} {Transport signatures of quasiparticle poisoning in a majorana island},\ }\href@noop {} {\bibfield  {journal} {\bibinfo  {journal} {Physical review letters}\ }\textbf {\bibinfo {volume} {118}},\ \bibinfo {pages} {137701} (\bibinfo {year} {2017})}\BibitemShut {NoStop}%
\bibitem [{\citenamefont {O~Farrell}\ \emph {et~al.}(2018)\citenamefont {O~Farrell}, \citenamefont {Drachmann}, \citenamefont {Hell}, \citenamefont {Fornieri}, \citenamefont {Whiticar}, \citenamefont {Hansen}, \citenamefont {Gronin}, \citenamefont {Gardner}, \citenamefont {Thomas}, \citenamefont {Manfra} \emph {et~al.}}]{o2018hybridization}%
  \BibitemOpen
  \bibfield  {author} {\bibinfo {author} {\bibfnamefont {E.}~\bibnamefont {O~Farrell}}, \bibinfo {author} {\bibfnamefont {A.}~\bibnamefont {Drachmann}}, \bibinfo {author} {\bibfnamefont {M.}~\bibnamefont {Hell}}, \bibinfo {author} {\bibfnamefont {A.}~\bibnamefont {Fornieri}}, \bibinfo {author} {\bibfnamefont {A.}~\bibnamefont {Whiticar}}, \bibinfo {author} {\bibfnamefont {E.}~\bibnamefont {Hansen}}, \bibinfo {author} {\bibfnamefont {S.}~\bibnamefont {Gronin}}, \bibinfo {author} {\bibfnamefont {G.}~\bibnamefont {Gardner}}, \bibinfo {author} {\bibfnamefont {C.}~\bibnamefont {Thomas}}, \bibinfo {author} {\bibfnamefont {M.}~\bibnamefont {Manfra}}, \emph {et~al.},\ }\bibfield  {title} {\bibinfo {title} {Hybridization of subgap states in one-dimensional superconductor-semiconductor coulomb islands},\ }\href@noop {} {\bibfield  {journal} {\bibinfo  {journal} {Physical review letters}\ }\textbf {\bibinfo {volume} {121}},\ \bibinfo {pages} {256803} (\bibinfo {year} {2018})}\BibitemShut {NoStop}%
\bibitem [{\citenamefont {Shen}\ \emph {et~al.}(2018)\citenamefont {Shen}, \citenamefont {Heedt}, \citenamefont {Borsoi}, \citenamefont {Van~Heck}, \citenamefont {Gazibegovic}, \citenamefont {het Veld}, \citenamefont {Car}, \citenamefont {Logan}, \citenamefont {Pendharkar}, \citenamefont {Ramakers} \emph {et~al.}}]{shen2018parity}%
  \BibitemOpen
  \bibfield  {author} {\bibinfo {author} {\bibfnamefont {J.}~\bibnamefont {Shen}}, \bibinfo {author} {\bibfnamefont {S.}~\bibnamefont {Heedt}}, \bibinfo {author} {\bibfnamefont {F.}~\bibnamefont {Borsoi}}, \bibinfo {author} {\bibfnamefont {B.}~\bibnamefont {Van~Heck}}, \bibinfo {author} {\bibfnamefont {S.}~\bibnamefont {Gazibegovic}}, \bibinfo {author} {\bibfnamefont {R.~L.~O.}\ \bibnamefont {het Veld}}, \bibinfo {author} {\bibfnamefont {D.}~\bibnamefont {Car}}, \bibinfo {author} {\bibfnamefont {J.~A.}\ \bibnamefont {Logan}}, \bibinfo {author} {\bibfnamefont {M.}~\bibnamefont {Pendharkar}}, \bibinfo {author} {\bibfnamefont {S.~J.}\ \bibnamefont {Ramakers}}, \emph {et~al.},\ }\bibfield  {title} {\bibinfo {title} {Parity transitions in the superconducting ground state of hybrid insb--al coulomb islands},\ }\href@noop {} {\bibfield  {journal} {\bibinfo  {journal} {Nature communications}\ }\textbf {\bibinfo {volume} {9}},\ \bibinfo {pages} {4801} (\bibinfo {year} {2018})}\BibitemShut {NoStop}%
\bibitem [{\citenamefont {Sherman}\ \emph {et~al.}(2017)\citenamefont {Sherman}, \citenamefont {Yodh}, \citenamefont {Albrecht}, \citenamefont {Nyg{\aa}rd}, \citenamefont {Krogstrup},\ and\ \citenamefont {Marcus}}]{sherman2017normal}%
  \BibitemOpen
  \bibfield  {author} {\bibinfo {author} {\bibfnamefont {D.}~\bibnamefont {Sherman}}, \bibinfo {author} {\bibfnamefont {J.}~\bibnamefont {Yodh}}, \bibinfo {author} {\bibfnamefont {S.~M.}\ \bibnamefont {Albrecht}}, \bibinfo {author} {\bibfnamefont {J.}~\bibnamefont {Nyg{\aa}rd}}, \bibinfo {author} {\bibfnamefont {P.}~\bibnamefont {Krogstrup}},\ and\ \bibinfo {author} {\bibfnamefont {C.~M.}\ \bibnamefont {Marcus}},\ }\bibfield  {title} {\bibinfo {title} {Normal, superconducting and topological regimes of hybrid double quantum dots},\ }\href@noop {} {\bibfield  {journal} {\bibinfo  {journal} {Nature nanotechnology}\ }\textbf {\bibinfo {volume} {12}},\ \bibinfo {pages} {212} (\bibinfo {year} {2017})}\BibitemShut {NoStop}%
\bibitem [{\citenamefont {Vaitiek{\.e}nas}\ \emph {et~al.}(2018)\citenamefont {Vaitiek{\.e}nas}, \citenamefont {Whiticar}, \citenamefont {Deng}, \citenamefont {Krizek}, \citenamefont {Sestoft}, \citenamefont {Palmstr{\o}m}, \citenamefont {Marti-Sanchez}, \citenamefont {Arbiol}, \citenamefont {Krogstrup}, \citenamefont {Casparis} \emph {et~al.}}]{vaitiekenas2018selective}%
  \BibitemOpen
  \bibfield  {author} {\bibinfo {author} {\bibfnamefont {S.}~\bibnamefont {Vaitiek{\.e}nas}}, \bibinfo {author} {\bibfnamefont {A.}~\bibnamefont {Whiticar}}, \bibinfo {author} {\bibfnamefont {M.-T.}\ \bibnamefont {Deng}}, \bibinfo {author} {\bibfnamefont {F.}~\bibnamefont {Krizek}}, \bibinfo {author} {\bibfnamefont {J.}~\bibnamefont {Sestoft}}, \bibinfo {author} {\bibfnamefont {C.}~\bibnamefont {Palmstr{\o}m}}, \bibinfo {author} {\bibfnamefont {S.}~\bibnamefont {Marti-Sanchez}}, \bibinfo {author} {\bibfnamefont {J.}~\bibnamefont {Arbiol}}, \bibinfo {author} {\bibfnamefont {P.}~\bibnamefont {Krogstrup}}, \bibinfo {author} {\bibfnamefont {L.}~\bibnamefont {Casparis}}, \emph {et~al.},\ }\bibfield  {title} {\bibinfo {title} {Selective-area-grown semiconductor-superconductor hybrids: A basis for topological networks},\ }\href@noop {} {\bibfield  {journal} {\bibinfo  {journal} {Physical review letters}\ }\textbf {\bibinfo {volume} {121}},\ \bibinfo {pages} {147701} (\bibinfo {year} {2018})}\BibitemShut {NoStop}%
\bibitem [{\citenamefont {Albrecht}\ \emph {et~al.}(2016)\citenamefont {Albrecht}, \citenamefont {Higginbotham}, \citenamefont {Madsen}, \citenamefont {Kuemmeth}, \citenamefont {Jespersen}, \citenamefont {Nyg{\aa}rd}, \citenamefont {Krogstrup},\ and\ \citenamefont {Marcus}}]{albrecht2016exponential}%
  \BibitemOpen
  \bibfield  {author} {\bibinfo {author} {\bibfnamefont {S.~M.}\ \bibnamefont {Albrecht}}, \bibinfo {author} {\bibfnamefont {A.~P.}\ \bibnamefont {Higginbotham}}, \bibinfo {author} {\bibfnamefont {M.}~\bibnamefont {Madsen}}, \bibinfo {author} {\bibfnamefont {F.}~\bibnamefont {Kuemmeth}}, \bibinfo {author} {\bibfnamefont {T.~S.}\ \bibnamefont {Jespersen}}, \bibinfo {author} {\bibfnamefont {J.}~\bibnamefont {Nyg{\aa}rd}}, \bibinfo {author} {\bibfnamefont {P.}~\bibnamefont {Krogstrup}},\ and\ \bibinfo {author} {\bibfnamefont {C.}~\bibnamefont {Marcus}},\ }\bibfield  {title} {\bibinfo {title} {Exponential protection of zero modes in majorana islands},\ }\href@noop {} {\bibfield  {journal} {\bibinfo  {journal} {Nature}\ }\textbf {\bibinfo {volume} {531}},\ \bibinfo {pages} {206} (\bibinfo {year} {2016})}\BibitemShut {NoStop}%
\bibitem [{\citenamefont {Yu}\ \emph {et~al.}(2021)\citenamefont {Yu}, \citenamefont {Chen}, \citenamefont {Gomanko}, \citenamefont {Badawy}, \citenamefont {Bakkers}, \citenamefont {Zuo}, \citenamefont {Mourik},\ and\ \citenamefont {Frolov}}]{Yu_2021}%
  \BibitemOpen
  \bibfield  {author} {\bibinfo {author} {\bibfnamefont {P.}~\bibnamefont {Yu}}, \bibinfo {author} {\bibfnamefont {J.}~\bibnamefont {Chen}}, \bibinfo {author} {\bibfnamefont {M.}~\bibnamefont {Gomanko}}, \bibinfo {author} {\bibfnamefont {G.}~\bibnamefont {Badawy}}, \bibinfo {author} {\bibfnamefont {E.~P. A.~M.}\ \bibnamefont {Bakkers}}, \bibinfo {author} {\bibfnamefont {K.}~\bibnamefont {Zuo}}, \bibinfo {author} {\bibfnamefont {V.}~\bibnamefont {Mourik}},\ and\ \bibinfo {author} {\bibfnamefont {S.~M.}\ \bibnamefont {Frolov}},\ }\bibfield  {title} {\bibinfo {title} {Non-majorana states yield nearly quantized conductance in proximatized nanowires},\ }\href {https://doi.org/10.1038/s41567-020-01107-w} {\bibfield  {journal} {\bibinfo  {journal} {Nature Physics}\ }\textbf {\bibinfo {volume} {17}},\ \bibinfo {pages} {482–488} (\bibinfo {year} {2021})}\BibitemShut {NoStop}%
\bibitem [{\citenamefont {Zhang}\ \emph {et~al.}(2021)\citenamefont {Zhang}, \citenamefont {de~Moor}, \citenamefont {Bommer}, \citenamefont {Xu}, \citenamefont {Wang}, \citenamefont {van Loo}, \citenamefont {Liu}, \citenamefont {Gazibegovic}, \citenamefont {Logan}, \citenamefont {Car}, \citenamefont {het Veld}, \citenamefont {van Veldhoven}, \citenamefont {Koelling}, \citenamefont {Verheijen}, \citenamefont {Pendharkar}, \citenamefont {Pennachio}, \citenamefont {Shojaei}, \citenamefont {Lee}, \citenamefont {Palmstrøm}, \citenamefont {Bakkers}, \citenamefont {Sarma},\ and\ \citenamefont {Kouwenhoven}}]{zhang2021}%
  \BibitemOpen
  \bibfield  {author} {\bibinfo {author} {\bibfnamefont {H.}~\bibnamefont {Zhang}}, \bibinfo {author} {\bibfnamefont {M.~W.~A.}\ \bibnamefont {de~Moor}}, \bibinfo {author} {\bibfnamefont {J.~D.~S.}\ \bibnamefont {Bommer}}, \bibinfo {author} {\bibfnamefont {D.}~\bibnamefont {Xu}}, \bibinfo {author} {\bibfnamefont {G.}~\bibnamefont {Wang}}, \bibinfo {author} {\bibfnamefont {N.}~\bibnamefont {van Loo}}, \bibinfo {author} {\bibfnamefont {C.-X.}\ \bibnamefont {Liu}}, \bibinfo {author} {\bibfnamefont {S.}~\bibnamefont {Gazibegovic}}, \bibinfo {author} {\bibfnamefont {J.~A.}\ \bibnamefont {Logan}}, \bibinfo {author} {\bibfnamefont {D.}~\bibnamefont {Car}}, \bibinfo {author} {\bibfnamefont {R.~L. M.~O.}\ \bibnamefont {het Veld}}, \bibinfo {author} {\bibfnamefont {P.~J.}\ \bibnamefont {van Veldhoven}}, \bibinfo {author} {\bibfnamefont {S.}~\bibnamefont {Koelling}}, \bibinfo {author} {\bibfnamefont {M.~A.}\ \bibnamefont {Verheijen}}, \bibinfo {author} {\bibfnamefont {M.}~\bibnamefont {Pendharkar}}, \bibinfo {author}
  {\bibfnamefont {D.~J.}\ \bibnamefont {Pennachio}}, \bibinfo {author} {\bibfnamefont {B.}~\bibnamefont {Shojaei}}, \bibinfo {author} {\bibfnamefont {J.~S.}\ \bibnamefont {Lee}}, \bibinfo {author} {\bibfnamefont {C.~J.}\ \bibnamefont {Palmstrøm}}, \bibinfo {author} {\bibfnamefont {E.~P. A.~M.}\ \bibnamefont {Bakkers}}, \bibinfo {author} {\bibfnamefont {S.~D.}\ \bibnamefont {Sarma}},\ and\ \bibinfo {author} {\bibfnamefont {L.~P.}\ \bibnamefont {Kouwenhoven}},\ }\href@noop {} {\bibinfo {title} {Large zero-bias peaks in insb-al hybrid semiconductor-superconductor nanowire devices}} (\bibinfo {year} {2021}),\ \Eprint {https://arxiv.org/abs/2101.11456} {arXiv:2101.11456 [cond-mat.mes-hall]} \BibitemShut {NoStop}%
\bibitem [{\citenamefont {Kells}\ \emph {et~al.}(2012)\citenamefont {Kells}, \citenamefont {Meidan},\ and\ \citenamefont {Brouwer}}]{kells2012near}%
  \BibitemOpen
  \bibfield  {author} {\bibinfo {author} {\bibfnamefont {G.}~\bibnamefont {Kells}}, \bibinfo {author} {\bibfnamefont {D.}~\bibnamefont {Meidan}},\ and\ \bibinfo {author} {\bibfnamefont {P.}~\bibnamefont {Brouwer}},\ }\bibfield  {title} {\bibinfo {title} {Near-zero-energy end states in topologically trivial spin-orbit coupled superconducting nanowires with a smooth confinement},\ }\href@noop {} {\bibfield  {journal} {\bibinfo  {journal} {Physical Review B}\ }\textbf {\bibinfo {volume} {86}},\ \bibinfo {pages} {100503} (\bibinfo {year} {2012})}\BibitemShut {NoStop}%
\bibitem [{\citenamefont {Aghaee}\ \emph {et~al.}(2023)\citenamefont {Aghaee}, \citenamefont {Akkala}, \citenamefont {Alam}, \citenamefont {Ali}, \citenamefont {Alcaraz~Ramirez}, \citenamefont {Andrzejczuk}, \citenamefont {Antipov}, \citenamefont {Aseev}, \citenamefont {Astafev}, \citenamefont {Bauer}, \citenamefont {Becker}, \citenamefont {Boddapati}, \citenamefont {Boekhout}, \citenamefont {Bommer}, \citenamefont {Bosma}, \citenamefont {Bourdet}, \citenamefont {Boutin}, \citenamefont {Caroff}, \citenamefont {Casparis}, \citenamefont {Cassidy}, \citenamefont {Chatoor}, \citenamefont {Christensen}, \citenamefont {Clay}, \citenamefont {Cole}, \citenamefont {Corsetti}, \citenamefont {Cui}, \citenamefont {Dalampiras}, \citenamefont {Dokania}, \citenamefont {de~Lange}, \citenamefont {de~Moor}, \citenamefont {Estrada Salda\~na}, \citenamefont {Fallahi}, \citenamefont {Fathabad}, \citenamefont {Gamble}, \citenamefont {Gardner}, \citenamefont {Govender}, \citenamefont {Griggio}, \citenamefont {Grigoryan},
  \citenamefont {Gronin}, \citenamefont {Gukelberger}, \citenamefont {Hansen}, \citenamefont {Heedt}, \citenamefont {Herranz~Zamorano}, \citenamefont {Ho}, \citenamefont {Holgaard}, \citenamefont {Ingerslev}, \citenamefont {Johansson}, \citenamefont {Jones}, \citenamefont {Kallaher}, \citenamefont {Karimi}, \citenamefont {Karzig}, \citenamefont {King}, \citenamefont {Kloster}, \citenamefont {Knapp}, \citenamefont {Kocon}, \citenamefont {Koski}, \citenamefont {Kostamo}, \citenamefont {Krogstrup}, \citenamefont {Kumar}, \citenamefont {Laeven}, \citenamefont {Larsen}, \citenamefont {Li}, \citenamefont {Lindemann}, \citenamefont {Love}, \citenamefont {Lutchyn}, \citenamefont {Madsen}, \citenamefont {Manfra}, \citenamefont {Markussen}, \citenamefont {Martinez}, \citenamefont {McNeil}, \citenamefont {Memisevic}, \citenamefont {Morgan}, \citenamefont {Mullally}, \citenamefont {Nayak}, \citenamefont {Nielsen}, \citenamefont {Nielsen}, \citenamefont {Nijholt}, \citenamefont {Nurmohamed}, \citenamefont {O'Farrell},
  \citenamefont {Otani}, \citenamefont {Pauka}, \citenamefont {Petersson}, \citenamefont {Petit}, \citenamefont {Pikulin}, \citenamefont {Preiss}, \citenamefont {Quintero-Perez}, \citenamefont {Rajpalke}, \citenamefont {Rasmussen}, \citenamefont {Razmadze}, \citenamefont {Reentila}, \citenamefont {Reilly}, \citenamefont {Rouse}, \citenamefont {Sadovskyy}, \citenamefont {Sainiemi}, \citenamefont {Schreppler}, \citenamefont {Sidorkin}, \citenamefont {Singh}, \citenamefont {Singh}, \citenamefont {Sinha}, \citenamefont {Sohr}, \citenamefont {Stankevi\ifmmode~\check{c}\else \v{c}\fi{}}, \citenamefont {Stek}, \citenamefont {Suominen}, \citenamefont {Suter}, \citenamefont {Svidenko}, \citenamefont {Teicher}, \citenamefont {Temuerhan}, \citenamefont {Thiyagarajah}, \citenamefont {Tholapi}, \citenamefont {Thomas}, \citenamefont {Toomey}, \citenamefont {Upadhyay}, \citenamefont {Urban}, \citenamefont {Vaitiek\ifmmode~\dot{e}\else \.{e}\fi{}nas}, \citenamefont {Van~Hoogdalem}, \citenamefont {Van~Woerkom}, \citenamefont
  {Viazmitinov}, \citenamefont {Vogel}, \citenamefont {Waddy}, \citenamefont {Watson}, \citenamefont {Weston}, \citenamefont {Winkler}, \citenamefont {Yang}, \citenamefont {Yau}, \citenamefont {Yi}, \citenamefont {Yucelen}, \citenamefont {Webster}, \citenamefont {Zeisel},\ and\ \citenamefont {Zhao}}]{PhysRevB.107.245423}%
  \BibitemOpen
  \bibfield  {author} {\bibinfo {author} {\bibfnamefont {M.}~\bibnamefont {Aghaee}}, \bibinfo {author} {\bibfnamefont {A.}~\bibnamefont {Akkala}}, \bibinfo {author} {\bibfnamefont {Z.}~\bibnamefont {Alam}}, \bibinfo {author} {\bibfnamefont {R.}~\bibnamefont {Ali}}, \bibinfo {author} {\bibfnamefont {A.}~\bibnamefont {Alcaraz~Ramirez}}, \bibinfo {author} {\bibfnamefont {M.}~\bibnamefont {Andrzejczuk}}, \bibinfo {author} {\bibfnamefont {A.~E.}\ \bibnamefont {Antipov}}, \bibinfo {author} {\bibfnamefont {P.}~\bibnamefont {Aseev}}, \bibinfo {author} {\bibfnamefont {M.}~\bibnamefont {Astafev}}, \bibinfo {author} {\bibfnamefont {B.}~\bibnamefont {Bauer}}, \bibinfo {author} {\bibfnamefont {J.}~\bibnamefont {Becker}}, \bibinfo {author} {\bibfnamefont {S.}~\bibnamefont {Boddapati}}, \bibinfo {author} {\bibfnamefont {F.}~\bibnamefont {Boekhout}}, \bibinfo {author} {\bibfnamefont {J.}~\bibnamefont {Bommer}}, \bibinfo {author} {\bibfnamefont {T.}~\bibnamefont {Bosma}}, \bibinfo {author} {\bibfnamefont {L.}~\bibnamefont
  {Bourdet}}, \bibinfo {author} {\bibfnamefont {S.}~\bibnamefont {Boutin}}, \bibinfo {author} {\bibfnamefont {P.}~\bibnamefont {Caroff}}, \bibinfo {author} {\bibfnamefont {L.}~\bibnamefont {Casparis}}, \bibinfo {author} {\bibfnamefont {M.}~\bibnamefont {Cassidy}}, \bibinfo {author} {\bibfnamefont {S.}~\bibnamefont {Chatoor}}, \bibinfo {author} {\bibfnamefont {A.~W.}\ \bibnamefont {Christensen}}, \bibinfo {author} {\bibfnamefont {N.}~\bibnamefont {Clay}}, \bibinfo {author} {\bibfnamefont {W.~S.}\ \bibnamefont {Cole}}, \bibinfo {author} {\bibfnamefont {F.}~\bibnamefont {Corsetti}}, \bibinfo {author} {\bibfnamefont {A.}~\bibnamefont {Cui}}, \bibinfo {author} {\bibfnamefont {P.}~\bibnamefont {Dalampiras}}, \bibinfo {author} {\bibfnamefont {A.}~\bibnamefont {Dokania}}, \bibinfo {author} {\bibfnamefont {G.}~\bibnamefont {de~Lange}}, \bibinfo {author} {\bibfnamefont {M.}~\bibnamefont {de~Moor}}, \bibinfo {author} {\bibfnamefont {J.~C.}\ \bibnamefont {Estrada Salda\~na}}, \bibinfo {author} {\bibfnamefont
  {S.}~\bibnamefont {Fallahi}}, \bibinfo {author} {\bibfnamefont {Z.~H.}\ \bibnamefont {Fathabad}}, \bibinfo {author} {\bibfnamefont {J.}~\bibnamefont {Gamble}}, \bibinfo {author} {\bibfnamefont {G.}~\bibnamefont {Gardner}}, \bibinfo {author} {\bibfnamefont {D.}~\bibnamefont {Govender}}, \bibinfo {author} {\bibfnamefont {F.}~\bibnamefont {Griggio}}, \bibinfo {author} {\bibfnamefont {R.}~\bibnamefont {Grigoryan}}, \bibinfo {author} {\bibfnamefont {S.}~\bibnamefont {Gronin}}, \bibinfo {author} {\bibfnamefont {J.}~\bibnamefont {Gukelberger}}, \bibinfo {author} {\bibfnamefont {E.~B.}\ \bibnamefont {Hansen}}, \bibinfo {author} {\bibfnamefont {S.}~\bibnamefont {Heedt}}, \bibinfo {author} {\bibfnamefont {J.}~\bibnamefont {Herranz~Zamorano}}, \bibinfo {author} {\bibfnamefont {S.}~\bibnamefont {Ho}}, \bibinfo {author} {\bibfnamefont {U.~L.}\ \bibnamefont {Holgaard}}, \bibinfo {author} {\bibfnamefont {H.}~\bibnamefont {Ingerslev}}, \bibinfo {author} {\bibfnamefont {L.}~\bibnamefont {Johansson}}, \bibinfo {author}
  {\bibfnamefont {J.}~\bibnamefont {Jones}}, \bibinfo {author} {\bibfnamefont {R.}~\bibnamefont {Kallaher}}, \bibinfo {author} {\bibfnamefont {F.}~\bibnamefont {Karimi}}, \bibinfo {author} {\bibfnamefont {T.}~\bibnamefont {Karzig}}, \bibinfo {author} {\bibfnamefont {C.}~\bibnamefont {King}}, \bibinfo {author} {\bibfnamefont {M.~E.}\ \bibnamefont {Kloster}}, \bibinfo {author} {\bibfnamefont {C.}~\bibnamefont {Knapp}}, \bibinfo {author} {\bibfnamefont {D.}~\bibnamefont {Kocon}}, \bibinfo {author} {\bibfnamefont {J.}~\bibnamefont {Koski}}, \bibinfo {author} {\bibfnamefont {P.}~\bibnamefont {Kostamo}}, \bibinfo {author} {\bibfnamefont {P.}~\bibnamefont {Krogstrup}}, \bibinfo {author} {\bibfnamefont {M.}~\bibnamefont {Kumar}}, \bibinfo {author} {\bibfnamefont {T.}~\bibnamefont {Laeven}}, \bibinfo {author} {\bibfnamefont {T.}~\bibnamefont {Larsen}}, \bibinfo {author} {\bibfnamefont {K.}~\bibnamefont {Li}}, \bibinfo {author} {\bibfnamefont {T.}~\bibnamefont {Lindemann}}, \bibinfo {author} {\bibfnamefont
  {J.}~\bibnamefont {Love}}, \bibinfo {author} {\bibfnamefont {R.}~\bibnamefont {Lutchyn}}, \bibinfo {author} {\bibfnamefont {M.~H.}\ \bibnamefont {Madsen}}, \bibinfo {author} {\bibfnamefont {M.}~\bibnamefont {Manfra}}, \bibinfo {author} {\bibfnamefont {S.}~\bibnamefont {Markussen}}, \bibinfo {author} {\bibfnamefont {E.}~\bibnamefont {Martinez}}, \bibinfo {author} {\bibfnamefont {R.}~\bibnamefont {McNeil}}, \bibinfo {author} {\bibfnamefont {E.}~\bibnamefont {Memisevic}}, \bibinfo {author} {\bibfnamefont {T.}~\bibnamefont {Morgan}}, \bibinfo {author} {\bibfnamefont {A.}~\bibnamefont {Mullally}}, \bibinfo {author} {\bibfnamefont {C.}~\bibnamefont {Nayak}}, \bibinfo {author} {\bibfnamefont {J.}~\bibnamefont {Nielsen}}, \bibinfo {author} {\bibfnamefont {W.~H.~P.}\ \bibnamefont {Nielsen}}, \bibinfo {author} {\bibfnamefont {B.}~\bibnamefont {Nijholt}}, \bibinfo {author} {\bibfnamefont {A.}~\bibnamefont {Nurmohamed}}, \bibinfo {author} {\bibfnamefont {E.}~\bibnamefont {O'Farrell}}, \bibinfo {author} {\bibfnamefont
  {K.}~\bibnamefont {Otani}}, \bibinfo {author} {\bibfnamefont {S.}~\bibnamefont {Pauka}}, \bibinfo {author} {\bibfnamefont {K.}~\bibnamefont {Petersson}}, \bibinfo {author} {\bibfnamefont {L.}~\bibnamefont {Petit}}, \bibinfo {author} {\bibfnamefont {D.~I.}\ \bibnamefont {Pikulin}}, \bibinfo {author} {\bibfnamefont {F.}~\bibnamefont {Preiss}}, \bibinfo {author} {\bibfnamefont {M.}~\bibnamefont {Quintero-Perez}}, \bibinfo {author} {\bibfnamefont {M.}~\bibnamefont {Rajpalke}}, \bibinfo {author} {\bibfnamefont {K.}~\bibnamefont {Rasmussen}}, \bibinfo {author} {\bibfnamefont {D.}~\bibnamefont {Razmadze}}, \bibinfo {author} {\bibfnamefont {O.}~\bibnamefont {Reentila}}, \bibinfo {author} {\bibfnamefont {D.}~\bibnamefont {Reilly}}, \bibinfo {author} {\bibfnamefont {R.}~\bibnamefont {Rouse}}, \bibinfo {author} {\bibfnamefont {I.}~\bibnamefont {Sadovskyy}}, \bibinfo {author} {\bibfnamefont {L.}~\bibnamefont {Sainiemi}}, \bibinfo {author} {\bibfnamefont {S.}~\bibnamefont {Schreppler}}, \bibinfo {author} {\bibfnamefont
  {V.}~\bibnamefont {Sidorkin}}, \bibinfo {author} {\bibfnamefont {A.}~\bibnamefont {Singh}}, \bibinfo {author} {\bibfnamefont {S.}~\bibnamefont {Singh}}, \bibinfo {author} {\bibfnamefont {S.}~\bibnamefont {Sinha}}, \bibinfo {author} {\bibfnamefont {P.}~\bibnamefont {Sohr}}, \bibinfo {author} {\bibfnamefont {T.~c.~v.}\ \bibnamefont {Stankevi\ifmmode~\check{c}\else \v{c}\fi{}}}, \bibinfo {author} {\bibfnamefont {L.}~\bibnamefont {Stek}}, \bibinfo {author} {\bibfnamefont {H.}~\bibnamefont {Suominen}}, \bibinfo {author} {\bibfnamefont {J.}~\bibnamefont {Suter}}, \bibinfo {author} {\bibfnamefont {V.}~\bibnamefont {Svidenko}}, \bibinfo {author} {\bibfnamefont {S.}~\bibnamefont {Teicher}}, \bibinfo {author} {\bibfnamefont {M.}~\bibnamefont {Temuerhan}}, \bibinfo {author} {\bibfnamefont {N.}~\bibnamefont {Thiyagarajah}}, \bibinfo {author} {\bibfnamefont {R.}~\bibnamefont {Tholapi}}, \bibinfo {author} {\bibfnamefont {M.}~\bibnamefont {Thomas}}, \bibinfo {author} {\bibfnamefont {E.}~\bibnamefont {Toomey}}, \bibinfo
  {author} {\bibfnamefont {S.}~\bibnamefont {Upadhyay}}, \bibinfo {author} {\bibfnamefont {I.}~\bibnamefont {Urban}}, \bibinfo {author} {\bibfnamefont {S.}~\bibnamefont {Vaitiek\ifmmode~\dot{e}\else \.{e}\fi{}nas}}, \bibinfo {author} {\bibfnamefont {K.}~\bibnamefont {Van~Hoogdalem}}, \bibinfo {author} {\bibfnamefont {D.}~\bibnamefont {Van~Woerkom}}, \bibinfo {author} {\bibfnamefont {D.~V.}\ \bibnamefont {Viazmitinov}}, \bibinfo {author} {\bibfnamefont {D.}~\bibnamefont {Vogel}}, \bibinfo {author} {\bibfnamefont {S.}~\bibnamefont {Waddy}}, \bibinfo {author} {\bibfnamefont {J.}~\bibnamefont {Watson}}, \bibinfo {author} {\bibfnamefont {J.}~\bibnamefont {Weston}}, \bibinfo {author} {\bibfnamefont {G.~W.}\ \bibnamefont {Winkler}}, \bibinfo {author} {\bibfnamefont {C.~K.}\ \bibnamefont {Yang}}, \bibinfo {author} {\bibfnamefont {S.}~\bibnamefont {Yau}}, \bibinfo {author} {\bibfnamefont {D.}~\bibnamefont {Yi}}, \bibinfo {author} {\bibfnamefont {E.}~\bibnamefont {Yucelen}}, \bibinfo {author} {\bibfnamefont
  {A.}~\bibnamefont {Webster}}, \bibinfo {author} {\bibfnamefont {R.}~\bibnamefont {Zeisel}},\ and\ \bibinfo {author} {\bibfnamefont {R.}~\bibnamefont {Zhao}} (\bibinfo {collaboration} {Microsoft Quantum}),\ }\bibfield  {title} {\bibinfo {title} {Inas-al hybrid devices passing the topological gap protocol},\ }\href {https://doi.org/10.1103/PhysRevB.107.245423} {\bibfield  {journal} {\bibinfo  {journal} {Phys. Rev. B}\ }\textbf {\bibinfo {volume} {107}},\ \bibinfo {pages} {245423} (\bibinfo {year} {2023})}\BibitemShut {NoStop}%
\bibitem [{\citenamefont {Mi}\ \emph {et~al.}(2014)\citenamefont {Mi}, \citenamefont {Pikulin}, \citenamefont {Marciani},\ and\ \citenamefont {Beenakker}}]{Mi2014}%
  \BibitemOpen
  \bibfield  {author} {\bibinfo {author} {\bibfnamefont {S.}~\bibnamefont {Mi}}, \bibinfo {author} {\bibfnamefont {D.~I.}\ \bibnamefont {Pikulin}}, \bibinfo {author} {\bibfnamefont {M.}~\bibnamefont {Marciani}},\ and\ \bibinfo {author} {\bibfnamefont {C.~W.~J.}\ \bibnamefont {Beenakker}},\ }\bibfield  {title} {\bibinfo {title} {X-shaped and y-shaped andreev resonance profiles in a superconducting quantum dot},\ }\href {https://doi.org/10.1134/S1063776114120176} {\bibfield  {journal} {\bibinfo  {journal} {Journal of Experimental and Theoretical Physics}\ }\textbf {\bibinfo {volume} {119}},\ \bibinfo {pages} {1018} (\bibinfo {year} {2014})}\BibitemShut {NoStop}%
\bibitem [{\citenamefont {Bagrets}\ and\ \citenamefont {Altland}(2012)}]{bagrets2012class}%
  \BibitemOpen
  \bibfield  {author} {\bibinfo {author} {\bibfnamefont {D.}~\bibnamefont {Bagrets}}\ and\ \bibinfo {author} {\bibfnamefont {A.}~\bibnamefont {Altland}},\ }\bibfield  {title} {\bibinfo {title} {Class d spectral peak in majorana quantum wires},\ }\href@noop {} {\bibfield  {journal} {\bibinfo  {journal} {Physical review letters}\ }\textbf {\bibinfo {volume} {109}},\ \bibinfo {pages} {227005} (\bibinfo {year} {2012})}\BibitemShut {NoStop}%
\bibitem [{\citenamefont {Pikulin}\ \emph {et~al.}(2012)\citenamefont {Pikulin}, \citenamefont {Dahlhaus}, \citenamefont {Wimmer}, \citenamefont {Schomerus},\ and\ \citenamefont {Beenakker}}]{pikulin2012zero}%
  \BibitemOpen
  \bibfield  {author} {\bibinfo {author} {\bibfnamefont {D.~I.}\ \bibnamefont {Pikulin}}, \bibinfo {author} {\bibfnamefont {J.}~\bibnamefont {Dahlhaus}}, \bibinfo {author} {\bibfnamefont {M.}~\bibnamefont {Wimmer}}, \bibinfo {author} {\bibfnamefont {H.}~\bibnamefont {Schomerus}},\ and\ \bibinfo {author} {\bibfnamefont {C.}~\bibnamefont {Beenakker}},\ }\bibfield  {title} {\bibinfo {title} {A zero-voltage conductance peak from weak antilocalization in a majorana nanowire},\ }\href@noop {} {\bibfield  {journal} {\bibinfo  {journal} {New Journal of Physics}\ }\textbf {\bibinfo {volume} {14}},\ \bibinfo {pages} {125011} (\bibinfo {year} {2012})}\BibitemShut {NoStop}%
\bibitem [{\citenamefont {Prada}\ \emph {et~al.}(2012)\citenamefont {Prada}, \citenamefont {San-Jose},\ and\ \citenamefont {Aguado}}]{prada2012transport}%
  \BibitemOpen
  \bibfield  {author} {\bibinfo {author} {\bibfnamefont {E.}~\bibnamefont {Prada}}, \bibinfo {author} {\bibfnamefont {P.}~\bibnamefont {San-Jose}},\ and\ \bibinfo {author} {\bibfnamefont {R.}~\bibnamefont {Aguado}},\ }\bibfield  {title} {\bibinfo {title} {Transport spectroscopy of n s nanowire junctions with majorana fermions},\ }\href@noop {} {\bibfield  {journal} {\bibinfo  {journal} {Physical Review B}\ }\textbf {\bibinfo {volume} {86}},\ \bibinfo {pages} {180503} (\bibinfo {year} {2012})}\BibitemShut {NoStop}%
\bibitem [{\citenamefont {Pan}\ and\ \citenamefont {Sarma}(2020)}]{pan2020physical}%
  \BibitemOpen
  \bibfield  {author} {\bibinfo {author} {\bibfnamefont {H.}~\bibnamefont {Pan}}\ and\ \bibinfo {author} {\bibfnamefont {S.~D.}\ \bibnamefont {Sarma}},\ }\bibfield  {title} {\bibinfo {title} {Physical mechanisms for zero-bias conductance peaks in majorana nanowires},\ }\href@noop {} {\bibfield  {journal} {\bibinfo  {journal} {Physical Review Research}\ }\textbf {\bibinfo {volume} {2}},\ \bibinfo {pages} {013377} (\bibinfo {year} {2020})}\BibitemShut {NoStop}%
\bibitem [{\citenamefont {Moore}\ \emph {et~al.}(2018{\natexlab{a}})\citenamefont {Moore}, \citenamefont {Stanescu},\ and\ \citenamefont {Tewari}}]{moore2018two}%
  \BibitemOpen
  \bibfield  {author} {\bibinfo {author} {\bibfnamefont {C.}~\bibnamefont {Moore}}, \bibinfo {author} {\bibfnamefont {T.~D.}\ \bibnamefont {Stanescu}},\ and\ \bibinfo {author} {\bibfnamefont {S.}~\bibnamefont {Tewari}},\ }\bibfield  {title} {\bibinfo {title} {Two-terminal charge tunneling: Disentangling majorana zero modes from partially separated andreev bound states in semiconductor-superconductor heterostructures},\ }\href@noop {} {\bibfield  {journal} {\bibinfo  {journal} {Physical Review B}\ }\textbf {\bibinfo {volume} {97}},\ \bibinfo {pages} {165302} (\bibinfo {year} {2018}{\natexlab{a}})}\BibitemShut {NoStop}%
\bibitem [{\citenamefont {Moore}\ \emph {et~al.}(2018{\natexlab{b}})\citenamefont {Moore}, \citenamefont {Zeng}, \citenamefont {Stanescu},\ and\ \citenamefont {Tewari}}]{Moore2018}%
  \BibitemOpen
  \bibfield  {author} {\bibinfo {author} {\bibfnamefont {C.}~\bibnamefont {Moore}}, \bibinfo {author} {\bibfnamefont {C.}~\bibnamefont {Zeng}}, \bibinfo {author} {\bibfnamefont {T.~D.}\ \bibnamefont {Stanescu}},\ and\ \bibinfo {author} {\bibfnamefont {S.}~\bibnamefont {Tewari}},\ }\bibfield  {title} {\bibinfo {title} {Quantized zero-bias conductance plateau in semiconductor-superconductor heterostructures without topological majorana zero modes},\ }\href {https://doi.org/10.1103/PhysRevB.98.155314} {\bibfield  {journal} {\bibinfo  {journal} {Phys. Rev. B}\ }\textbf {\bibinfo {volume} {98}},\ \bibinfo {pages} {155314} (\bibinfo {year} {2018}{\natexlab{b}})}\BibitemShut {NoStop}%
\bibitem [{\citenamefont {Vuik}\ \emph {et~al.}(2018)\citenamefont {Vuik}, \citenamefont {Nijholt}, \citenamefont {Akhmerov},\ and\ \citenamefont {Wimmer}}]{vuik2018reproducing}%
  \BibitemOpen
  \bibfield  {author} {\bibinfo {author} {\bibfnamefont {A.}~\bibnamefont {Vuik}}, \bibinfo {author} {\bibfnamefont {B.}~\bibnamefont {Nijholt}}, \bibinfo {author} {\bibfnamefont {A.~R.}\ \bibnamefont {Akhmerov}},\ and\ \bibinfo {author} {\bibfnamefont {M.}~\bibnamefont {Wimmer}},\ }\bibfield  {title} {\bibinfo {title} {Reproducing topological properties with quasi-majorana states},\ }\href@noop {} {\bibfield  {journal} {\bibinfo  {journal} {arXiv preprint arXiv:1806.02801}\ } (\bibinfo {year} {2018})}\BibitemShut {NoStop}%
\bibitem [{\citenamefont {Stanescu}\ and\ \citenamefont {Tewari}(2019)}]{stanescu2019robust}%
  \BibitemOpen
  \bibfield  {author} {\bibinfo {author} {\bibfnamefont {T.~D.}\ \bibnamefont {Stanescu}}\ and\ \bibinfo {author} {\bibfnamefont {S.}~\bibnamefont {Tewari}},\ }\bibfield  {title} {\bibinfo {title} {Robust low-energy andreev bound states in semiconductor-superconductor structures: Importance of partial separation of component majorana bound states},\ }\href@noop {} {\bibfield  {journal} {\bibinfo  {journal} {Physical Review B}\ }\textbf {\bibinfo {volume} {100}},\ \bibinfo {pages} {155429} (\bibinfo {year} {2019})}\BibitemShut {NoStop}%
\bibitem [{\citenamefont {Reeg}\ \emph {et~al.}(2018)\citenamefont {Reeg}, \citenamefont {Dmytruk}, \citenamefont {Chevallier}, \citenamefont {Loss},\ and\ \citenamefont {Klinovaja}}]{added_Loss_2018prb_abs}%
  \BibitemOpen
  \bibfield  {author} {\bibinfo {author} {\bibfnamefont {C.}~\bibnamefont {Reeg}}, \bibinfo {author} {\bibfnamefont {O.}~\bibnamefont {Dmytruk}}, \bibinfo {author} {\bibfnamefont {D.}~\bibnamefont {Chevallier}}, \bibinfo {author} {\bibfnamefont {D.}~\bibnamefont {Loss}},\ and\ \bibinfo {author} {\bibfnamefont {J.}~\bibnamefont {Klinovaja}},\ }\bibfield  {title} {\bibinfo {title} {Zero-energy andreev bound states from quantum dots in proximitized rashba nanowires},\ }\href {https://doi.org/10.1103/PhysRevB.98.245407} {\bibfield  {journal} {\bibinfo  {journal} {Phys. Rev. B}\ }\textbf {\bibinfo {volume} {98}},\ \bibinfo {pages} {245407} (\bibinfo {year} {2018})}\BibitemShut {NoStop}%
\bibitem [{\citenamefont {San-Jose}\ \emph {et~al.}(2016)\citenamefont {San-Jose}, \citenamefont {Cayao}, \citenamefont {Prada},\ and\ \citenamefont {Aguado}}]{san2016majorana}%
  \BibitemOpen
  \bibfield  {author} {\bibinfo {author} {\bibfnamefont {P.}~\bibnamefont {San-Jose}}, \bibinfo {author} {\bibfnamefont {J.}~\bibnamefont {Cayao}}, \bibinfo {author} {\bibfnamefont {E.}~\bibnamefont {Prada}},\ and\ \bibinfo {author} {\bibfnamefont {R.}~\bibnamefont {Aguado}},\ }\bibfield  {title} {\bibinfo {title} {Majorana bound states from exceptional points in non-topological superconductors},\ }\href@noop {} {\bibfield  {journal} {\bibinfo  {journal} {Scientific reports}\ }\textbf {\bibinfo {volume} {6}},\ \bibinfo {pages} {21427} (\bibinfo {year} {2016})}\BibitemShut {NoStop}%
\bibitem [{\citenamefont {Avila}\ \emph {et~al.}(2019)\citenamefont {Avila}, \citenamefont {Peñaranda}, \citenamefont {Prada}, \citenamefont {San-Jose},\ and\ \citenamefont {Aguado}}]{ramon2019nonhermitian}%
  \BibitemOpen
  \bibfield  {author} {\bibinfo {author} {\bibfnamefont {J.}~\bibnamefont {Avila}}, \bibinfo {author} {\bibfnamefont {F.}~\bibnamefont {Peñaranda}}, \bibinfo {author} {\bibfnamefont {E.}~\bibnamefont {Prada}}, \bibinfo {author} {\bibfnamefont {P.}~\bibnamefont {San-Jose}},\ and\ \bibinfo {author} {\bibfnamefont {R.}~\bibnamefont {Aguado}},\ }\bibfield  {title} {\bibinfo {title} {Non-hermitian topology as a unifying framework for the andreev versus majorana states controversy},\ }\bibfield  {journal} {\bibinfo  {journal} {Communications Physics}\ }\textbf {\bibinfo {volume} {2}},\ \href {https://doi.org/10.1038/s42005-019-0231-8} {10.1038/s42005-019-0231-8} (\bibinfo {year} {2019})\BibitemShut {NoStop}%
\bibitem [{\citenamefont {Awoga}\ \emph {et~al.}(2019)\citenamefont {Awoga}, \citenamefont {Cayao},\ and\ \citenamefont {Black-Schaffer}}]{Jorge2019supercurrent}%
  \BibitemOpen
  \bibfield  {author} {\bibinfo {author} {\bibfnamefont {O.~A.}\ \bibnamefont {Awoga}}, \bibinfo {author} {\bibfnamefont {J.}~\bibnamefont {Cayao}},\ and\ \bibinfo {author} {\bibfnamefont {A.~M.}\ \bibnamefont {Black-Schaffer}},\ }\bibfield  {title} {\bibinfo {title} {Supercurrent detection of topologically trivial zero-energy states in nanowire junctions},\ }\href {https://doi.org/10.1103/PhysRevLett.123.117001} {\bibfield  {journal} {\bibinfo  {journal} {Phys. Rev. Lett.}\ }\textbf {\bibinfo {volume} {123}},\ \bibinfo {pages} {117001} (\bibinfo {year} {2019})}\BibitemShut {NoStop}%
\bibitem [{\citenamefont {Prada}\ \emph {et~al.}(2020)\citenamefont {Prada}, \citenamefont {San-Jose}, \citenamefont {de~Moor}, \citenamefont {Geresdi}, \citenamefont {Lee}, \citenamefont {Klinovaja}, \citenamefont {Loss}, \citenamefont {Nygård}, \citenamefont {Aguado},\ and\ \citenamefont {Kouwenhoven}}]{ramon2020from}%
  \BibitemOpen
  \bibfield  {author} {\bibinfo {author} {\bibfnamefont {E.}~\bibnamefont {Prada}}, \bibinfo {author} {\bibfnamefont {P.}~\bibnamefont {San-Jose}}, \bibinfo {author} {\bibfnamefont {M.~W.~A.}\ \bibnamefont {de~Moor}}, \bibinfo {author} {\bibfnamefont {A.}~\bibnamefont {Geresdi}}, \bibinfo {author} {\bibfnamefont {E.~J.~H.}\ \bibnamefont {Lee}}, \bibinfo {author} {\bibfnamefont {J.}~\bibnamefont {Klinovaja}}, \bibinfo {author} {\bibfnamefont {D.}~\bibnamefont {Loss}}, \bibinfo {author} {\bibfnamefont {J.}~\bibnamefont {Nygård}}, \bibinfo {author} {\bibfnamefont {R.}~\bibnamefont {Aguado}},\ and\ \bibinfo {author} {\bibfnamefont {L.~P.}\ \bibnamefont {Kouwenhoven}},\ }\bibfield  {title} {\bibinfo {title} {From andreev to majorana bound states in hybrid superconductor–semiconductor nanowires},\ }\bibfield  {journal} {\bibinfo  {journal} {Nature Reviews Physics}\ }\href {https://doi.org/10.1038/s42254-020-0228-y} {10.1038/s42254-020-0228-y} (\bibinfo {year} {2020})\BibitemShut {NoStop}%
\bibitem [{\citenamefont {Cayao}\ and\ \citenamefont {Black-Schaffer}(2021)}]{Jorge2021distinguishing}%
  \BibitemOpen
  \bibfield  {author} {\bibinfo {author} {\bibfnamefont {J.}~\bibnamefont {Cayao}}\ and\ \bibinfo {author} {\bibfnamefont {A.~M.}\ \bibnamefont {Black-Schaffer}},\ }\bibfield  {title} {\bibinfo {title} {Distinguishing trivial and topological zero-energy states in long nanowire junctions},\ }\bibfield  {journal} {\bibinfo  {journal} {Physical Review B}\ }\textbf {\bibinfo {volume} {104}},\ \href {https://doi.org/10.1103/physrevb.104.l020501} {10.1103/physrevb.104.l020501} (\bibinfo {year} {2021})\BibitemShut {NoStop}%
\bibitem [{\citenamefont {Das~Sarma}\ \emph {et~al.}(2023)\citenamefont {Das~Sarma}, \citenamefont {Sau},\ and\ \citenamefont {Stanescu}}]{sarma2023spectral}%
  \BibitemOpen
  \bibfield  {author} {\bibinfo {author} {\bibfnamefont {S.}~\bibnamefont {Das~Sarma}}, \bibinfo {author} {\bibfnamefont {J.~D.}\ \bibnamefont {Sau}},\ and\ \bibinfo {author} {\bibfnamefont {T.~D.}\ \bibnamefont {Stanescu}},\ }\bibfield  {title} {\bibinfo {title} {Spectral properties, topological patches, and effective phase diagrams of finite disordered majorana nanowires},\ }\href {https://doi.org/10.1103/PhysRevB.108.085416} {\bibfield  {journal} {\bibinfo  {journal} {Phys. Rev. B}\ }\textbf {\bibinfo {volume} {108}},\ \bibinfo {pages} {085416} (\bibinfo {year} {2023})}\BibitemShut {NoStop}%
\bibitem [{\citenamefont {Stanescu}\ and\ \citenamefont {Das~Sarma}(2017)}]{PhysRevB.96.014510}%
  \BibitemOpen
  \bibfield  {author} {\bibinfo {author} {\bibfnamefont {T.~D.}\ \bibnamefont {Stanescu}}\ and\ \bibinfo {author} {\bibfnamefont {S.}~\bibnamefont {Das~Sarma}},\ }\bibfield  {title} {\bibinfo {title} {Proximity-induced low-energy renormalization in hybrid semiconductor-superconductor majorana structures},\ }\href {https://doi.org/10.1103/PhysRevB.96.014510} {\bibfield  {journal} {\bibinfo  {journal} {Phys. Rev. B}\ }\textbf {\bibinfo {volume} {96}},\ \bibinfo {pages} {014510} (\bibinfo {year} {2017})}\BibitemShut {NoStop}%
\bibitem [{\citenamefont {Stanescu}\ and\ \citenamefont {Das~Sarma}(2022)}]{PhysRevB.106.085429}%
  \BibitemOpen
  \bibfield  {author} {\bibinfo {author} {\bibfnamefont {T.~D.}\ \bibnamefont {Stanescu}}\ and\ \bibinfo {author} {\bibfnamefont {S.}~\bibnamefont {Das~Sarma}},\ }\bibfield  {title} {\bibinfo {title} {Proximity-induced superconductivity generated by thin films: Effects of fermi surface mismatch and disorder in the superconductor},\ }\href {https://doi.org/10.1103/PhysRevB.106.085429} {\bibfield  {journal} {\bibinfo  {journal} {Phys. Rev. B}\ }\textbf {\bibinfo {volume} {106}},\ \bibinfo {pages} {085429} (\bibinfo {year} {2022})}\BibitemShut {NoStop}%
\bibitem [{\citenamefont {Stanescu}\ \emph {et~al.}(2011)\citenamefont {Stanescu}, \citenamefont {Lutchyn},\ and\ \citenamefont {Das~Sarma}}]{PhysRevB.84.144522}%
  \BibitemOpen
  \bibfield  {author} {\bibinfo {author} {\bibfnamefont {T.~D.}\ \bibnamefont {Stanescu}}, \bibinfo {author} {\bibfnamefont {R.~M.}\ \bibnamefont {Lutchyn}},\ and\ \bibinfo {author} {\bibfnamefont {S.}~\bibnamefont {Das~Sarma}},\ }\bibfield  {title} {\bibinfo {title} {Majorana fermions in semiconductor nanowires},\ }\href {https://doi.org/10.1103/PhysRevB.84.144522} {\bibfield  {journal} {\bibinfo  {journal} {Phys. Rev. B}\ }\textbf {\bibinfo {volume} {84}},\ \bibinfo {pages} {144522} (\bibinfo {year} {2011})}\BibitemShut {NoStop}%
\bibitem [{\citenamefont {Sau}\ \emph {et~al.}(2010{\natexlab{c}})\citenamefont {Sau}, \citenamefont {Tewari}, \citenamefont {Lutchyn}, \citenamefont {Stanescu},\ and\ \citenamefont {Das~Sarma}}]{PhysRevB.82.214509}%
  \BibitemOpen
  \bibfield  {author} {\bibinfo {author} {\bibfnamefont {J.~D.}\ \bibnamefont {Sau}}, \bibinfo {author} {\bibfnamefont {S.}~\bibnamefont {Tewari}}, \bibinfo {author} {\bibfnamefont {R.~M.}\ \bibnamefont {Lutchyn}}, \bibinfo {author} {\bibfnamefont {T.~D.}\ \bibnamefont {Stanescu}},\ and\ \bibinfo {author} {\bibfnamefont {S.}~\bibnamefont {Das~Sarma}},\ }\bibfield  {title} {\bibinfo {title} {Non-abelian quantum order in spin-orbit-coupled semiconductors: Search for topological majorana particles in solid-state systems},\ }\href {https://doi.org/10.1103/PhysRevB.82.214509} {\bibfield  {journal} {\bibinfo  {journal} {Phys. Rev. B}\ }\textbf {\bibinfo {volume} {82}},\ \bibinfo {pages} {214509} (\bibinfo {year} {2010}{\natexlab{c}})}\BibitemShut {NoStop}%
\bibitem [{\citenamefont {Sau}\ \emph {et~al.}(2010{\natexlab{d}})\citenamefont {Sau}, \citenamefont {Lutchyn}, \citenamefont {Tewari},\ and\ \citenamefont {Das~Sarma}}]{PhysRevB.82.094522}%
  \BibitemOpen
  \bibfield  {author} {\bibinfo {author} {\bibfnamefont {J.~D.}\ \bibnamefont {Sau}}, \bibinfo {author} {\bibfnamefont {R.~M.}\ \bibnamefont {Lutchyn}}, \bibinfo {author} {\bibfnamefont {S.}~\bibnamefont {Tewari}},\ and\ \bibinfo {author} {\bibfnamefont {S.}~\bibnamefont {Das~Sarma}},\ }\bibfield  {title} {\bibinfo {title} {Robustness of majorana fermions in proximity-induced superconductors},\ }\href {https://doi.org/10.1103/PhysRevB.82.094522} {\bibfield  {journal} {\bibinfo  {journal} {Phys. Rev. B}\ }\textbf {\bibinfo {volume} {82}},\ \bibinfo {pages} {094522} (\bibinfo {year} {2010}{\natexlab{d}})}\BibitemShut {NoStop}%
\bibitem [{\citenamefont {Potter}\ and\ \citenamefont {Lee}(2011)}]{PhysRevB.83.094525}%
  \BibitemOpen
  \bibfield  {author} {\bibinfo {author} {\bibfnamefont {A.~C.}\ \bibnamefont {Potter}}\ and\ \bibinfo {author} {\bibfnamefont {P.~A.}\ \bibnamefont {Lee}},\ }\bibfield  {title} {\bibinfo {title} {Majorana end states in multiband microstructures with rashba spin-orbit coupling},\ }\href {https://doi.org/10.1103/PhysRevB.83.094525} {\bibfield  {journal} {\bibinfo  {journal} {Phys. Rev. B}\ }\textbf {\bibinfo {volume} {83}},\ \bibinfo {pages} {094525} (\bibinfo {year} {2011})}\BibitemShut {NoStop}%
\bibitem [{\citenamefont {Stanescu}\ \emph {et~al.}(2010)\citenamefont {Stanescu}, \citenamefont {Sau}, \citenamefont {Lutchyn},\ and\ \citenamefont {Das~Sarma}}]{PhysRevB.81.241310}%
  \BibitemOpen
  \bibfield  {author} {\bibinfo {author} {\bibfnamefont {T.~D.}\ \bibnamefont {Stanescu}}, \bibinfo {author} {\bibfnamefont {J.~D.}\ \bibnamefont {Sau}}, \bibinfo {author} {\bibfnamefont {R.~M.}\ \bibnamefont {Lutchyn}},\ and\ \bibinfo {author} {\bibfnamefont {S.}~\bibnamefont {Das~Sarma}},\ }\bibfield  {title} {\bibinfo {title} {Proximity effect at the superconductor--topological insulator interface},\ }\href {https://doi.org/10.1103/PhysRevB.81.241310} {\bibfield  {journal} {\bibinfo  {journal} {Phys. Rev. B}\ }\textbf {\bibinfo {volume} {81}},\ \bibinfo {pages} {241310} (\bibinfo {year} {2010})}\BibitemShut {NoStop}%
\bibitem [{\citenamefont {Khaymovich}\ \emph {et~al.}(2011)\citenamefont {Khaymovich}, \citenamefont {Chtchelkatchev},\ and\ \citenamefont {Vinokur}}]{PhysRevB.84.075142}%
  \BibitemOpen
  \bibfield  {author} {\bibinfo {author} {\bibfnamefont {I.~M.}\ \bibnamefont {Khaymovich}}, \bibinfo {author} {\bibfnamefont {N.~M.}\ \bibnamefont {Chtchelkatchev}},\ and\ \bibinfo {author} {\bibfnamefont {V.~M.}\ \bibnamefont {Vinokur}},\ }\bibfield  {title} {\bibinfo {title} {Instability of topological order and localization of edge states in hgte quantum wells coupled to $s$-wave superconductor},\ }\href {https://doi.org/10.1103/PhysRevB.84.075142} {\bibfield  {journal} {\bibinfo  {journal} {Phys. Rev. B}\ }\textbf {\bibinfo {volume} {84}},\ \bibinfo {pages} {075142} (\bibinfo {year} {2011})}\BibitemShut {NoStop}%
\bibitem [{\citenamefont {Grein}\ \emph {et~al.}(2012)\citenamefont {Grein}, \citenamefont {Michelsen},\ and\ \citenamefont {Eschrig}}]{Grein_2012}%
  \BibitemOpen
  \bibfield  {author} {\bibinfo {author} {\bibfnamefont {R.}~\bibnamefont {Grein}}, \bibinfo {author} {\bibfnamefont {J.}~\bibnamefont {Michelsen}},\ and\ \bibinfo {author} {\bibfnamefont {M.}~\bibnamefont {Eschrig}},\ }\bibfield  {title} {\bibinfo {title} {A numerical study of the superconducting proximity effect in topological surface states},\ }\href {https://doi.org/10.1088/1742-6596/391/1/012149} {\bibfield  {journal} {\bibinfo  {journal} {Journal of Physics: Conference Series}\ }\textbf {\bibinfo {volume} {391}},\ \bibinfo {pages} {012149} (\bibinfo {year} {2012})}\BibitemShut {NoStop}%
\bibitem [{\citenamefont {Sau}\ \emph {et~al.}(2012)\citenamefont {Sau}, \citenamefont {Tewari},\ and\ \citenamefont {Das~Sarma}}]{PhysRevB.85.064512}%
  \BibitemOpen
  \bibfield  {author} {\bibinfo {author} {\bibfnamefont {J.~D.}\ \bibnamefont {Sau}}, \bibinfo {author} {\bibfnamefont {S.}~\bibnamefont {Tewari}},\ and\ \bibinfo {author} {\bibfnamefont {S.}~\bibnamefont {Das~Sarma}},\ }\bibfield  {title} {\bibinfo {title} {Experimental and materials considerations for the topological superconducting state in electron- and hole-doped semiconductors: Searching for non-abelian majorana modes in 1d nanowires and 2d heterostructures},\ }\href {https://doi.org/10.1103/PhysRevB.85.064512} {\bibfield  {journal} {\bibinfo  {journal} {Phys. Rev. B}\ }\textbf {\bibinfo {volume} {85}},\ \bibinfo {pages} {064512} (\bibinfo {year} {2012})}\BibitemShut {NoStop}%
\bibitem [{\citenamefont {{MacKinnon}}(1985)}]{1985ZPhyB..59..385M}%
  \BibitemOpen
  \bibfield  {author} {\bibinfo {author} {\bibfnamefont {A.}~\bibnamefont {{MacKinnon}}},\ }\bibfield  {title} {\bibinfo {title} {{The calculation of transport properties and density of states of disordered solids}},\ }\href {https://doi.org/10.1007/BF01328846} {\bibfield  {journal} {\bibinfo  {journal} {Zeitschrift fur Physik B Condensed Matter}\ }\textbf {\bibinfo {volume} {59}},\ \bibinfo {pages} {385} (\bibinfo {year} {1985})}\BibitemShut {NoStop}%
\bibitem [{\citenamefont {Lewenkopf}\ and\ \citenamefont {Mucciolo}(2013)}]{lewenkopf2013recursive}%
  \BibitemOpen
  \bibfield  {author} {\bibinfo {author} {\bibfnamefont {C.~H.}\ \bibnamefont {Lewenkopf}}\ and\ \bibinfo {author} {\bibfnamefont {E.~R.}\ \bibnamefont {Mucciolo}},\ }\bibfield  {title} {\bibinfo {title} {The recursive green’s function method for graphene},\ }\href@noop {} {\bibfield  {journal} {\bibinfo  {journal} {Journal of Computational Electronics}\ }\textbf {\bibinfo {volume} {12}},\ \bibinfo {pages} {203} (\bibinfo {year} {2013})}\BibitemShut {NoStop}%
\bibitem [{\citenamefont {Zeng}\ \emph {et~al.}(2022)\citenamefont {Zeng}, \citenamefont {Sharma}, \citenamefont {Tewari},\ and\ \citenamefont {Stanescu}}]{PhysRevB.105.205122}%
  \BibitemOpen
  \bibfield  {author} {\bibinfo {author} {\bibfnamefont {C.}~\bibnamefont {Zeng}}, \bibinfo {author} {\bibfnamefont {G.}~\bibnamefont {Sharma}}, \bibinfo {author} {\bibfnamefont {S.}~\bibnamefont {Tewari}},\ and\ \bibinfo {author} {\bibfnamefont {T.}~\bibnamefont {Stanescu}},\ }\bibfield  {title} {\bibinfo {title} {Partially separated majorana modes in a disordered medium},\ }\href {https://doi.org/10.1103/PhysRevB.105.205122} {\bibfield  {journal} {\bibinfo  {journal} {Phys. Rev. B}\ }\textbf {\bibinfo {volume} {105}},\ \bibinfo {pages} {205122} (\bibinfo {year} {2022})}\BibitemShut {NoStop}%
\bibitem [{\citenamefont {Motrunich}\ \emph {et~al.}(2001)\citenamefont {Motrunich}, \citenamefont {Damle},\ and\ \citenamefont {Huse}}]{Motrunich2001}%
  \BibitemOpen
  \bibfield  {author} {\bibinfo {author} {\bibfnamefont {O.}~\bibnamefont {Motrunich}}, \bibinfo {author} {\bibfnamefont {K.}~\bibnamefont {Damle}},\ and\ \bibinfo {author} {\bibfnamefont {D.~A.}\ \bibnamefont {Huse}},\ }\bibfield  {title} {\bibinfo {title} {Griffiths effects and quantum critical points in dirty superconductors without spin-rotation invariance: One-dimensional examples},\ }\href {https://doi.org/10.1103/PhysRevB.63.224204} {\bibfield  {journal} {\bibinfo  {journal} {Phys. Rev. B}\ }\textbf {\bibinfo {volume} {63}},\ \bibinfo {pages} {224204} (\bibinfo {year} {2001})}\BibitemShut {NoStop}%
\bibitem [{\citenamefont {Huang}\ \emph {et~al.}(2018)\citenamefont {Huang}, \citenamefont {Pan}, \citenamefont {Liu}, \citenamefont {Sau}, \citenamefont {Stanescu},\ and\ \citenamefont {Das~Sarma}}]{Huang2018}%
  \BibitemOpen
  \bibfield  {author} {\bibinfo {author} {\bibfnamefont {Y.}~\bibnamefont {Huang}}, \bibinfo {author} {\bibfnamefont {H.}~\bibnamefont {Pan}}, \bibinfo {author} {\bibfnamefont {C.-X.}\ \bibnamefont {Liu}}, \bibinfo {author} {\bibfnamefont {J.~D.}\ \bibnamefont {Sau}}, \bibinfo {author} {\bibfnamefont {T.~D.}\ \bibnamefont {Stanescu}},\ and\ \bibinfo {author} {\bibfnamefont {S.}~\bibnamefont {Das~Sarma}},\ }\bibfield  {title} {\bibinfo {title} {Metamorphosis of andreev bound states into majorana bound states in pristine nanowires},\ }\href {https://doi.org/10.1103/PhysRevB.98.144511} {\bibfield  {journal} {\bibinfo  {journal} {Phys. Rev. B}\ }\textbf {\bibinfo {volume} {98}},\ \bibinfo {pages} {144511} (\bibinfo {year} {2018})}\BibitemShut {NoStop}%
\bibitem [{\citenamefont {Woods}\ \emph {et~al.}(2021)\citenamefont {Woods}, \citenamefont {Das~Sarma},\ and\ \citenamefont {Stanescu}}]{Woods2021}%
  \BibitemOpen
  \bibfield  {author} {\bibinfo {author} {\bibfnamefont {B.~D.}\ \bibnamefont {Woods}}, \bibinfo {author} {\bibfnamefont {S.}~\bibnamefont {Das~Sarma}},\ and\ \bibinfo {author} {\bibfnamefont {T.~D.}\ \bibnamefont {Stanescu}},\ }\bibfield  {title} {\bibinfo {title} {Charge-impurity effects in hybrid majorana nanowires},\ }\href {https://doi.org/10.1103/PhysRevApplied.16.054053} {\bibfield  {journal} {\bibinfo  {journal} {Phys. Rev. Appl.}\ }\textbf {\bibinfo {volume} {16}},\ \bibinfo {pages} {054053} (\bibinfo {year} {2021})}\BibitemShut {NoStop}%
\end{thebibliography}%
\end{document}